\documentclass[aps,twocolumn,superscriptaddress,floatfix,longbibliography]{revtex4-1}

\usepackage{graphicx}
\usepackage{amsmath,amssymb}
\usepackage{mathtools}
\usepackage[english]{babel}
\babelprovide[import=en]{English}
\usepackage{xr-hyper}
\usepackage[hidelinks]{hyperref}
\usepackage{color}
\usepackage{stmaryrd}
\usepackage{xcolor}
\usepackage{pifont}

\begin{document}

\title{Designing Strong and Broadband Nonreciprocal Thermal Radiation in Magnetic Topological Materials}

\author{Yiyang Jiang}
\affiliation{Department of Physics, The Pennsylvania State University, University Park, Pennsylvania 16802, USA}
\author{Yufei Zhao}
\affiliation{Department of Physics, The Pennsylvania State University, University Park, Pennsylvania 16802, USA}
\affiliation{Department of Condensed Matter Physics, Weizmann Institute of Science, Rehovot 7610001, Israel}
\author{Linxiao Zhu}
\affiliation{Department of Mechanical Engineering, The Pennsylvania State University, University Park, Pennsylvania 16802, USA}
\author{Binghai Yan$^{*,}$}
\affiliation{Department of Physics, The Pennsylvania State University, University Park, Pennsylvania 16802, USA}
\affiliation{Department of Condensed Matter Physics, Weizmann Institute of Science, Rehovot 7610001, Israel}
\affiliation{The Center for Theory of Emergent Quantum Matter, The Pennsylvania State University, University Park, Pennsylvania 16802, USA}
\date{\today}

\begin{abstract}

\noindent$^{*}$Corresponding author. Email: binghai.yan@psu.edu\par\smallskip

\textbf{ABSTRACT:} Breaking reciprocity in thermal radiation opens opportunities for energy harvesting, sensing, and thermal management. Traditional nonreciprocal radiative semiconductor devices need external magnetic field. In this work, we predict a series of magnetic topological materials for magnetic-field-free nonreciprocal thermal radiation in the infrared regime, by combining first-principles calculations with Maxwell electrodynamics. We find strong and broadband nonreciprocity in magnetic Weyl semimetals (e.g., Co$_3$Sn$_2$S$_2$), outperforming the conventional semiconductor such as InAs. Furthermore, we propose material design recipes: strong nonreciprocity requires a large anomalous Hall response relative to the optical loss, whereas the broadband response favors large optical loss and small dielectric dispersion. Our work establishes a predictive materials-discovery framework and quantitative design rules for next-generation magnet-free nonreciprocal thermal devices.


\end{abstract}
\maketitle

The ability to manipulate the absorption and emission of thermal radiation is central to energy harvesting, sensing, and thermal management~\cite{hadad2016breaking, siddharth2018thermodynamic,park2021reaching,zhang2022nonreciprocal,Mittapally2023,yang2024nonreciprocal}. Conventionally, Kirchhoff’s law~\cite{Kirchhoff1860On, Landsberg1980Thermodynamic,Siegel2001Thermal, Bergman2011Fundamentals} enforces a strict equality between directional absorptivity ($\alpha$) and emissivity ($e$), which constrains the efficiency of thermal photonic devices~\cite{liu2011taming,fan2017thermal,li2018nanophotonic}. Violating this constraint requires breaking the Lorentz reciprocity through Hall-type optical response induced by either external magnetic field or intrinsic magnetization~\cite{Ries1983Complete,Miller2017Universal,khandekar2020new,zhang2022nonreciprocal,yang2024nonreciprocal}.
Breaking this constraint enables nonreciprocal thermal radiation, opening opportunities for energy harvesting at thermodynamic limits, infrared camouflage, and advanced heat flux control ~\cite{green2012time,Buddhiraju2020Photonic,Park2022Nonreciprocal,zhang2022nonreciprocal}.

\begin{figure*}[tbp]
  \centering
  \includegraphics[width=0.7\textwidth]{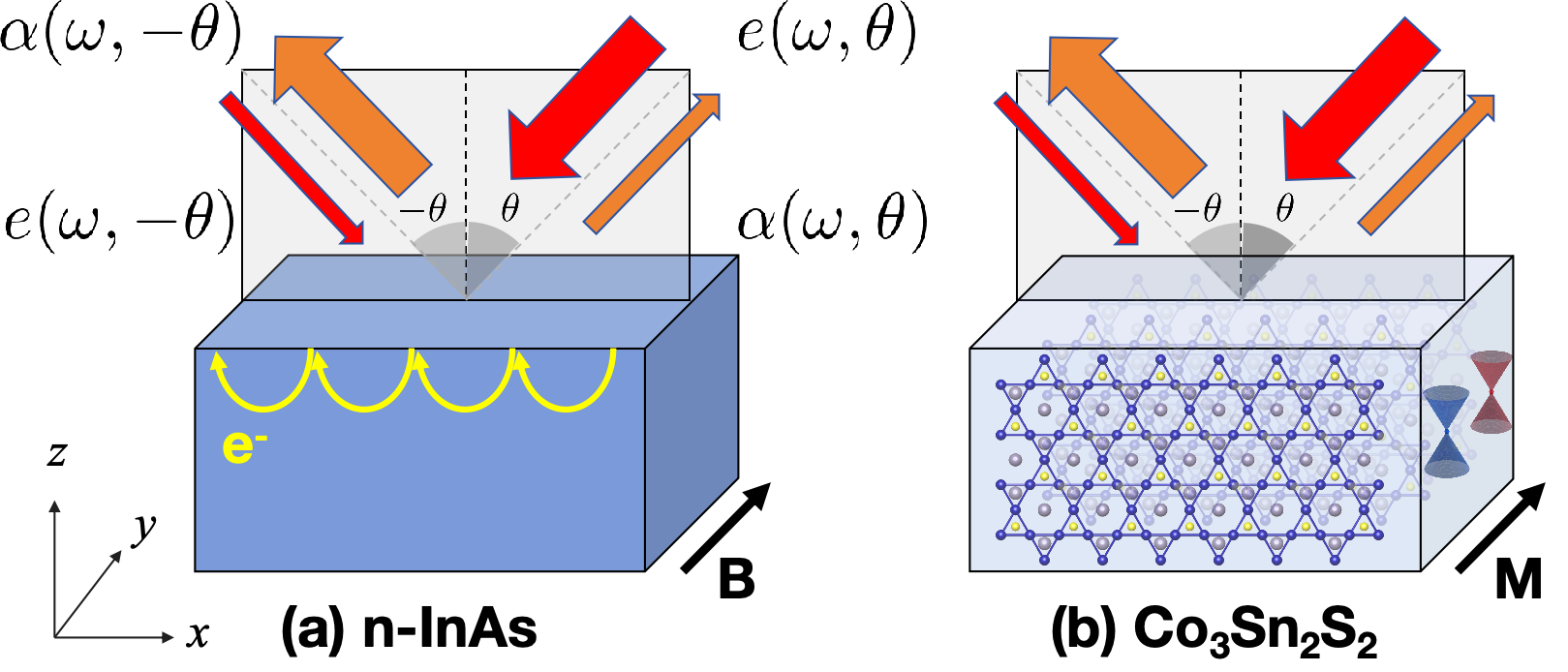}
  \caption{Intrinsic nonreciprocal thermal photonics in a magnetic topological material.
    (a)~Conventional platform: An $n$-type InAs slab requires an
    external magnetic field $\mathbf{B}$ to break time-reversal symmetry. Heuristically, the Hall-effect-induced boundary flow (yellow arrows) breaks the balance between directional absorptivity $\alpha(\omega,\theta)$ and emissivity $e(\omega,\theta)$.
    The resulting nonreciprocity,
    $\alpha(\omega,\theta) \neq e(\omega,\theta)$, is visualized by the varied arrow size.
    (b)~Current work: A magnetic Weyl semimetal, spontaneously breaks time-reversal symmetry via its intrinsic magnetization $\mathbf{M}$ arising from the topological Weyl fermion band structure. The anomalous Hall conductivity directly generates an off-diagonal dielectric component $\varepsilon_{xz} = -\varepsilon_{zx}$, producing the nonreciprocity without an external magnetic field.%
    }
  \label{fig:schematic}
\end{figure*}

Recent theory~\cite{zhu2014near,Wang2018Nonreciprocal,Zhao2019Near,ZHANG2020JQSRT, Liu2021Evolution,Wu2021Strong,Wu2021Nearcomplete,zhang2023broadband} and experiments~\cite{shayegan2022nonreciprocal,Liu2023Broadband,shayegan2023direct,shayegan2024broadband,zhang2025observation,nabavi2025high,pajovic2025nonreciprocal_reflection} have demonstrated nonreciprocal thermal emission and absorption. Current experimental realizations rely almost exclusively on semiconductors (e.g., InAs, Fig.~\ref{fig:schematic}(a)) driven by an external magnetic field~\cite{shayegan2023direct,shayegan2024broadband,zhang2025observation}, making practical implementation challenging. Moreover, the resulting nonreciprocity is often confined to narrow spectral windows, so substantial performance typically requires additional photonic structuring~\cite{shayegan2023direct,shayegan2024broadband,zhang2025observation}, further reducing the prospects for compact, broadband implementations. At the same time, existing theory has largely linked strong nonreciprocity near the plasma frequency region, or equivalently epsilon-near-zero (ENZ) region~\cite{Liu2021Evolution,Liu2023Broadband,zhang2023broadband}, but has not yet established a quantitative recipe for the nonreciprocity strength or a design principle for the operational bandwidth.

Inspired by their giant anomalous Hall effect (AHE), magnetic Weyl semimetals~\cite{Yan2017,Armitage2018} were recently proposed to realize nonreciprocal thermal radiation without requiring an external magnetic field~\cite{zhao2020axion, tsurimaki2020large, pajovic2020intrinsic, guo2020radiative}. These pioneering studies established Berry curvature as the intrinsic origin of nonreciprocity based on toy model Hamiltonian and simplified dielectric tensors, but have not yet predicted materials by considering realistic band structures and anisotropic optical properties.

In this Letter, we discovered strong nonreciprocal thermal radiation in magnetic topological materials by a first-principles predictive framework. We demonstrate that topological Weyl semimetals, exemplified by Co$_3$Sn$_2$S$_2$~\cite{liu2018giant}, can exhibit giant and broadband nonreciprocal thermal radiation in the mid-infrared range (Fig.~\ref{fig:schematic}(b)). Building on these results, we further develop an analytical theory that yields material design rules to optimize two key performance targets of nonreciprocal thermal radiation, the maximal nonreciprocity and operational bandwidth. These rules reveal how anomalous Hall response, optical loss, and dielectric dispersion govern the nonreciprocity, thereby providing quantitative guidance for realistic material discovery and device optimization.

To quantitatively predict intrinsic nonreciprocal thermal radiation in real materials, we developed an integrated DFT–Kubo–Maxwell framework beyond the approximations of effective medium or single-band models.

First, from the atomic structure of the material, we perform density functional theory (DFT) calculations including spin-orbit coupling (SOC) and magnetic order to obtain the ground-state electronic structure. We construct maximally localized Wannier functions to generate a tight-binding Hamiltonian $H(\mathbf{k})$ that accurately reproduces the multiband dispersion near the Fermi energy. \label{par:computational_pointer}Computational details are given in Sec.~\ref{SI-app:computational} of the Supporting Information.

Second, we compute the full frequency-dependent conductivity tensor $\sigma_{ij}(\omega)$ using the Kubo-Greenwood formula. The explicit expression is given by~\cite{komissarov2024quantum}
:
\begin{equation}
\begin{aligned}
    \sigma^{ab}(\omega) & = \frac{i}{\omega + i\eta} \int_{\mathbf{k}} \left( \sum_n f_n w_{nn}^{ab} + \sum_{n,m} \frac{f_{nm} v_{nm}^a v_{mn}^b}{\omega + \varepsilon_{nm} + i\eta} \right) \\
    & = \frac{i}{\omega + i\eta} \int_{\mathbf{k}} \left( \sum_n f_n D_n^{ab} + \sum_{n,m} \frac{ f_{nm} Q^{ab}_{nm} \varepsilon_{nm} }{\omega + \varepsilon_{nm} + i\eta} \right),
\end{aligned}
\label{eq:optical_conductivity}
\end{equation}
where, $\int_{\mathbf{k}} \equiv \int \frac{d^3 \mathbf{k}}{(2\pi)^3}$ denotes
the Brillouin-zone integration and $|u_n\rangle$ are the cell-periodic Bloch
states. The velocity and second-derivative matrix elements are
$v_{nm}^a = \langle u_n | \partial_{k_a} H(\mathbf{k}) | u_m \rangle$ and
$w_{nn}^{ab} = \langle u_n | \partial_{k_a}\partial_{k_b} H(\mathbf{k}) |
u_n \rangle$;
$f_n$ is the Fermi--Dirac distribution,
$f_{nm} \equiv f_n - f_m$, $\varepsilon_{nm} \equiv \varepsilon_n -
\varepsilon_m$, and $\eta=0.01~$eV is a phenomenological broadening
representing the inverse quasiparticle lifetime. \label{par:eta_thermal}The same $\eta$ is used for all materials so that their optical responses are compared at a common broadening. Thermal effects enter the optical response principally through this scattering rate: $\eta$ is the finite-temperature quasiparticle scattering rate set by phonon and disorder scattering at the operating temperature.
\label{par:magnetic_temperature}Here, we assume that the magnetic order persists at the operating temperature; thermal suppression of this order would reduce the anomalous Hall response and consequently weaken the nonreciprocity.

In the equivalent geometric form on the second line, $D_n^{ab} \equiv \partial_{k_a}\partial_{k_b}\varepsilon_n$ represents the Drude weight, and $Q^{ab}_{nm} \equiv r^a_{nm}\, r^b_{mn} = \mathcal{G}^{ab}_{nm} +
\tfrac{i}{2}\Omega^{ab}_{nm}$ (with $n \neq m$) represents the band-resolved quantum geometric tensor, where $r^a_{nm} = i\langle u_n|\partial_{k_a} u_m\rangle = v^a_{nm}/(i\varepsilon_{nm})$ is the interband Berry connection. Its symmetric (real) part $\mathcal{G}^{ab}_{nm}$ is the band-resolved quantum metric, and its antisymmetric (imaginary) part $\Omega^{ab}_{nm}$ is the band-resolved Berry curvature, the latter generating the off-diagonal components that drives nonreciprocal thermal radiation. When $\omega \rightarrow 0$, Eq.~\ref{eq:optical_conductivity} reduces to usual anomalous Hall conductivity that is proportional to Berry curvature. This captures both the intraband Drude response and the critical interband transitions driven by Berry curvature. The dielectric tensor is then obtained as
\begin{equation}
\varepsilon_{ij}(\omega) = \varepsilon_{\infty}\delta_{ij} + i\frac{\sigma_{ij}(\omega)}{\varepsilon_0 \omega},
\end{equation}
retaining all symmetry-allowed off-diagonal (Hall-type) components responsible for nonreciprocity.

Finally, we compute the angle-resolved nonreciprocity on the material surface by solving the Maxwell's equations. Here, we input the first-principles anisotropic permittivity tensor $\varepsilon_{ij}(\omega)$ into a generalized $4\times4$ Berreman matrix solver. This solver deals with Maxwell's equations for a stratified medium, accounting for arbitrary optical anisotropy and polarization mixing (see details in Sec.~\ref{SI-app:Maxwell} of the Supporting Information). The nonreciprocity is quantified by the contrast between the directional emissivity $e(\omega, \theta)$ and absorptivity $\alpha(\omega, \theta)$
\begin{equation}
  \Delta(\omega, \theta)
  \equiv
  e(\omega, \theta) - \alpha(\omega, \theta),
  \label{eq:delta_definition}
\end{equation}
which is the physically observable nonreciprocity of a thermal emitter.

\label{par:bare_scope}To distinguish material properties from effects of photonic structures, we compare all materials in the same bare, semi-infinite geometry. The resulting $\Delta_{\max}$ only applies to this geometry, rather than an optimized-device upper bound or an application-performance threshold.
Furthermore, we set the magnetization to the $y$-direction and calculate the nonreciprocity in $p$ (transverse-magnetic, TM) polarizations, because this combination gives the maximum nonreciprocity. In this case, $\Delta(\omega)$ is rigorously obtained from the $p$-polarized reflectivity at $\pm\theta$. In this work, we only considered bulk electronic states. \label{par:surface_estimate}For Co$_3$Sn$_2$S$_2$, surface states raise the density of states of the outermost unit cell to at most about three times the bulk value, but only within about $1$~nm, whereas mid-infrared light penetrates about $180$~nm ($670$ unit cells); their contribution to the nonreciprocity is therefore negligible (Sec.~\ref{SI-app:surface} of the Supporting Information), consistent with the small surface contribution found in an earlier model calculation~\cite{tsurimaki2020large}.

\begin{figure*}[tbp]
  \centering
  \includegraphics[width=0.85\textwidth]{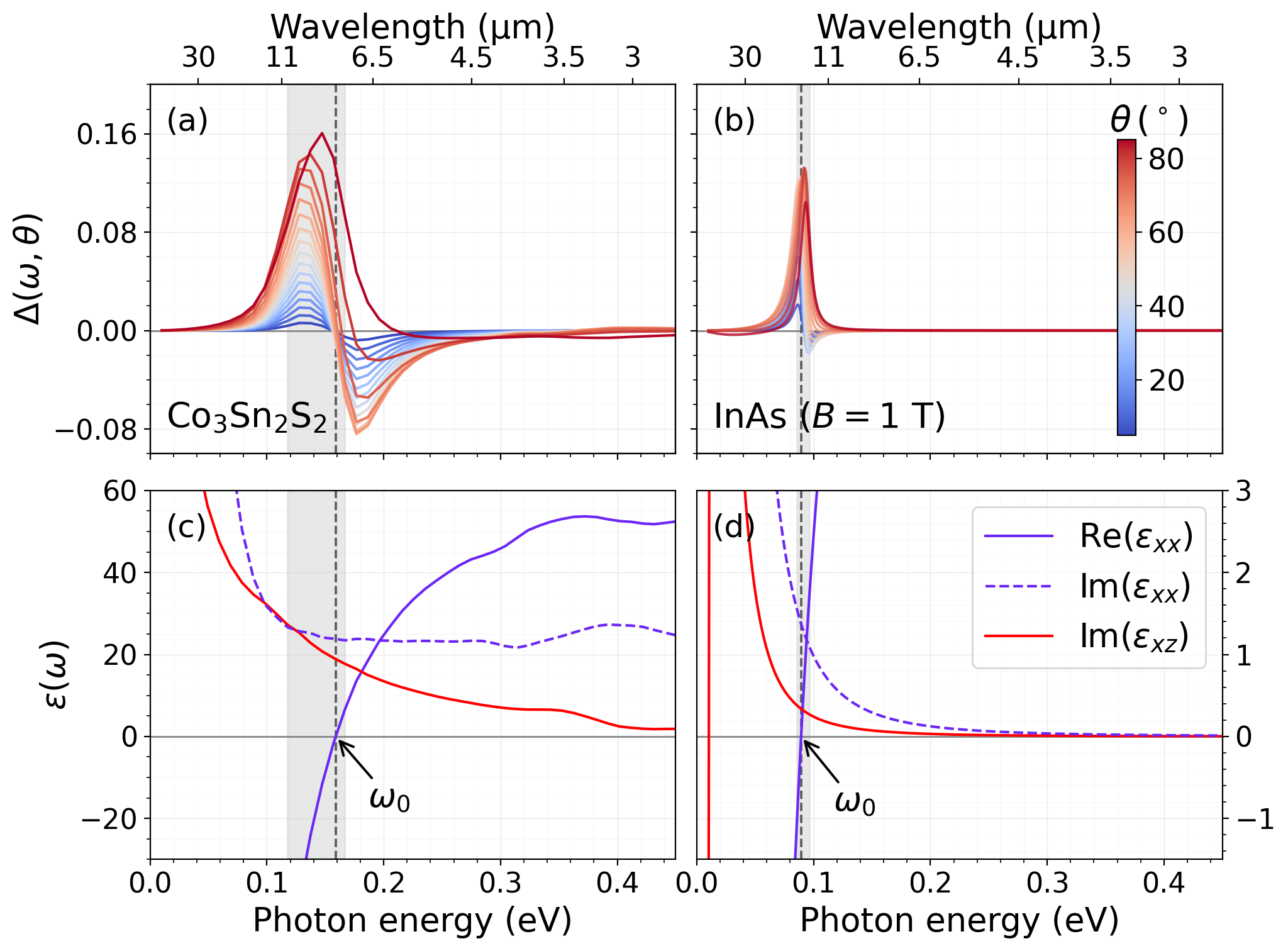}
\caption{Intrinsic nonreciprocity of magnetic Weyl semimetal Co$_3$Sn$_2$S$_2$ versus the conventional InAs semiconductor setup. (a) Angle- and frequency-resolved nonreciprocity $\Delta(\omega,\theta) = e(\omega,\theta) - \alpha(\omega,\theta)$ for Co$_3$Sn$_2$S$_2$. Colors encode incidence angle; the $x$-axis is the photon energy (or the equivalent wavelength of photons). (b) Same for $n$-doped InAs under $B = 1\,\mathrm{T}$, computed with the identical Berreman solver using experimental parameters~\cite{zhang2025observation, shayegan2024broadband}. (c,d) Corresponding dielectric functions for each material: solid blue lines show $\mathrm{Re}\,\varepsilon_{xx}(\omega)$, dashed blue lines show $\mathrm{Im}\,\varepsilon_{xx}(\omega)$ (optical loss), and red lines show the imaginary part of the off-diagonal (anomalous-Hall) component $\mathrm{Im}\,\varepsilon_{xz}(\omega)$. Dashed vertical lines mark the plasma frequency $\omega_0$, where $\mathrm{Re}\,\varepsilon_{xx}(\omega_0) = 0$, and
$\Delta(\omega,\theta)$ exhibits a peak at $\omega_0$. The shaded region marks the half-maximum interval defining $\Delta\omega$.
}
  \label{fig:Co3Sn2S2_vs_InAS}
\end{figure*}

We now apply the DFT--Kubo--Maxwell methods to study magnetic topological materials and benchmark them against the traditional InAs platform with an external magnetic field~\cite{zhang2025observation, shayegan2024broadband}. Throughout the Letter, our main figure of merit is the directional emissivity--absorptivity contrast spectrum $\Delta(\omega, \theta)$.
The magnitude of $\Delta(\omega)$ is between 0 and 1 because the emissivity $e$ and absorptivity $\alpha$ are all between 0 and 1.
To be more specific, we focus on the amplitude $\Delta_{\text{max}}$, which indicates the maximal nonreciprocity one material can reach in a semi-infinite plane platform, and the bandwidth $\Delta \omega$, \label{par:bandwidth_definition}defined as the frequency interval half of its maximum nonreciprocity.

Figure~\ref{fig:Co3Sn2S2_vs_InAS} shows $\Delta(\omega, \theta)$ of a magnetic Weyl semimetal Co$_3$Sn$_2$S$_2$~\cite{liu2018giant} at a certain Fermi energy ($\mu=-0.17$\,eV): the horizontal axis is the photon energy (or wavelength), the vertical axis is the $\Delta(\omega,\theta)$, sweeping through different angles. Figure~\ref{fig:Co3Sn2S2_vs_InAS}(b) shows the same quantity for our benchmark, $n$-doped InAs under $1\,\mathrm{T}$ magnetic field~\cite{zhang2025observation,shayegan2024broadband}.

Comparing the two spectra in Fig.~\ref{fig:Co3Sn2S2_vs_InAS}, Co$_3$Sn$_2$S$_2$ nonreciprocity is not only stronger in magnitude but also substantially broader in bandwidth than InAs. InAs exhibits a single narrow peak at epsilon-near-zero condition ($\Delta\omega \approx 0.011$~eV or $\Delta\lambda \approx 1.6~\mu$m around $\lambda_0 \approx 13.9~\mu$m), whereas Co$_3$Sn$_2$S$_2$ spans a wide mid-infrared window ($\Delta\omega \approx 0.049$~eV or $\Delta\lambda \approx 3.1~\mu$m covering $\lambda \approx 7.4\text{--}10.5~\mu$m) around $\lambda_0 \approx 7.8~\mu$m, a factor of $4.5$ broader, without an external magnetic field. For magneto-optical materials based on doped semiconductors, it usually needs gradient-doped multilayer to reach a bandwidth~\cite{zhang2025observation,shayegan2024broadband} comparable to Co$_3$Sn$_2$S$_2$.
\label{par:benchmark_pointer}Section~\ref{SI-app:benchmark} compares the nominal-composition longitudinal and Hall optical responses with published calculations and discusses discrepancies with measured spectra~\cite{okamura2020giant}. The working point $\mu=-0.17$~eV is a rigid-band extrapolation.

Inspecting their dielectric functions for both materials (Fig.~\ref{fig:Co3Sn2S2_vs_InAS}(c,d)), we identify that the $\Delta(\omega,\theta)$ peak lies at the plasma frequency or equivalently the epsilon-near-zero frequency ($\omega_0$) where the real part of the diagonal dielectric tensor crosses zero,
\begin{equation}
  \mathrm{Re}\,\varepsilon_{xx}(\omega_0) \;=\; 0.
  \label{eq:ENZ}
\end{equation}
The intuition is that the nonreciprocity is driven by the ratio of the asymmetric off-diagonal component ($\varepsilon_{xz}$) to the symmetric diagonal component, $|\varepsilon_{xz}| / |\mathrm{Re}\,\varepsilon_{xx}|$, which diverges at $\mathrm{Re}\,\varepsilon_{xx}\to 0$ at the plasma frequency and hence predicts a strong enhancement there\cite{zhao2020axion,yang2024nonreciprocal,Liu2023Broadband}.

The conventional recipe would select materials by maximizing the anomalous Hall response and chasing the $|\varepsilon_{xz}|/|\mathrm{Re}\, \varepsilon_{xx}|$ divergence at plasma frequency. Though it predicts the correct frequency $\omega_0$ for maximal nonreciprocity, it cannot quantitatively predict the nonreciprocal contrast. Then the only comparable parameter becomes the (anomalous) Hall strength $\text{Im} \varepsilon_{xz}$. However, $\text{Im} \varepsilon_{xz}$ itself cannot indicate the nonreciprocity between different materials. For example, though well-known antiferromagnetic Weyl semimetals, Mn$_3$Ge and Mn$_3$Sn~\cite{zhang2017strong}, exhibit large anomalous Hall response, our calculation shows only weak nonreciprocity about 1\% in Mn$_3$Ge and Mn$_3$Sn. In contrast, the Eu$_3$In$_2$As$_4$ family magnets \cite{zhao2024hybrid} have a smaller anomalous Hall effect than Mn$_3$Ge or Mn$_3$Sn but push $\Delta_{\max}$ to $\sim 35\%$ (FMa phase, $\mu=-0.1$\,eV. See details in Sec.~\ref{SI-app:material_parameter} of the Supporting Information).
The limitation of the conventional recipe motivated us to search for new descriptors, as described below.

\begin{figure*}[tbp]
    \centering
    \includegraphics[width=\linewidth]{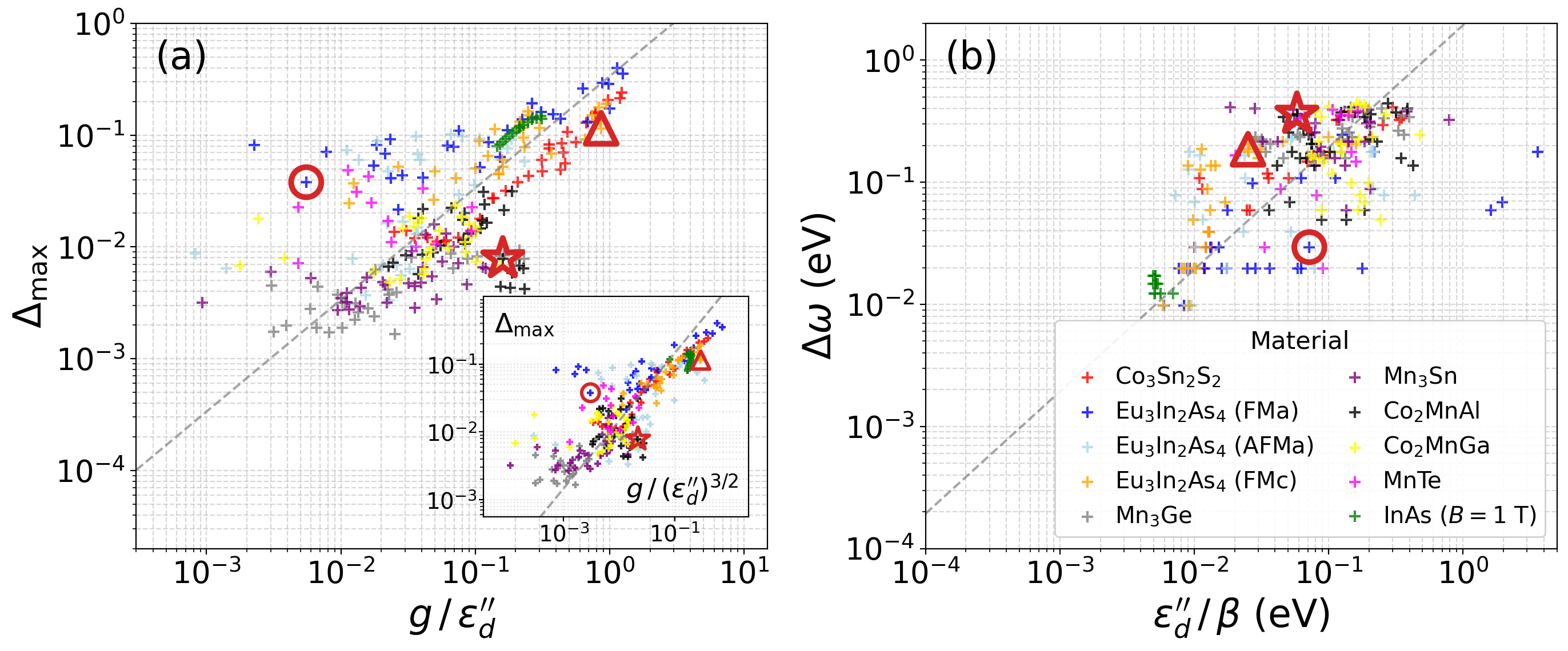}
    \caption{Scaling laws for intrinsic nonreciprocal thermal radiation, tested across all candidate magnetic topological materials and benchmarked against InAs under an external magnetic field ($B = 1$~T). (a)~Nonreciprocity strength scaling [Eq.~\eqref{eq:strength_scaling}]: peak nonreciprocal contrast $\Delta_{\max}$ vs.\ the anomalous Hall figure of merit $x = g/\varepsilon_{d}'' = \mathrm{Im}\,\varepsilon_{i\neq j}/\mathrm{Im}\,\varepsilon_{ii}$ evaluated at the plasma/ENZ frequency $\omega_0$, corresponding to the low-loss semiconductor limit ($p \to 1$) of the general scaling $\Delta_{\max}\propto g/(\varepsilon_d'')^{p}$ with $1\le p\le 3/2$. The bottom-right inset of (a) shows the same data re-plotted against the complementary high-loss metal limit ($p \to 3/2$), $g/(\varepsilon_d'')^{3/2}$, demonstrating that the two endpoints of the analytical bracket produce comparably tight collapses across the screened material set. (b)~Bandwidth scaling [Eq.~\eqref{eq:bandwidth_scaling}]: spectral bandwidth $\Delta\omega$ (eV) vs.\ the bandwidth figure of merit $\varepsilon_{d}''/\beta = \mathrm{Im}\,\varepsilon_{ii}/\frac{\partial \mathrm{Re}\,\varepsilon_{ii}}{\partial \omega}$ (eV) at $\omega_0$. In both main panels the dashed line indicates the predicted linear scaling, and data from structurally distinct materials follow it closely, identifying $x$ and $\varepsilon_{d}''/\beta$ as the leading material descriptors of nonreciprocity strength and operational bandwidth, respectively, within the conditions stated in the text. The external-field InAs point ($B=1$~T) is included only as a conventional benchmark and follows the same trend. Three representative outliers, analyzed in Sec.~\ref{SI-app:outliers} of the Supporting Information, are highlighted with red open symbols (and likewise in the $p=3/2$ inset): a circle ($\bigcirc$) for diagonal optical anisotropy [Eu$_3$In$_2$As$_4$ (FMa)], a triangle ($\triangle$) for multiple ENZ crossings [Eu$_3$In$_2$As$_4$ (FMc)], and a star (\ding{73}) for a real-part-dominated off-diagonal response [Co$_2$MnAl]. Full database including dielectric spectra is available at Ref.~\cite{nrtp_material_database_2026}.}
    \label{fig:scaling_laws}
\end{figure*}

To extract the fundamental material parameters that determine the intrinsic nonreciprocity, we use a model to represent a generic gyrotropic dielectric tensor,
\begin{equation}
  \varepsilon(\omega) = \begin{pmatrix} \varepsilon_d(\omega) & 0 & i g(\omega) \\ 0 & \varepsilon_d(\omega) & 0 \\ -ig(\omega) & 0 & \varepsilon_d(\omega) \end{pmatrix}
  \label{eq:gyrotropic}
\end{equation}
where the diagonal permittivity $\varepsilon_{d}(\omega) = \varepsilon_d'(\omega) + i\varepsilon_d''(\omega)$ and off-diagonal component $\varepsilon_{xz}(\omega) = ig(\omega)$, $g(\omega)$ stands for gyrotropy, equivalently the AHE response at optical frequency regime.
At the plasma frequency ($\omega_0$) when $\varepsilon_d'(\omega)$ crosses zero, we further simplify their frequency dependence as,
\begin{equation}
  \varepsilon_d'(\omega) = (\omega - \omega_0)\beta ,\ \varepsilon_d''(\omega) = \varepsilon_d''(\omega_0),\ g(\omega) = g(\omega_0)
  \label{eq:simplification}
\end{equation}
Here, $\beta$ represents the linear frequency-dependence of the diagonal dielectric tensor, which we call \emph{dielectric dispersion} at plasma frequency, since this quantity characterizes how fast the dielectric tensor changes, similar to optical dispersion that describes how refractive index change with respect to light frequency. $ \varepsilon_d''(\omega_0)$ and $g(\omega_0)$ represent the optical loss and (anomalous) Hall strength at plasma frequency. For simplicity, we omit the argument $\omega_0$ for the rest of the Letter without specific implication.

The nonreciprocity $\Delta$ can be shown to be factorized cleanly into independent shape factor $f(\omega)$ and angular dependent part $G(\theta)$ after simplification (see Sec.~\ref{SI-subsec:factorization_derivation} of the Supporting Information for the derivation and approximations),
\begin{equation}
  \Delta(\omega, \theta) \propto f(\omega) G(\theta),
  \label{eq:factorization}
\end{equation}
where $f(\omega) = -\mathrm{Re}(n_{v}^{2}\gamma^{*})/|n_{v}|^{4}$, $n_{v} = \sqrt{(\varepsilon_{d}^{2}-g^{2})/\varepsilon_{d}}$ is the effective Voigt refractive index~\cite{wallis1974surface}, $\gamma = ig/\varepsilon_{d}$ is the complex off-diagonal to diagonal ratio of the dielectric tensor, and $G(\theta) = \sin\theta\cos\theta/(\cos\theta+\kappa)^{2}$ with $\kappa = 2\,\mathrm{Re}(n_v)/|n_v|^2$ (strictly, $\kappa$ is itself frequency dependent; because the nonreciprocal feature is confined to a narrow window about the ENZ frequency $\omega_0$, we evaluate it there and write $G(\theta)\approx G(\kappa(\omega_0),\theta)$, while the explicit $\kappa$ dependence is retained in Sec.~\ref{SI-app:factorization} of the Supporting Information).

By analyzing the $\omega$ and $\theta$ dependence and the maximum value of $f(\omega)$ and $G(\theta)$ (see Sec.~\ref{SI-app:factorization} of the Supporting Information), we can obtain two approximate scaling laws for designing nonreciprocal thermophotonic materials:
\begin{equation}
  \Delta_{\max} \propto \frac{g}{(\varepsilon_{d}'')^{p}},
  \label{eq:strength_scaling}
\end{equation}
\begin{equation}
  \Delta\omega \propto \frac{\varepsilon_{d}''}{\beta}.
  \label{eq:bandwidth_scaling}
\end{equation}

Equations~\eqref{eq:strength_scaling} and~\eqref{eq:bandwidth_scaling} are the main results of our new material design principles. \label{par:scaling_assumptions}These laws assume $|\mathrm{Re}\,\varepsilon_{xz}|\ll|\mathrm{Im}\,\varepsilon_{xz}|$, $\varepsilon_{xx}\simeq\varepsilon_{zz}$, and a single ENZ crossing; the bandwidth proportionality further assumes $|g|/\varepsilon_d''\ll1$, where its dimensionless width factor approaches a constant (Sec.~\ref{SI-subsec:bandwidth} of the Supporting Information). Deviations are examined in Sec.~\ref{SI-app:outliers} of the Supporting Information. The strength law refines the older $|\varepsilon_{xz}|/|\mathrm{Re}\,\varepsilon_{xx}|$ figure of merit~\cite{zhao2020axion,yang2024nonreciprocal,Liu2023Broadband}: at the plasma frequency $\mathrm{Re}\,\varepsilon_{xx}$ vanishes, so the only finite dielectric scale left for the off-diagonal anomalous-Hall response $g$ to be measured against is the optical loss $\varepsilon_{d}''$, and the dimensionless ratio $ g/\varepsilon_{d}''$---through $\Delta_{\max}\propto g/(\varepsilon_{d}'')^{p}$---becomes the natural figure of merit for nonreciprocity strength; the exponent interpolates between the low-loss topological-semiconductor limit ($p\to1$, $\varepsilon_{d}''\ll1$) and the high-loss topological-metal limit ($p\to3/2$, $\varepsilon_{d}''\gg1$). The bandwidth law combines two independent broadening mechanisms: a small dielectric dispersion $\beta$ keeps the material near $\mathrm{Re}\,\varepsilon_{xx}=0$ over a wider spectral window, while larger optical loss broadens the resonance lineshape. Their ratio $\varepsilon_{d}''/\beta$, which is exactly the plasma linewidth~\cite{Raether1980}, determines the bandwidth of nonreciprocal radiation spectrum. These intuitive arguments are placed on rigorous mathematical footing in Sec.~\ref{SI-app:factorization} of the Supporting Information.

To benchmark these laws, we screen nine experimentally realized magnetic topological materials (phases): the kagome Weyl ferromagnet Co$_3$Sn$_2$S$_2$~\cite{liu2018giant}; three magnetic configurations (FMa/AFMa/FMc) of the hybrid topological semimetal Eu$_3$In$_2$As$_4$~\cite{song2024topotaxial, zhao2024hybrid}; the Heusler topological ferromagnets Co$_2$MnAl~\cite{Li2020Giant} and Co$_2$MnGa~\cite{Sakai2018bc,Guin2019Anomalous}; the noncollinear antiferromagnets Mn$_3$Ge and Mn$_3$Sn~\cite{nakatsuji2015large,nayak2016large,zhang2017strong}, which carry a large anomalous Hall conductivity despite a vanishing net magnetization; and the altermagnet MnTe~\cite{gonzalezbetancourt2023spontaneous, zhou2026surface}.
Throughout this work, we use the transverse (Voigt) geometry, with the magnetization or AHE parallel to surface and perpendicular to the plane of incidence, so that the gyrotropic off-diagonal $\varepsilon_{xz}$ drives a nonreciprocity in TM polarization.
For example, the kagome plane of Co$_3$Sn$_2$S$_2$ lies in the xz plane (see Fig. 1b) while the kagome plane of Mn$_3$Ge or Mn$_3$Sn aligns in the xy plane.
For each material (phase) we compute the full dielectric tensor with varied chemical potential $\mu$ across a $\pm0.5$\,eV window about charge neutrality. Sweeping $\mu$ provides both experimental doping advice and independent $(g/\varepsilon_{d}'',\Delta_{\max})$ and $(\varepsilon_{d}''/\beta,\Delta\omega)$ data to examine our scaling laws. \label{par:mu_window}For Co$_3$Sn$_2$S$_2$, $\Delta_{\max}=0.11$--$0.22$ within $\pm50$~meV of the working point (Sec.~\ref{SI-app:material_parameter} of the Supporting Information), and experiments have shifted its Fermi level by about $-150$~meV (In substitution) and $+200$~meV (Li-ion gating)~\cite{zhou2020enhanced,lohani2023electronic,matsuoka2025electron}. Figure~\ref{fig:scaling_laws} shows the outcome: on both the strength and bandwidth panels, most data points fall on the gray dashed slope-one reference line, confirming Eqs.~\eqref{eq:strength_scaling} and~\eqref{eq:bandwidth_scaling} as material design rules.

The fit is best for the InAs benchmark and Co$_3$Sn$_2$S$_2$, whose off-diagonal response is dominated by its imaginary part ($|\mathrm{Re}\,\varepsilon_{xz}|\ll|\mathrm{Im}\,\varepsilon_{xz}|$) and corresponds to the reduced model (Eq.~\ref{eq:gyrotropic}), which is related to the breakdown of our assumptions in special cases.

These design rules rationalize the material-by-material hierarchy
established by the first-principles results discussed above. Equation~\eqref{eq:strength_scaling} immediately explains why the celebrated large-anomalous-Hall magnets
Mn$_{3}$Ge and Mn$_{3}$Sn~\cite{zhang2017strong} fail to deliver large
$\Delta_{\max}$ despite their gyrotropy ($g$): their strongly metallic
interband background drives $\varepsilon_{d}''(\omega_{0}) \gtrsim 60$,
collapsing the dimensionless ratio $g/\varepsilon_{d}''$ that
actually sets the strength and capping $\Delta_{\max} \lesssim 2\%$.
Co$_2$MnGa and Co$_2$MnAl are similar to the case of Mn$_3$Ge/Mn$_3$Sn.
The same rule identifies Eu$_{3}$In$_{2}$As$_{4}$ \cite{zhao2024hybrid} as the strength leader of our material set: a
substantial Berry-curvature-driven gyrotropy combined with a
comparatively moderate $\varepsilon_{d}''$ at $\omega_{0}$ enhance the ratio
$g/\varepsilon_{d}''$, predicting the largest $\Delta_{\max} \sim 35\%$ (see Sec.~\ref{SI-app:material_parameter} of the Supporting Information).

Regarding the bandwidth,
Eq.~\eqref{eq:bandwidth_scaling} rationalizes the striking observation
that nearly all intrinsic magnetic materials studied here exceed the InAs benchmark regardless of Fermi-level position. As a low-loss semiconductor, InAs carries $\varepsilon_{d}''(\omega_{0}) \lesssim 7$ even under $B = 1$\,T, fixing
the bandwidth figure of merit at
$\varepsilon_{d}''/\beta \approx 5 \times 10^{-3}$\,eV across every
carrier density and capping $\Delta\omega \lesssim 0.02$\,eV. The
intrinsic magnetic semimetals/metals carry $\varepsilon_{d}''$ of order
$10$--$10^{2}$ at $\omega_{0}$, raising $\varepsilon_{d}''/\beta$ by one
to two decades and broadening the spectral window accordingly. This
inverts the conventional photonic intuition: within the
nonreciprocal-emission framework, large optical loss is a bandwidth
\emph{resource}, traded against strength through the loss-dependent scaling of Eq.~\eqref{eq:strength_scaling}.

\label{par:tradeoff}Equations~\eqref{eq:strength_scaling} and~\eqref{eq:bandwidth_scaling} indicate the dilemma between nonreciprocal strength and bandwidth. The former favors low optical loss (absorption) while the latter prefers high loss. Therefore, metals with high density of states (e.g., Mn$_3$Sn or Co$_2$MnAl) show large $\Delta \omega$ but small $\Delta_{max}$. In contrast, low carrier semiconductors (e.g., InAs and MnTe) exhibit moderate $\Delta_{max}$ but very small $\Delta \omega$. Therefore, materials with moderate carrier density and large AHE (e.g. Co$_3$Sn$_2$S$_2$) are promising candidates for both strong and broadband emission within the stated model conditions. The product $\Delta_{\max}\Delta\omega\propto(g/\beta)(\varepsilon_d'')^{1-p}$ quantifies the loss trade-off in the weak-gyrotropy limit. Figure~\ref{SI-fig:si_tradeoff} of the Supporting Information shows this trade-off in the strength--bandwidth plane: for both Co$_3$Sn$_2$S$_2$ and InAs, $\Delta_{\max}$ decreases as $\Delta\omega$ increases across chemical potentials, while Co$_3$Sn$_2$S$_2$ extends to much broader bandwidths. It is worth noting that these recipes apply to both conventional semiconductors and magnetic anomalous Hall materials.

Some datapoints deviate from the scaling law, as marked by red open symbols in Fig.~\ref{fig:scaling_laws} and detailed in Sec.~\ref{SI-app:outliers} of the Supporting Information.
Three mechanisms are at play. (i) We assumed the off-diagonal $\varepsilon_{xz}$ or $ig$ imaginary in Eq.~\ref{eq:gyrotropic}. When $|\mathrm{Re}\,\varepsilon_{xz}|\gg|\mathrm{Im}\,\varepsilon_{xz}|$ , however, the resulting $\Delta_{max}$ is smaller and $\Delta \omega$ is larger than the scaling relations from Eqs.~\ref{eq:strength_scaling} and \ref{eq:bandwidth_scaling}, such as Co$_2$MnAl/Co$_2$MnGa. We attribute this to material's metallic nature since $|\mathrm{Re}\,\varepsilon_{xz}|$ is dissipative while $|\mathrm{Im}\,\varepsilon_{xz}|$ is non-dissipative.
(ii) We assumed an isotropic dielectric function in Eq.~\ref{eq:gyrotropic}. When diagonal anisotropy is strong (e.g., $\varepsilon_{xx}\neq\varepsilon_{zz}$), the scaling relations can break down (e.g., Eu$_3$In$_2$As$_4$ FMa).
(iii) We assumed only one ENZ crossing in Eq.~\ref{eq:gyrotropic}, yet multiple $\mathrm{Re}\,\varepsilon_{xx}=0$ crossings (e.g.\ Eu$_3$In$_2$As$_4$ FMc) could superpose resonances and broaden the
bandwidth. Additionally, materials whose diagonal permittivity never crosses zero in the energy window investigated (e.g.\ MnTe at high hole doping) have no plasma frequency and thus no reduced-model coordinate.

\label{par:outlook}In summary, our work establishes a quantitative, first-principles framework for nonreciprocal thermal radiation in magnetic topological materials, built on two co-equal contributions: predictive DFT--Kubo--Maxwell calculations that identify Co$_3$Sn$_2$S$_2$ and Eu$_3$In$_2$As$_4$ as materials outperforming the InAs benchmark in the nonreciprocity strength, and closed-form analytical design rules that convert first-principles trends into material-selection criteria for both nonreciprocity strength and operational bandwidth. We bridge the gap between toy models and realistic materials discovery by developing a predictive tool for intrinsic nonreciprocal thermal radiation. For emitters involving magnetic topological materials, photonic structures such as gratings~\cite{zhao2020axion}, guided-mode resonators~\cite{wu2025midinfrared}, and thin films~\cite{Wu2021Nearcomplete,zhang2023broadband}\ could be used to further enhance the nonreciprocity contrast. Therefore, the material figures of merit $g/\epsilon_d''$ and $\varepsilon_d''/\beta$ derived here allow for pre-screening candidates of magnetic topological materials for nonreciprocal thermal emission, which can be used with photonic structures to further enhance the nonreciprocity contrast.

\section*{Supporting Information}
$4\times4$ Berreman matrix method for bianisotropic and nonreciprocal media; tabulated material parameters and chemical-potential dependence; surface-state contribution; permittivity model for the InAs benchmark; analytical derivations of the factorization, material figure of merit, and scaling laws; outliers of the scaling laws; strength--bandwidth trade-off; computational details; comparison with published optical responses; bare-interface contrast, application metrics, and photonic structuring (PDF)

\begin{acknowledgements}

We thank helpful discussions with Chaoxing Liu, Kamal Das, Hengxin Tan, Pu Xiao, You-Chiuan Chen, Yiqi Zhao and Rui Jiang. B.Y. acknowledges the financial support by the Penn State Materials Research Science and Engineering Center for Nanoscale Science (MRSEC) under National Science Foundation (NSF) award DMR-2011839 and by the NSF DMREF Award 2522898. L.Z. acknowledges support from NSF award CBET-2238927.

\end{acknowledgements}

\renewcommand{\bibsection}{\section*{References}}%
%\bibliography{reference_photonics,references-chiral,nonlinear-literature,nonlinear-reference}

\begin{thebibliography}{60}%
\makeatletter
\providecommand \@ifxundefined [1]{%
 \@ifx{#1\undefined}
}%
\providecommand \@ifnum [1]{%
 \ifnum #1\expandafter \@firstoftwo
 \else \expandafter \@secondoftwo
 \fi
}%
\providecommand \@ifx [1]{%
 \ifx #1\expandafter \@firstoftwo
 \else \expandafter \@secondoftwo
 \fi
}%
\providecommand \natexlab [1]{#1}%
\providecommand \enquote  [1]{``#1''}%
\providecommand \bibnamefont  [1]{#1}%
\providecommand \bibfnamefont [1]{#1}%
\providecommand \citenamefont [1]{#1}%
\providecommand \href@noop [0]{\@secondoftwo}%
\providecommand \href [0]{\begingroup \@sanitize@url \@href}%
\providecommand \@href[1]{\@@startlink{#1}\@@href}%
\providecommand \@@href[1]{\endgroup#1\@@endlink}%
\providecommand \@sanitize@url [0]{\catcode `\\12\catcode `\$12\catcode
  `\&12\catcode `\#12\catcode `\^12\catcode `\_12\catcode `\%12\relax}%
\providecommand \@@startlink[1]{}%
\providecommand \@@endlink[0]{}%
\providecommand \url  [0]{\begingroup\@sanitize@url \@url }%
\providecommand \@url [1]{\endgroup\@href {#1}{\urlprefix }}%
\providecommand \urlprefix  [0]{URL }%
\providecommand \Eprint [0]{\href }%
\providecommand \doibase [0]{http://dx.doi.org/}%
\providecommand \selectlanguage [0]{\@gobble}%
\providecommand \bibinfo  [0]{\@secondoftwo}%
\providecommand \bibfield  [0]{\@secondoftwo}%
\providecommand \translation [1]{[#1]}%
\providecommand \BibitemOpen [0]{}%
\providecommand \bibitemStop [0]{}%
\providecommand \bibitemNoStop [0]{.\EOS\space}%
\providecommand \EOS [0]{\spacefactor3000\relax}%
\providecommand \BibitemShut  [1]{\csname bibitem#1\endcsname}%
\let\auto@bib@innerbib\@empty
%</preamble>
\bibitem [{\citenamefont {Hadad}\ \emph {et~al.}(2016)\citenamefont {Hadad},
  \citenamefont {Soric},\ and\ \citenamefont {Alu}}]{hadad2016breaking}%
  \BibitemOpen
  \bibfield  {author} {\bibinfo {author} {\bibfnamefont {Yakir}\ \bibnamefont
  {Hadad}}, \bibinfo {author} {\bibfnamefont {Jason~C.}\ \bibnamefont {Soric}},
  \ and\ \bibinfo {author} {\bibfnamefont {Andrea}\ \bibnamefont {Alu}},\
  }\bibfield  {title} {\enquote {\bibinfo {title} {Breaking temporal symmetries
  for emission and absorption},}\ }\href {\doibase 10.1073/pnas.1517363113}
  {\bibfield  {journal} {\bibinfo  {journal} {Proceedings of the National
  Academy of Sciences}\ }\textbf {\bibinfo {volume} {113}},\ \bibinfo {pages}
  {3471--3475} (\bibinfo {year} {2016})}\BibitemShut {NoStop}%
\bibitem [{\citenamefont {Buddhiraju}\ \emph {et~al.}(2018)\citenamefont
  {Buddhiraju}, \citenamefont {Santhanam},\ and\ \citenamefont
  {Fan}}]{siddharth2018thermodynamic}%
  \BibitemOpen
  \bibfield  {author} {\bibinfo {author} {\bibfnamefont {Siddharth}\
  \bibnamefont {Buddhiraju}}, \bibinfo {author} {\bibfnamefont {Parthiban}\
  \bibnamefont {Santhanam}}, \ and\ \bibinfo {author} {\bibfnamefont {Shanhui}\
  \bibnamefont {Fan}},\ }\bibfield  {title} {\enquote {\bibinfo {title}
  {Thermodynamic limits of energy harvesting from outgoing thermal
  radiation},}\ }\href {\doibase 10.1073/pnas.1717595115} {\bibfield  {journal}
  {\bibinfo  {journal} {Proceedings of the National Academy of Sciences}\
  }\textbf {\bibinfo {volume} {115}},\ \bibinfo {pages} {E3609--E3615}
  (\bibinfo {year} {2018})}\BibitemShut {NoStop}%
\bibitem [{\citenamefont {Park}\ \emph
  {et~al.}(2022{\natexlab{a}})\citenamefont {Park}, \citenamefont {Zhao},\ and\
  \citenamefont {Fan}}]{park2021reaching}%
  \BibitemOpen
  \bibfield  {author} {\bibinfo {author} {\bibfnamefont {Yubin}\ \bibnamefont
  {Park}}, \bibinfo {author} {\bibfnamefont {Bo}~\bibnamefont {Zhao}}, \ and\
  \bibinfo {author} {\bibfnamefont {Shanhui}\ \bibnamefont {Fan}},\ }\bibfield
  {title} {\enquote {\bibinfo {title} {Reaching the ultimate efficiency of
  solar energy harvesting with a nonreciprocal multijunction solar cell},}\
  }\href {\doibase 10.1021/acs.nanolett.1c04288} {\bibfield  {journal}
  {\bibinfo  {journal} {Nano Letters}\ }\textbf {\bibinfo {volume} {22}},\
  \bibinfo {pages} {448--452} (\bibinfo {year}
  {2022}{\natexlab{a}})}\BibitemShut {NoStop}%
\bibitem [{\citenamefont {Zhang}\ and\ \citenamefont
  {Zhu}(2022)}]{zhang2022nonreciprocal}%
  \BibitemOpen
  \bibfield  {author} {\bibinfo {author} {\bibfnamefont {Zhenong}\ \bibnamefont
  {Zhang}}\ and\ \bibinfo {author} {\bibfnamefont {Linxiao}\ \bibnamefont
  {Zhu}},\ }\bibfield  {title} {\enquote {\bibinfo {title} {Nonreciprocal
  thermal photonics for energy conversion and radiative heat transfer},}\
  }\href {\doibase 10.1103/PhysRevApplied.18.027001} {\bibfield  {journal}
  {\bibinfo  {journal} {Phys. Rev. Appl.}\ }\textbf {\bibinfo {volume} {18}},\
  \bibinfo {pages} {027001} (\bibinfo {year} {2022})}\BibitemShut {NoStop}%
\bibitem [{\citenamefont {Mittapally}\ \emph {et~al.}(2023)\citenamefont
  {Mittapally}, \citenamefont {Majumder}, \citenamefont {Reddy},\ and\
  \citenamefont {Meyhofer}}]{Mittapally2023}%
  \BibitemOpen
  \bibfield  {author} {\bibinfo {author} {\bibfnamefont {Rohith}\ \bibnamefont
  {Mittapally}}, \bibinfo {author} {\bibfnamefont {Ayan}\ \bibnamefont
  {Majumder}}, \bibinfo {author} {\bibfnamefont {Pramod}\ \bibnamefont
  {Reddy}}, \ and\ \bibinfo {author} {\bibfnamefont {Edgar}\ \bibnamefont
  {Meyhofer}},\ }\bibfield  {title} {\enquote {\bibinfo {title} {Near-field
  thermophotovoltaic energy conversion: Progress and opportunities},}\ }\href
  {\doibase 10.1103/PhysRevApplied.19.037002} {\bibfield  {journal} {\bibinfo
  {journal} {Phys. Rev. Appl.}\ }\textbf {\bibinfo {volume} {19}},\ \bibinfo
  {pages} {037002} (\bibinfo {year} {2023})}\BibitemShut {NoStop}%
\bibitem [{\citenamefont {Yang}\ \emph {et~al.}(2024)\citenamefont {Yang},
  \citenamefont {Liu}, \citenamefont {Zhao}, \citenamefont {Fan},\ and\
  \citenamefont {Qiu}}]{yang2024nonreciprocal}%
  \BibitemOpen
  \bibfield  {author} {\bibinfo {author} {\bibfnamefont {Shuihua}\ \bibnamefont
  {Yang}}, \bibinfo {author} {\bibfnamefont {Mengqi}\ \bibnamefont {Liu}},
  \bibinfo {author} {\bibfnamefont {Changying}\ \bibnamefont {Zhao}}, \bibinfo
  {author} {\bibfnamefont {Shanhui}\ \bibnamefont {Fan}}, \ and\ \bibinfo
  {author} {\bibfnamefont {Cheng-Wei}\ \bibnamefont {Qiu}},\ }\bibfield
  {title} {\enquote {\bibinfo {title} {Nonreciprocal thermal photonics},}\
  }\href {\doibase 10.1038/s41566-024-01409-y} {\bibfield  {journal} {\bibinfo
  {journal} {Nature Photonics}\ }\textbf {\bibinfo {volume} {18}},\ \bibinfo
  {pages} {412--424} (\bibinfo {year} {2024})}\BibitemShut {NoStop}%
\bibitem [{\citenamefont {Kirchhoff}(1860)}]{Kirchhoff1860On}%
  \BibitemOpen
  \bibfield  {author} {\bibinfo {author} {\bibfnamefont {G.}~\bibnamefont
  {Kirchhoff}},\ }\bibfield  {title} {\enquote {\bibinfo {title} {On the
  relation between the radiating and absorbing powers of different bodies for
  light and heat},}\ }\href {\doibase 10.1080/14786446008642901} {\bibfield
  {journal} {\bibinfo  {journal} {Philosophical Magazine and Journal of
  Science}\ }\textbf {\bibinfo {volume} {20}},\ \bibinfo {pages} {1--21}
  (\bibinfo {year} {1860})}\BibitemShut {NoStop}%
\bibitem [{\citenamefont {Landsberg}\ and\ \citenamefont
  {Tonge}(1980)}]{Landsberg1980Thermodynamic}%
  \BibitemOpen
  \bibfield  {author} {\bibinfo {author} {\bibfnamefont {P.~T.}\ \bibnamefont
  {Landsberg}}\ and\ \bibinfo {author} {\bibfnamefont {G.}~\bibnamefont
  {Tonge}},\ }\bibfield  {title} {\enquote {\bibinfo {title} {Thermodynamic
  energy conversion efficiencies},}\ }\href {\doibase 10.1063/1.328187}
  {\bibfield  {journal} {\bibinfo  {journal} {Journal of Applied Physics}\
  }\textbf {\bibinfo {volume} {51}},\ \bibinfo {pages} {R1--R20} (\bibinfo
  {year} {1980})}\BibitemShut {NoStop}%
\bibitem [{\citenamefont {Siegel}\ and\ \citenamefont
  {Howell}(2001)}]{Siegel2001Thermal}%
  \BibitemOpen
  \bibfield  {author} {\bibinfo {author} {\bibfnamefont {Robert}\ \bibnamefont
  {Siegel}}\ and\ \bibinfo {author} {\bibfnamefont {John~R.}\ \bibnamefont
  {Howell}},\ }\href@noop {} {\emph {\bibinfo {title} {Thermal Radiation Heat
  Transfer}}},\ \bibinfo {edition} {4th}\ ed.\ (\bibinfo  {publisher} {Taylor
  \& Francis},\ \bibinfo {year} {2001})\BibitemShut {NoStop}%
\bibitem [{\citenamefont {Bergman}\ \emph {et~al.}(2011)\citenamefont
  {Bergman}, \citenamefont {Lavine}, \citenamefont {Incropera},\ and\
  \citenamefont {DeWitt}}]{Bergman2011Fundamentals}%
  \BibitemOpen
  \bibfield  {author} {\bibinfo {author} {\bibfnamefont {Theodore~L.}\
  \bibnamefont {Bergman}}, \bibinfo {author} {\bibfnamefont {Adrienne~S.}\
  \bibnamefont {Lavine}}, \bibinfo {author} {\bibfnamefont {Frank~P.}\
  \bibnamefont {Incropera}}, \ and\ \bibinfo {author} {\bibfnamefont
  {David~P.}\ \bibnamefont {DeWitt}},\ }\href@noop {} {\emph {\bibinfo {title}
  {Fundamentals of Heat and Mass Transfer}}},\ \bibinfo {edition} {7th}\ ed.\
  (\bibinfo  {publisher} {John Wiley \& Sons},\ \bibinfo {year}
  {2011})\BibitemShut {NoStop}%
\bibitem [{\citenamefont {Liu}\ \emph {et~al.}(2011)\citenamefont {Liu},
  \citenamefont {Tyler}, \citenamefont {Starr}, \citenamefont {Starr},
  \citenamefont {Jokerst},\ and\ \citenamefont {Padilla}}]{liu2011taming}%
  \BibitemOpen
  \bibfield  {author} {\bibinfo {author} {\bibfnamefont {Xianliang}\
  \bibnamefont {Liu}}, \bibinfo {author} {\bibfnamefont {Talmage}\ \bibnamefont
  {Tyler}}, \bibinfo {author} {\bibfnamefont {Tatiana}\ \bibnamefont {Starr}},
  \bibinfo {author} {\bibfnamefont {Anthony~F.}\ \bibnamefont {Starr}},
  \bibinfo {author} {\bibfnamefont {Nan~Marie}\ \bibnamefont {Jokerst}}, \ and\
  \bibinfo {author} {\bibfnamefont {Willie~J.}\ \bibnamefont {Padilla}},\
  }\bibfield  {title} {\enquote {\bibinfo {title} {Taming the blackbody with
  infrared metamaterials as selective thermal emitters},}\ }\href {\doibase
  10.1103/PhysRevLett.107.045901} {\bibfield  {journal} {\bibinfo  {journal}
  {Phys. Rev. Lett.}\ }\textbf {\bibinfo {volume} {107}},\ \bibinfo {pages}
  {045901} (\bibinfo {year} {2011})}\BibitemShut {NoStop}%
\bibitem [{\citenamefont {Fan}(2017)}]{fan2017thermal}%
  \BibitemOpen
  \bibfield  {author} {\bibinfo {author} {\bibfnamefont {Shanhui}\ \bibnamefont
  {Fan}},\ }\bibfield  {title} {\enquote {\bibinfo {title} {Thermal photonics
  and energy applications},}\ }\href {\doibase 10.1016/j.joule.2017.07.012}
  {\bibfield  {journal} {\bibinfo  {journal} {Joule}\ }\textbf {\bibinfo
  {volume} {1}},\ \bibinfo {pages} {264--273} (\bibinfo {year}
  {2017})}\BibitemShut {NoStop}%
\bibitem [{\citenamefont {Li}\ and\ \citenamefont
  {Fan}(2018)}]{li2018nanophotonic}%
  \BibitemOpen
  \bibfield  {author} {\bibinfo {author} {\bibfnamefont {Wei}\ \bibnamefont
  {Li}}\ and\ \bibinfo {author} {\bibfnamefont {Shanhui}\ \bibnamefont {Fan}},\
  }\bibfield  {title} {\enquote {\bibinfo {title} {Nanophotonic control of
  thermal radiation for energy applications (invited)},}\ }\href {\doibase
  10.1364/OE.26.015995} {\bibfield  {journal} {\bibinfo  {journal} {Opt.
  Express}\ }\textbf {\bibinfo {volume} {26}},\ \bibinfo {pages} {15995--16021}
  (\bibinfo {year} {2018})}\BibitemShut {NoStop}%
\bibitem [{\citenamefont {Ries}(1983)}]{Ries1983Complete}%
  \BibitemOpen
  \bibfield  {author} {\bibinfo {author} {\bibfnamefont {H.}~\bibnamefont
  {Ries}},\ }\bibfield  {title} {\enquote {\bibinfo {title} {Complete and
  reversible absorption of radiation},}\ }\href {\doibase 10.1007/BF00688821}
  {\bibfield  {journal} {\bibinfo  {journal} {Applied Physics B}\ }\textbf
  {\bibinfo {volume} {32}},\ \bibinfo {pages} {153--156} (\bibinfo {year}
  {1983})}\BibitemShut {NoStop}%
\bibitem [{\citenamefont {Miller}\ \emph {et~al.}(2017)\citenamefont {Miller},
  \citenamefont {Zhu},\ and\ \citenamefont {Fan}}]{Miller2017Universal}%
  \BibitemOpen
  \bibfield  {author} {\bibinfo {author} {\bibfnamefont {D.~A.~B.}\
  \bibnamefont {Miller}}, \bibinfo {author} {\bibfnamefont {L.}~\bibnamefont
  {Zhu}}, \ and\ \bibinfo {author} {\bibfnamefont {S.}~\bibnamefont {Fan}},\
  }\bibfield  {title} {\enquote {\bibinfo {title} {Universal modal radiation
  laws for all thermal emitters},}\ }\href {\doibase 10.1073/pnas.1701606114}
  {\bibfield  {journal} {\bibinfo  {journal} {Proceedings of the National
  Academy of Sciences}\ }\textbf {\bibinfo {volume} {114}},\ \bibinfo {pages}
  {4336--4341} (\bibinfo {year} {2017})}\BibitemShut {NoStop}%
\bibitem [{\citenamefont {Khandekar}\ \emph {et~al.}(2020)\citenamefont
  {Khandekar}, \citenamefont {Khosravi}, \citenamefont {Li},\ and\
  \citenamefont {Jacob}}]{khandekar2020new}%
  \BibitemOpen
  \bibfield  {author} {\bibinfo {author} {\bibfnamefont {Chinmay}\ \bibnamefont
  {Khandekar}}, \bibinfo {author} {\bibfnamefont {Farhad}\ \bibnamefont
  {Khosravi}}, \bibinfo {author} {\bibfnamefont {Zhou}\ \bibnamefont {Li}}, \
  and\ \bibinfo {author} {\bibfnamefont {Zubin}\ \bibnamefont {Jacob}},\
  }\bibfield  {title} {\enquote {\bibinfo {title} {New spin-resolved thermal
  radiation laws for nonreciprocal bianisotropic media},}\ }\href {\doibase
  10.1088/1367-2630/abc988} {\bibfield  {journal} {\bibinfo  {journal} {New
  Journal of Physics}\ }\textbf {\bibinfo {volume} {22}},\ \bibinfo {pages}
  {123005} (\bibinfo {year} {2020})}\BibitemShut {NoStop}%
\bibitem [{\citenamefont {Green}(2012)}]{green2012time}%
  \BibitemOpen
  \bibfield  {author} {\bibinfo {author} {\bibfnamefont {Martin~A.}\
  \bibnamefont {Green}},\ }\bibfield  {title} {\enquote {\bibinfo {title}
  {Time-asymmetric photovoltaics},}\ }\href {\doibase 10.1021/nl3034784}
  {\bibfield  {journal} {\bibinfo  {journal} {Nano Letters}\ }\textbf {\bibinfo
  {volume} {12}},\ \bibinfo {pages} {5985--5988} (\bibinfo {year}
  {2012})}\BibitemShut {NoStop}%
\bibitem [{\citenamefont {Buddhiraju}\ \emph {et~al.}(2020)\citenamefont
  {Buddhiraju}, \citenamefont {Li},\ and\ \citenamefont
  {Fan}}]{Buddhiraju2020Photonic}%
  \BibitemOpen
  \bibfield  {author} {\bibinfo {author} {\bibfnamefont {Siddharth}\
  \bibnamefont {Buddhiraju}}, \bibinfo {author} {\bibfnamefont {Wei}\
  \bibnamefont {Li}}, \ and\ \bibinfo {author} {\bibfnamefont {Shanhui}\
  \bibnamefont {Fan}},\ }\bibfield  {title} {\enquote {\bibinfo {title}
  {Photonic refrigeration from time-modulated thermal emission},}\ }\href
  {\doibase 10.1103/PhysRevLett.124.077402} {\bibfield  {journal} {\bibinfo
  {journal} {Physical Review Letters}\ }\textbf {\bibinfo {volume} {124}},\
  \bibinfo {pages} {077402} (\bibinfo {year} {2020})}\BibitemShut {NoStop}%
\bibitem [{\citenamefont {Park}\ \emph
  {et~al.}(2022{\natexlab{b}})\citenamefont {Park}, \citenamefont {Omair},\
  and\ \citenamefont {Fan}}]{Park2022Nonreciprocal}%
  \BibitemOpen
  \bibfield  {author} {\bibinfo {author} {\bibfnamefont {Y.}~\bibnamefont
  {Park}}, \bibinfo {author} {\bibfnamefont {Z.}~\bibnamefont {Omair}}, \ and\
  \bibinfo {author} {\bibfnamefont {S.}~\bibnamefont {Fan}},\ }\bibfield
  {title} {\enquote {\bibinfo {title} {Nonreciprocal thermophotovoltaic
  systems},}\ }\href {\doibase 10.1021/acsphotonics.2c01308} {\bibfield
  {journal} {\bibinfo  {journal} {ACS Photonics}\ }\textbf {\bibinfo {volume}
  {9}},\ \bibinfo {pages} {3943--3949} (\bibinfo {year}
  {2022}{\natexlab{b}})}\BibitemShut {NoStop}%
\bibitem [{\citenamefont {Zhu}\ and\ \citenamefont {Fan}(2014)}]{zhu2014near}%
  \BibitemOpen
  \bibfield  {author} {\bibinfo {author} {\bibfnamefont {Linxiao}\ \bibnamefont
  {Zhu}}\ and\ \bibinfo {author} {\bibfnamefont {Shanhui}\ \bibnamefont
  {Fan}},\ }\bibfield  {title} {\enquote {\bibinfo {title} {Near-complete
  violation of detailed balance in thermal radiation},}\ }\href {\doibase
  10.1103/PhysRevB.90.220301} {\bibfield  {journal} {\bibinfo  {journal} {Phys.
  Rev. B}\ }\textbf {\bibinfo {volume} {90}},\ \bibinfo {pages} {220301}
  (\bibinfo {year} {2014})}\BibitemShut {NoStop}%
\bibitem [{\citenamefont {Wang}\ \emph {et~al.}(2018)\citenamefont {Wang},
  \citenamefont {Wu},\ and\ \citenamefont {Zhou}}]{Wang2018Nonreciprocal}%
  \BibitemOpen
  \bibfield  {author} {\bibinfo {author} {\bibfnamefont {H.}~\bibnamefont
  {Wang}}, \bibinfo {author} {\bibfnamefont {H.}~\bibnamefont {Wu}}, \ and\
  \bibinfo {author} {\bibfnamefont {J.~Q.}\ \bibnamefont {Zhou}},\ }\bibfield
  {title} {\enquote {\bibinfo {title} {Nonreciprocal optical properties based
  on magneto-optical materials: n-{InAs}, {GaAs} and {HgCdTe}},}\ }\href
  {\doibase 10.1016/j.jqsrt.2017.11.015} {\bibfield  {journal} {\bibinfo
  {journal} {Journal of Quantitative Spectroscopy and Radiative Transfer}\
  }\textbf {\bibinfo {volume} {206}},\ \bibinfo {pages} {254--263} (\bibinfo
  {year} {2018})}\BibitemShut {NoStop}%
\bibitem [{\citenamefont {Zhao}\ \emph {et~al.}(2019)\citenamefont {Zhao},
  \citenamefont {Shi}, \citenamefont {Wang}, \citenamefont {Zhao},
  \citenamefont {Zhao},\ and\ \citenamefont {Fan}}]{Zhao2019Near}%
  \BibitemOpen
  \bibfield  {author} {\bibinfo {author} {\bibfnamefont {Bo}~\bibnamefont
  {Zhao}}, \bibinfo {author} {\bibfnamefont {Yu}~\bibnamefont {Shi}}, \bibinfo
  {author} {\bibfnamefont {Jiahui}\ \bibnamefont {Wang}}, \bibinfo {author}
  {\bibfnamefont {Zhexin}\ \bibnamefont {Zhao}}, \bibinfo {author}
  {\bibfnamefont {Nathan}\ \bibnamefont {Zhao}}, \ and\ \bibinfo {author}
  {\bibfnamefont {Shanhui}\ \bibnamefont {Fan}},\ }\bibfield  {title} {\enquote
  {\bibinfo {title} {Near-complete violation of {Kirchhoff}'s law of thermal
  radiation with a 0.3 {T} magnetic field},}\ }\href {\doibase
  10.1364/OL.44.004203} {\bibfield  {journal} {\bibinfo  {journal} {Optics
  Letters}\ }\textbf {\bibinfo {volume} {44}},\ \bibinfo {pages} {4203--4206}
  (\bibinfo {year} {2019})}\BibitemShut {NoStop}%
\bibitem [{\citenamefont {Zhang}\ \emph {et~al.}(2020)\citenamefont {Zhang},
  \citenamefont {Wu},\ and\ \citenamefont {Fu}}]{ZHANG2020JQSRT}%
  \BibitemOpen
  \bibfield  {author} {\bibinfo {author} {\bibfnamefont {Zhuomin~M.}\
  \bibnamefont {Zhang}}, \bibinfo {author} {\bibfnamefont {Xiaohu}\
  \bibnamefont {Wu}}, \ and\ \bibinfo {author} {\bibfnamefont {Ceji}\
  \bibnamefont {Fu}},\ }\bibfield  {title} {\enquote {\bibinfo {title}
  {Validity of {Kirchhoff}'s law for semitransparent films made of anisotropic
  materials},}\ }\href {\doibase https://doi.org/10.1016/j.jqsrt.2020.106904}
  {\bibfield  {journal} {\bibinfo  {journal} {Journal of Quantitative
  Spectroscopy and Radiative Transfer}\ }\textbf {\bibinfo {volume} {245}},\
  \bibinfo {pages} {106904} (\bibinfo {year} {2020})}\BibitemShut {NoStop}%
\bibitem [{\citenamefont {Liu}\ \emph {et~al.}(2021)\citenamefont {Liu},
  \citenamefont {Zhao}, \citenamefont {Zeng}, \citenamefont {Chen},
  \citenamefont {Zhao},\ and\ \citenamefont {Qiu}}]{Liu2021Evolution}%
  \BibitemOpen
  \bibfield  {author} {\bibinfo {author} {\bibfnamefont {Mengqi}\ \bibnamefont
  {Liu}}, \bibinfo {author} {\bibfnamefont {Chen}\ \bibnamefont {Zhao}},
  \bibinfo {author} {\bibfnamefont {Yixuan}\ \bibnamefont {Zeng}}, \bibinfo
  {author} {\bibfnamefont {Yang}\ \bibnamefont {Chen}}, \bibinfo {author}
  {\bibfnamefont {Changying}\ \bibnamefont {Zhao}}, \ and\ \bibinfo {author}
  {\bibfnamefont {Cheng-Wei}\ \bibnamefont {Qiu}},\ }\bibfield  {title}
  {\enquote {\bibinfo {title} {Evolution and nonreciprocity of loss-induced
  topological phase singularity pairs},}\ }\href {\doibase
  10.1103/PhysRevLett.127.266101} {\bibfield  {journal} {\bibinfo  {journal}
  {Physical Review Letters}\ }\textbf {\bibinfo {volume} {127}},\ \bibinfo
  {pages} {266101} (\bibinfo {year} {2021})}\BibitemShut {NoStop}%
\bibitem [{\citenamefont {Wu}\ \emph {et~al.}(2021{\natexlab{a}})\citenamefont
  {Wu}, \citenamefont {Liu}, \citenamefont {Yu},\ and\ \citenamefont
  {Wu}}]{Wu2021Strong}%
  \BibitemOpen
  \bibfield  {author} {\bibinfo {author} {\bibfnamefont {X.~H.}\ \bibnamefont
  {Wu}}, \bibinfo {author} {\bibfnamefont {R.~Y.}\ \bibnamefont {Liu}},
  \bibinfo {author} {\bibfnamefont {H.~Y.}\ \bibnamefont {Yu}}, \ and\ \bibinfo
  {author} {\bibfnamefont {B.~Y.}\ \bibnamefont {Wu}},\ }\bibfield  {title}
  {\enquote {\bibinfo {title} {Strong non-reciprocal radiation in
  magnetophotonic crystals},}\ }\href {\doibase 10.1016/j.jqsrt.2021.107794}
  {\bibfield  {journal} {\bibinfo  {journal} {Journal of Quantitative
  Spectroscopy and Radiative Transfer}\ }\textbf {\bibinfo {volume} {272}},\
  \bibinfo {pages} {107794} (\bibinfo {year} {2021}{\natexlab{a}})}\BibitemShut
  {NoStop}%
\bibitem [{\citenamefont {Wu}\ \emph {et~al.}(2021{\natexlab{b}})\citenamefont
  {Wu}, \citenamefont {Wang}, \citenamefont {Zhai}, \citenamefont {Shi},
  \citenamefont {Wu},\ and\ \citenamefont {Wu}}]{Wu2021Nearcomplete}%
  \BibitemOpen
  \bibfield  {author} {\bibinfo {author} {\bibfnamefont {Jun}\ \bibnamefont
  {Wu}}, \bibinfo {author} {\bibfnamefont {Zhongmin}\ \bibnamefont {Wang}},
  \bibinfo {author} {\bibfnamefont {Han}\ \bibnamefont {Zhai}}, \bibinfo
  {author} {\bibfnamefont {Zhangxing}\ \bibnamefont {Shi}}, \bibinfo {author}
  {\bibfnamefont {Xiaohu}\ \bibnamefont {Wu}}, \ and\ \bibinfo {author}
  {\bibfnamefont {Feng}\ \bibnamefont {Wu}},\ }\bibfield  {title} {\enquote
  {\bibinfo {title} {Near-complete violation of {Kirchhoff}'s law of thermal
  radiation in ultrathin magnetic {Weyl} semimetal films},}\ }\href {\doibase
  10.1364/OME.444308} {\bibfield  {journal} {\bibinfo  {journal} {Opt. Mater.
  Express}\ }\textbf {\bibinfo {volume} {11}},\ \bibinfo {pages} {4058--4066}
  (\bibinfo {year} {2021}{\natexlab{b}})}\BibitemShut {NoStop}%
\bibitem [{\citenamefont {Zhang}\ and\ \citenamefont
  {Zhu}(2023)}]{zhang2023broadband}%
  \BibitemOpen
  \bibfield  {author} {\bibinfo {author} {\bibfnamefont {Zhenong}\ \bibnamefont
  {Zhang}}\ and\ \bibinfo {author} {\bibfnamefont {Linxiao}\ \bibnamefont
  {Zhu}},\ }\bibfield  {title} {\enquote {\bibinfo {title} {Broadband
  nonreciprocal thermal emission},}\ }\href {\doibase
  10.1103/PhysRevApplied.19.014013} {\bibfield  {journal} {\bibinfo  {journal}
  {Phys. Rev. Appl.}\ }\textbf {\bibinfo {volume} {19}},\ \bibinfo {pages}
  {014013} (\bibinfo {year} {2023})}\BibitemShut {NoStop}%
\bibitem [{\citenamefont {Shayegan}\ \emph {et~al.}(2022)\citenamefont
  {Shayegan}, \citenamefont {Zhao}, \citenamefont {Kim}, \citenamefont {Fan},\
  and\ \citenamefont {Atwater}}]{shayegan2022nonreciprocal}%
  \BibitemOpen
  \bibfield  {author} {\bibinfo {author} {\bibfnamefont {K.~J.}\ \bibnamefont
  {Shayegan}}, \bibinfo {author} {\bibfnamefont {B.}~\bibnamefont {Zhao}},
  \bibinfo {author} {\bibfnamefont {Y.}~\bibnamefont {Kim}}, \bibinfo {author}
  {\bibfnamefont {S.}~\bibnamefont {Fan}}, \ and\ \bibinfo {author}
  {\bibfnamefont {H.~A.}\ \bibnamefont {Atwater}},\ }\bibfield  {title}
  {\enquote {\bibinfo {title} {Nonreciprocal infrared absorption via resonant
  magneto-optical coupling to {InAs}},}\ }\href {\doibase
  10.1126/sciadv.abm4308} {\bibfield  {journal} {\bibinfo  {journal} {Science
  Advances}\ }\textbf {\bibinfo {volume} {8}},\ \bibinfo {pages} {eabm4308}
  (\bibinfo {year} {2022})}\BibitemShut {NoStop}%
\bibitem [{\citenamefont {Liu}\ \emph {et~al.}(2023)\citenamefont {Liu},
  \citenamefont {Xia}, \citenamefont {Wan}, \citenamefont {Qin}, \citenamefont
  {Li}, \citenamefont {Zhao}, \citenamefont {Bi},\ and\ \citenamefont
  {Qiu}}]{Liu2023Broadband}%
  \BibitemOpen
  \bibfield  {author} {\bibinfo {author} {\bibfnamefont {Mengqi}\ \bibnamefont
  {Liu}}, \bibinfo {author} {\bibfnamefont {Shuang}\ \bibnamefont {Xia}},
  \bibinfo {author} {\bibfnamefont {Wenjian}\ \bibnamefont {Wan}}, \bibinfo
  {author} {\bibfnamefont {Jun}\ \bibnamefont {Qin}}, \bibinfo {author}
  {\bibfnamefont {Hua}\ \bibnamefont {Li}}, \bibinfo {author} {\bibfnamefont
  {Changying}\ \bibnamefont {Zhao}}, \bibinfo {author} {\bibfnamefont {Lei}\
  \bibnamefont {Bi}}, \ and\ \bibinfo {author} {\bibfnamefont {Cheng-Wei}\
  \bibnamefont {Qiu}},\ }\bibfield  {title} {\enquote {\bibinfo {title}
  {Broadband mid-infrared non-reciprocal absorption using magnetized gradient
  epsilon-near-zero thin films},}\ }\href {\doibase 10.1038/s41563-023-01635-9}
  {\bibfield  {journal} {\bibinfo  {journal} {Nature Materials}\ }\textbf
  {\bibinfo {volume} {22}},\ \bibinfo {pages} {1196--1202} (\bibinfo {year}
  {2023})}\BibitemShut {NoStop}%
\bibitem [{\citenamefont {Shayegan}\ \emph {et~al.}(2023)\citenamefont
  {Shayegan}, \citenamefont {Biswas}, \citenamefont {Zhao}, \citenamefont
  {Fan},\ and\ \citenamefont {Atwater}}]{shayegan2023direct}%
  \BibitemOpen
  \bibfield  {author} {\bibinfo {author} {\bibfnamefont {Komron~J.}\
  \bibnamefont {Shayegan}}, \bibinfo {author} {\bibfnamefont {Souvik}\
  \bibnamefont {Biswas}}, \bibinfo {author} {\bibfnamefont {Bo}~\bibnamefont
  {Zhao}}, \bibinfo {author} {\bibfnamefont {Shanhui}\ \bibnamefont {Fan}}, \
  and\ \bibinfo {author} {\bibfnamefont {Harry~A.}\ \bibnamefont {Atwater}},\
  }\bibfield  {title} {\enquote {\bibinfo {title} {Direct observation of the
  violation of {Kirchhoff}'s law of thermal radiation},}\ }\href {\doibase
  10.1038/s41566-023-01261-6} {\bibfield  {journal} {\bibinfo  {journal}
  {Nature Photonics}\ }\textbf {\bibinfo {volume} {17}},\ \bibinfo {pages}
  {891--896} (\bibinfo {year} {2023})}\BibitemShut {NoStop}%
\bibitem [{\citenamefont {Shayegan}\ \emph {et~al.}(2024)\citenamefont
  {Shayegan}, \citenamefont {Hwang}, \citenamefont {Zhao}, \citenamefont
  {Raman},\ and\ \citenamefont {Atwater}}]{shayegan2024broadband}%
  \BibitemOpen
  \bibfield  {author} {\bibinfo {author} {\bibfnamefont {Komron~J.}\
  \bibnamefont {Shayegan}}, \bibinfo {author} {\bibfnamefont {Jae~S.}\
  \bibnamefont {Hwang}}, \bibinfo {author} {\bibfnamefont {Bo}~\bibnamefont
  {Zhao}}, \bibinfo {author} {\bibfnamefont {Aaswath~P.}\ \bibnamefont
  {Raman}}, \ and\ \bibinfo {author} {\bibfnamefont {Harry~A.}\ \bibnamefont
  {Atwater}},\ }\bibfield  {title} {\enquote {\bibinfo {title} {Broadband
  nonreciprocal thermal emissivity and absorptivity},}\ }\href {\doibase
  10.1038/s41377-024-01520-3} {\bibfield  {journal} {\bibinfo  {journal}
  {Light: Science \& Applications}\ }\textbf {\bibinfo {volume} {13}},\
  \bibinfo {pages} {176} (\bibinfo {year} {2024})}\BibitemShut {NoStop}%
\bibitem [{\citenamefont {Zhang}\ \emph {et~al.}(2025)\citenamefont {Zhang},
  \citenamefont {Kalantari~Dehaghi}, \citenamefont {Ghosh},\ and\ \citenamefont
  {Zhu}}]{zhang2025observation}%
  \BibitemOpen
  \bibfield  {author} {\bibinfo {author} {\bibfnamefont {Zhenong}\ \bibnamefont
  {Zhang}}, \bibinfo {author} {\bibfnamefont {Alireza}\ \bibnamefont
  {Kalantari~Dehaghi}}, \bibinfo {author} {\bibfnamefont {Pramit}\ \bibnamefont
  {Ghosh}}, \ and\ \bibinfo {author} {\bibfnamefont {Linxiao}\ \bibnamefont
  {Zhu}},\ }\bibfield  {title} {\enquote {\bibinfo {title} {Observation of
  strong nonreciprocal thermal emission},}\ }\href {\doibase
  10.1103/PhysRevLett.135.016901} {\bibfield  {journal} {\bibinfo  {journal}
  {Phys. Rev. Lett.}\ }\textbf {\bibinfo {volume} {135}},\ \bibinfo {pages}
  {016901} (\bibinfo {year} {2025})}\BibitemShut {NoStop}%
\bibitem [{\citenamefont {Nabavi}\ \emph {et~al.}(2025)\citenamefont {Nabavi},
  \citenamefont {Jafari~Ghalekohneh}, \citenamefont {Shayegan}, \citenamefont
  {Tervo}, \citenamefont {Atwater},\ and\ \citenamefont
  {Zhao}}]{nabavi2025high}%
  \BibitemOpen
  \bibfield  {author} {\bibinfo {author} {\bibfnamefont {Bardia}\ \bibnamefont
  {Nabavi}}, \bibinfo {author} {\bibfnamefont {Sina}\ \bibnamefont
  {Jafari~Ghalekohneh}}, \bibinfo {author} {\bibfnamefont {Komron~J.}\
  \bibnamefont {Shayegan}}, \bibinfo {author} {\bibfnamefont {Eric~J.}\
  \bibnamefont {Tervo}}, \bibinfo {author} {\bibfnamefont {Harry}\ \bibnamefont
  {Atwater}}, \ and\ \bibinfo {author} {\bibfnamefont {Bo}~\bibnamefont
  {Zhao}},\ }\bibfield  {title} {\enquote {\bibinfo {title} {High-temperature
  strong nonreciprocal thermal radiation from semiconductors},}\ }\href
  {\doibase 10.1021/acsphotonics.5c00365} {\bibfield  {journal} {\bibinfo
  {journal} {ACS Photonics}\ }\textbf {\bibinfo {volume} {12}},\ \bibinfo
  {pages} {2767--2774} (\bibinfo {year} {2025})}\BibitemShut {NoStop}%
\bibitem [{\citenamefont {Pajovic}\ \emph {et~al.}(2025)\citenamefont
  {Pajovic}, \citenamefont {Tsurimaki}, \citenamefont {Qian}, \citenamefont
  {Chen},\ and\ \citenamefont
  {Boriskina}}]{pajovic2025nonreciprocal_reflection}%
  \BibitemOpen
  \bibfield  {author} {\bibinfo {author} {\bibfnamefont {Simo}\ \bibnamefont
  {Pajovic}}, \bibinfo {author} {\bibfnamefont {Yoichiro}\ \bibnamefont
  {Tsurimaki}}, \bibinfo {author} {\bibfnamefont {Xin}\ \bibnamefont {Qian}},
  \bibinfo {author} {\bibfnamefont {Gang}\ \bibnamefont {Chen}}, \ and\
  \bibinfo {author} {\bibfnamefont {Svetlana~V.}\ \bibnamefont {Boriskina}},\
  }\bibfield  {title} {\enquote {\bibinfo {title} {Nonreciprocal reflection of
  mid-infrared light by highly doped {InAs} at low magnetic fields},}\ }\href
  {\doibase 10.1364/OE.544496} {\bibfield  {journal} {\bibinfo  {journal}
  {Optics Express}\ }\textbf {\bibinfo {volume} {33}},\ \bibinfo {pages}
  {8661--8674} (\bibinfo {year} {2025})}\BibitemShut {NoStop}%
\bibitem [{\citenamefont {Yan}\ and\ \citenamefont {Felser}(2017)}]{Yan2017}%
  \BibitemOpen
  \bibfield  {author} {\bibinfo {author} {\bibfnamefont {Binghai}\ \bibnamefont
  {Yan}}\ and\ \bibinfo {author} {\bibfnamefont {Claudia}\ \bibnamefont
  {Felser}},\ }\bibfield  {title} {\enquote {\bibinfo {title} {{Topological
  Materials: Weyl Semimetals}},}\ }\href {\doibase
  10.1146/annurev-conmatphys-031016-025458} {\bibfield  {journal} {\bibinfo
  {journal} {Annu. Rev. Cond. Mat. Phys.}\ }\textbf {\bibinfo {volume} {8}},\
  \bibinfo {pages} {337 -- 354} (\bibinfo {year} {2017})}\BibitemShut {NoStop}%
\bibitem [{\citenamefont {{Armitage}}\ \emph {et~al.}(2018)\citenamefont
  {{Armitage}}, \citenamefont {{Mele}},\ and\ \citenamefont
  {{Vishwanath}}}]{Armitage2018}%
  \BibitemOpen
  \bibfield  {author} {\bibinfo {author} {\bibfnamefont {N.~P.}\ \bibnamefont
  {{Armitage}}}, \bibinfo {author} {\bibfnamefont {E.~J.}\ \bibnamefont
  {{Mele}}}, \ and\ \bibinfo {author} {\bibfnamefont {A.}~\bibnamefont
  {{Vishwanath}}},\ }\bibfield  {title} {\enquote {\bibinfo {title} {{Weyl and
  Dirac semimetals in three-dimensional solids}},}\ }\href {\doibase
  10.1103/RevModPhys.90.015001} {\bibfield  {journal} {\bibinfo  {journal}
  {Rev. Mod. Phys.}\ }\textbf {\bibinfo {volume} {90}},\ \bibinfo {pages}
  {015001} (\bibinfo {year} {2018})}\BibitemShut {NoStop}%
\bibitem [{\citenamefont {Zhao}\ \emph {et~al.}(2020)\citenamefont {Zhao},
  \citenamefont {Guo}, \citenamefont {Garcia}, \citenamefont {Narang},\ and\
  \citenamefont {Fan}}]{zhao2020axion}%
  \BibitemOpen
  \bibfield  {author} {\bibinfo {author} {\bibfnamefont {Bo}~\bibnamefont
  {Zhao}}, \bibinfo {author} {\bibfnamefont {Cheng}\ \bibnamefont {Guo}},
  \bibinfo {author} {\bibfnamefont {Christina A.~C.}\ \bibnamefont {Garcia}},
  \bibinfo {author} {\bibfnamefont {Prineha}\ \bibnamefont {Narang}}, \ and\
  \bibinfo {author} {\bibfnamefont {Shanhui}\ \bibnamefont {Fan}},\ }\bibfield
  {title} {\enquote {\bibinfo {title} {Axion-field-enabled nonreciprocal
  thermal radiation in {Weyl} semimetals},}\ }\href {\doibase
  10.1021/acs.nanolett.9b05179} {\bibfield  {journal} {\bibinfo  {journal}
  {Nano Letters}\ }\textbf {\bibinfo {volume} {20}},\ \bibinfo {pages}
  {1923--1927} (\bibinfo {year} {2020})}\BibitemShut {NoStop}%
\bibitem [{\citenamefont {Tsurimaki}\ \emph {et~al.}(2020)\citenamefont
  {Tsurimaki}, \citenamefont {Qian}, \citenamefont {Pajovic}, \citenamefont
  {Han}, \citenamefont {Li},\ and\ \citenamefont {Chen}}]{tsurimaki2020large}%
  \BibitemOpen
  \bibfield  {author} {\bibinfo {author} {\bibfnamefont {Yoichiro}\
  \bibnamefont {Tsurimaki}}, \bibinfo {author} {\bibfnamefont {Xin}\
  \bibnamefont {Qian}}, \bibinfo {author} {\bibfnamefont {Simo}\ \bibnamefont
  {Pajovic}}, \bibinfo {author} {\bibfnamefont {Fei}\ \bibnamefont {Han}},
  \bibinfo {author} {\bibfnamefont {Mingda}\ \bibnamefont {Li}}, \ and\
  \bibinfo {author} {\bibfnamefont {Gang}\ \bibnamefont {Chen}},\ }\bibfield
  {title} {\enquote {\bibinfo {title} {Large nonreciprocal absorption and
  emission of radiation in type-{I} {Weyl} semimetals with time reversal
  symmetry breaking},}\ }\href {\doibase 10.1103/PhysRevB.101.165426}
  {\bibfield  {journal} {\bibinfo  {journal} {Phys. Rev. B}\ }\textbf {\bibinfo
  {volume} {101}},\ \bibinfo {pages} {165426} (\bibinfo {year}
  {2020})}\BibitemShut {NoStop}%
\bibitem [{\citenamefont {Pajovic}\ \emph {et~al.}(2020)\citenamefont
  {Pajovic}, \citenamefont {Tsurimaki}, \citenamefont {Qian},\ and\
  \citenamefont {Chen}}]{pajovic2020intrinsic}%
  \BibitemOpen
  \bibfield  {author} {\bibinfo {author} {\bibfnamefont {Simo}\ \bibnamefont
  {Pajovic}}, \bibinfo {author} {\bibfnamefont {Yoichiro}\ \bibnamefont
  {Tsurimaki}}, \bibinfo {author} {\bibfnamefont {Xin}\ \bibnamefont {Qian}}, \
  and\ \bibinfo {author} {\bibfnamefont {Gang}\ \bibnamefont {Chen}},\
  }\bibfield  {title} {\enquote {\bibinfo {title} {Intrinsic nonreciprocal
  reflection and violation of {Kirchhoff}'s law of radiation in planar type-{I}
  magnetic {Weyl} semimetal surfaces},}\ }\href {\doibase
  10.1103/PhysRevB.102.165417} {\bibfield  {journal} {\bibinfo  {journal}
  {Phys. Rev. B}\ }\textbf {\bibinfo {volume} {102}},\ \bibinfo {pages}
  {165417} (\bibinfo {year} {2020})}\BibitemShut {NoStop}%
\bibitem [{\citenamefont {Guo}\ \emph {et~al.}(2020)\citenamefont {Guo},
  \citenamefont {Zhao}, \citenamefont {Huang},\ and\ \citenamefont
  {Fan}}]{guo2020radiative}%
  \BibitemOpen
  \bibfield  {author} {\bibinfo {author} {\bibfnamefont {Cheng}\ \bibnamefont
  {Guo}}, \bibinfo {author} {\bibfnamefont {Bo}~\bibnamefont {Zhao}}, \bibinfo
  {author} {\bibfnamefont {Danhong}\ \bibnamefont {Huang}}, \ and\ \bibinfo
  {author} {\bibfnamefont {Shanhui}\ \bibnamefont {Fan}},\ }\bibfield  {title}
  {\enquote {\bibinfo {title} {Radiative thermal router based on tunable
  magnetic {Weyl} semimetals},}\ }\href {\doibase 10.1021/acsphotonics.0c01376}
  {\bibfield  {journal} {\bibinfo  {journal} {ACS Photonics}\ }\textbf
  {\bibinfo {volume} {7}},\ \bibinfo {pages} {3257--3263} (\bibinfo {year}
  {2020})}\BibitemShut {NoStop}%
\bibitem [{\citenamefont {Liu}\ \emph {et~al.}(2018)\citenamefont {Liu},
  \citenamefont {Sun}, \citenamefont {Kumar}, \citenamefont {Muechler},
  \citenamefont {Sun}, \citenamefont {Jiao}, \citenamefont {Yang},
  \citenamefont {Liu}, \citenamefont {Liang}, \citenamefont {Xu}, \citenamefont
  {Kroder}, \citenamefont {S{\"u}{\ss}}, \citenamefont {Borrmann},
  \citenamefont {Shekhar}, \citenamefont {Wang}, \citenamefont {Xi},
  \citenamefont {Wang}, \citenamefont {Schnelle}, \citenamefont {Wirth},
  \citenamefont {Chen}, \citenamefont {Goennenwein},\ and\ \citenamefont
  {Felser}}]{liu2018giant}%
  \BibitemOpen
  \bibfield  {author} {\bibinfo {author} {\bibfnamefont {Enke}\ \bibnamefont
  {Liu}}, \bibinfo {author} {\bibfnamefont {Yan}\ \bibnamefont {Sun}}, \bibinfo
  {author} {\bibfnamefont {Nitesh}\ \bibnamefont {Kumar}}, \bibinfo {author}
  {\bibfnamefont {Lukas}\ \bibnamefont {Muechler}}, \bibinfo {author}
  {\bibfnamefont {Aili}\ \bibnamefont {Sun}}, \bibinfo {author} {\bibfnamefont
  {Lin}\ \bibnamefont {Jiao}}, \bibinfo {author} {\bibfnamefont {Shuo-Ying}\
  \bibnamefont {Yang}}, \bibinfo {author} {\bibfnamefont {Defa}\ \bibnamefont
  {Liu}}, \bibinfo {author} {\bibfnamefont {Aiji}\ \bibnamefont {Liang}},
  \bibinfo {author} {\bibfnamefont {Qiunan}\ \bibnamefont {Xu}}, \bibinfo
  {author} {\bibfnamefont {Johannes}\ \bibnamefont {Kroder}}, \bibinfo {author}
  {\bibfnamefont {Vicky}\ \bibnamefont {S{\"u}{\ss}}}, \bibinfo {author}
  {\bibfnamefont {Horst}\ \bibnamefont {Borrmann}}, \bibinfo {author}
  {\bibfnamefont {Chandra}\ \bibnamefont {Shekhar}}, \bibinfo {author}
  {\bibfnamefont {Zhaosheng}\ \bibnamefont {Wang}}, \bibinfo {author}
  {\bibfnamefont {Chuanying}\ \bibnamefont {Xi}}, \bibinfo {author}
  {\bibfnamefont {Wenhong}\ \bibnamefont {Wang}}, \bibinfo {author}
  {\bibfnamefont {Walter}\ \bibnamefont {Schnelle}}, \bibinfo {author}
  {\bibfnamefont {Steffen}\ \bibnamefont {Wirth}}, \bibinfo {author}
  {\bibfnamefont {Yulin}\ \bibnamefont {Chen}}, \bibinfo {author}
  {\bibfnamefont {Sebastian T.~B.}\ \bibnamefont {Goennenwein}}, \ and\
  \bibinfo {author} {\bibfnamefont {Claudia}\ \bibnamefont {Felser}},\
  }\bibfield  {title} {\enquote {\bibinfo {title} {Giant anomalous {Hall}
  effect in a ferromagnetic kagome-lattice semimetal},}\ }\href {\doibase
  10.1038/s41567-018-0234-5} {\bibfield  {journal} {\bibinfo  {journal} {Nature
  Physics}\ }\textbf {\bibinfo {volume} {14}},\ \bibinfo {pages} {1125--1131}
  (\bibinfo {year} {2018})}\BibitemShut {NoStop}%
\bibitem [{\citenamefont {Komissarov}\ \emph {et~al.}(2024)\citenamefont
  {Komissarov}, \citenamefont {Holder},\ and\ \citenamefont
  {Queiroz}}]{komissarov2024quantum}%
  \BibitemOpen
  \bibfield  {author} {\bibinfo {author} {\bibfnamefont {Ilia}\ \bibnamefont
  {Komissarov}}, \bibinfo {author} {\bibfnamefont {Tobias}\ \bibnamefont
  {Holder}}, \ and\ \bibinfo {author} {\bibfnamefont {Raquel}\ \bibnamefont
  {Queiroz}},\ }\bibfield  {title} {\enquote {\bibinfo {title} {The quantum
  geometric origin of capacitance in insulators},}\ }\href {\doibase
  10.1038/s41467-024-48808-x} {\bibfield  {journal} {\bibinfo  {journal}
  {Nature Communications}\ }\textbf {\bibinfo {volume} {15}},\ \bibinfo {pages}
  {4621} (\bibinfo {year} {2024})}\BibitemShut {NoStop}%
\bibitem [{\citenamefont {Okamura}\ \emph {et~al.}(2020)\citenamefont
  {Okamura}, \citenamefont {Minami}, \citenamefont {Kato}, \citenamefont
  {Fujishiro}, \citenamefont {Kaneko}, \citenamefont {Ikeda}, \citenamefont
  {Muramoto}, \citenamefont {Kaneko}, \citenamefont {Ueda}, \citenamefont
  {Kocsis}, \citenamefont {Kanazawa}, \citenamefont {Taguchi}, \citenamefont
  {Koretsune}, \citenamefont {Fujiwara}, \citenamefont {Tsukazaki},
  \citenamefont {Arita}, \citenamefont {Tokura},\ and\ \citenamefont
  {Takahashi}}]{okamura2020giant}%
  \BibitemOpen
  \bibfield  {author} {\bibinfo {author} {\bibfnamefont {Y.}~\bibnamefont
  {Okamura}}, \bibinfo {author} {\bibfnamefont {S.}~\bibnamefont {Minami}},
  \bibinfo {author} {\bibfnamefont {Y.}~\bibnamefont {Kato}}, \bibinfo {author}
  {\bibfnamefont {Y.}~\bibnamefont {Fujishiro}}, \bibinfo {author}
  {\bibfnamefont {Y.}~\bibnamefont {Kaneko}}, \bibinfo {author} {\bibfnamefont
  {J.}~\bibnamefont {Ikeda}}, \bibinfo {author} {\bibfnamefont
  {J.}~\bibnamefont {Muramoto}}, \bibinfo {author} {\bibfnamefont
  {R.}~\bibnamefont {Kaneko}}, \bibinfo {author} {\bibfnamefont
  {K.}~\bibnamefont {Ueda}}, \bibinfo {author} {\bibfnamefont {V.}~\bibnamefont
  {Kocsis}}, \bibinfo {author} {\bibfnamefont {N.}~\bibnamefont {Kanazawa}},
  \bibinfo {author} {\bibfnamefont {Y.}~\bibnamefont {Taguchi}}, \bibinfo
  {author} {\bibfnamefont {T.}~\bibnamefont {Koretsune}}, \bibinfo {author}
  {\bibfnamefont {K.}~\bibnamefont {Fujiwara}}, \bibinfo {author}
  {\bibfnamefont {A.}~\bibnamefont {Tsukazaki}}, \bibinfo {author}
  {\bibfnamefont {R.}~\bibnamefont {Arita}}, \bibinfo {author} {\bibfnamefont
  {Y.}~\bibnamefont {Tokura}}, \ and\ \bibinfo {author} {\bibfnamefont
  {Y.}~\bibnamefont {Takahashi}},\ }\bibfield  {title} {\enquote {\bibinfo
  {title} {Giant magneto-optical responses in magnetic {W}eyl semimetal
  {C}o$_3${S}n$_2${S}$_2$},}\ }\href {\doibase 10.1038/s41467-020-18470-0}
  {\bibfield  {journal} {\bibinfo  {journal} {Nature Communications}\ }\textbf
  {\bibinfo {volume} {11}},\ \bibinfo {pages} {4619} (\bibinfo {year}
  {2020})}\BibitemShut {NoStop}%
\bibitem [{\citenamefont {Zhang}\ \emph {et~al.}(2017)\citenamefont {Zhang},
  \citenamefont {Sun}, \citenamefont {Yang}, \citenamefont
  {\ifmmode~\check{Z}\else \v{Z}\fi{}elezn\'y}, \citenamefont {Parkin},
  \citenamefont {Felser},\ and\ \citenamefont {Yan}}]{zhang2017strong}%
  \BibitemOpen
  \bibfield  {author} {\bibinfo {author} {\bibfnamefont {Yang}\ \bibnamefont
  {Zhang}}, \bibinfo {author} {\bibfnamefont {Yan}\ \bibnamefont {Sun}},
  \bibinfo {author} {\bibfnamefont {Hao}\ \bibnamefont {Yang}}, \bibinfo
  {author} {\bibfnamefont {Jakub}\ \bibnamefont {\ifmmode~\check{Z}\else
  \v{Z}\fi{}elezn\'y}}, \bibinfo {author} {\bibfnamefont {Stuart P.~P.}\
  \bibnamefont {Parkin}}, \bibinfo {author} {\bibfnamefont {Claudia}\
  \bibnamefont {Felser}}, \ and\ \bibinfo {author} {\bibfnamefont {Binghai}\
  \bibnamefont {Yan}},\ }\bibfield  {title} {\enquote {\bibinfo {title} {Strong
  anisotropic anomalous {Hall} effect and spin {Hall} effect in the chiral
  antiferromagnetic compounds {${\mathrm{Mn}}_{3}X$ ($X=\mathrm{Ge}$, Sn, Ga,
  Ir, Rh, and Pt)}},}\ }\href {\doibase 10.1103/PhysRevB.95.075128} {\bibfield
  {journal} {\bibinfo  {journal} {Phys. Rev. B}\ }\textbf {\bibinfo {volume}
  {95}},\ \bibinfo {pages} {075128} (\bibinfo {year} {2017})}\BibitemShut
  {NoStop}%
\bibitem [{\citenamefont {Zhao}\ \emph {et~al.}(2024)\citenamefont {Zhao},
  \citenamefont {Jiang}, \citenamefont {Bae}, \citenamefont {Das},
  \citenamefont {Li}, \citenamefont {Liu},\ and\ \citenamefont
  {Yan}}]{zhao2024hybrid}%
  \BibitemOpen
  \bibfield  {author} {\bibinfo {author} {\bibfnamefont {Yufei}\ \bibnamefont
  {Zhao}}, \bibinfo {author} {\bibfnamefont {Yiyang}\ \bibnamefont {Jiang}},
  \bibinfo {author} {\bibfnamefont {Hyeonhu}\ \bibnamefont {Bae}}, \bibinfo
  {author} {\bibfnamefont {Kamal}\ \bibnamefont {Das}}, \bibinfo {author}
  {\bibfnamefont {Yongkang}\ \bibnamefont {Li}}, \bibinfo {author}
  {\bibfnamefont {Chao-Xing}\ \bibnamefont {Liu}}, \ and\ \bibinfo {author}
  {\bibfnamefont {Binghai}\ \bibnamefont {Yan}},\ }\bibfield  {title} {\enquote
  {\bibinfo {title} {Hybrid-order topology in unconventional magnets of
  {Eu}-based {Zintl} compounds with surface-dependent quantum geometry},}\
  }\href {\doibase 10.1103/PhysRevB.110.205111} {\bibfield  {journal} {\bibinfo
   {journal} {Phys. Rev. B}\ }\textbf {\bibinfo {volume} {110}},\ \bibinfo
  {pages} {205111} (\bibinfo {year} {2024})}\BibitemShut {NoStop}%
\bibitem [{nrt()}]{nrtp_material_database_2026}%
  \BibitemOpen
  \href@noop {} {}\bibinfo {note} {Nonreciprocal Thermal Radiation Database.
  \url{https://okongoyango.github.io/Nonreciprocal_Thermal_Material_Database/}
  (accessed 2026-10-11), interactive viewer of the first-principles material
  parameters and dielectric tensors}\BibitemShut {NoStop}%
\bibitem [{\citenamefont {Wallis}\ \emph {et~al.}(1974)\citenamefont {Wallis},
  \citenamefont {Brion}, \citenamefont {Burstein},\ and\ \citenamefont
  {Hartstein}}]{wallis1974surface}%
  \BibitemOpen
  \bibfield  {author} {\bibinfo {author} {\bibfnamefont {R.~F.}\ \bibnamefont
  {Wallis}}, \bibinfo {author} {\bibfnamefont {J.~J.}\ \bibnamefont {Brion}},
  \bibinfo {author} {\bibfnamefont {E.}~\bibnamefont {Burstein}}, \ and\
  \bibinfo {author} {\bibfnamefont {A.}~\bibnamefont {Hartstein}},\ }\bibfield
  {title} {\enquote {\bibinfo {title} {Theory of surface polaritons in
  anisotropic dielectric media with application to surface magnetoplasmons in
  semiconductors},}\ }\href {\doibase 10.1103/PhysRevB.9.3424} {\bibfield
  {journal} {\bibinfo  {journal} {Phys. Rev. B}\ }\textbf {\bibinfo {volume}
  {9}},\ \bibinfo {pages} {3424--3437} (\bibinfo {year} {1974})}\BibitemShut
  {NoStop}%
\bibitem [{\citenamefont {Raether}(1980)}]{Raether1980}%
  \BibitemOpen
  \bibfield  {author} {\bibinfo {author} {\bibfnamefont {Heinz}\ \bibnamefont
  {Raether}},\ }\href {\doibase 10.1007/BFb0045951} {\emph {\bibinfo {title}
  {Excitation of Plasmons and Interband Transitions by Electrons}}},\ \bibinfo
  {series} {Springer Tracts in Modern Physics}, Vol.~\bibinfo {volume} {88}\
  (\bibinfo  {publisher} {Springer},\ \bibinfo {address} {Berlin, Heidelberg},\
  \bibinfo {year} {1980})\BibitemShut {NoStop}%
\bibitem [{\citenamefont {Song}\ \emph {et~al.}(2024)\citenamefont {Song},
  \citenamefont {Houben}, \citenamefont {Zhao}, \citenamefont {Bae},
  \citenamefont {Rothem}, \citenamefont {Gupta}, \citenamefont {Yan},
  \citenamefont {Kalisky}, \citenamefont {Zaluska-Kotur}, \citenamefont
  {Kacman}, \citenamefont {Shtrikman},\ and\ \citenamefont
  {Beidenkopf}}]{song2024topotaxial}%
  \BibitemOpen
  \bibfield  {author} {\bibinfo {author} {\bibfnamefont {Man~Suk}\ \bibnamefont
  {Song}}, \bibinfo {author} {\bibfnamefont {Lothar}\ \bibnamefont {Houben}},
  \bibinfo {author} {\bibfnamefont {Yufei}\ \bibnamefont {Zhao}}, \bibinfo
  {author} {\bibfnamefont {Hyeonhu}\ \bibnamefont {Bae}}, \bibinfo {author}
  {\bibfnamefont {Nadav}\ \bibnamefont {Rothem}}, \bibinfo {author}
  {\bibfnamefont {Ambikesh}\ \bibnamefont {Gupta}}, \bibinfo {author}
  {\bibfnamefont {Binghai}\ \bibnamefont {Yan}}, \bibinfo {author}
  {\bibfnamefont {Beena}\ \bibnamefont {Kalisky}}, \bibinfo {author}
  {\bibfnamefont {Magdalena}\ \bibnamefont {Zaluska-Kotur}}, \bibinfo {author}
  {\bibfnamefont {Perla}\ \bibnamefont {Kacman}}, \bibinfo {author}
  {\bibfnamefont {Hadas}\ \bibnamefont {Shtrikman}}, \ and\ \bibinfo {author}
  {\bibfnamefont {Haim}\ \bibnamefont {Beidenkopf}},\ }\bibfield  {title}
  {\enquote {\bibinfo {title} {Topotaxial mutual-exchange growth of magnetic
  {Zintl} {Eu$_3$In$_2$As$_4$} nanowires with axion insulator
  classification},}\ }\href {\doibase 10.1038/s41565-024-01762-7} {\bibfield
  {journal} {\bibinfo  {journal} {Nature Nanotechnology}\ }\textbf {\bibinfo
  {volume} {19}},\ \bibinfo {pages} {1796--1803} (\bibinfo {year}
  {2024})}\BibitemShut {NoStop}%
\bibitem [{\citenamefont {Li}\ \emph {et~al.}(2020)\citenamefont {Li},
  \citenamefont {Koo}, \citenamefont {Ning}, \citenamefont {Li}, \citenamefont
  {Miao}, \citenamefont {Min}, \citenamefont {Zhu}, \citenamefont {Wang},
  \citenamefont {Alem}, \citenamefont {Liu} \emph {et~al.}}]{Li2020Giant}%
  \BibitemOpen
  \bibfield  {author} {\bibinfo {author} {\bibfnamefont {Peigang}\ \bibnamefont
  {Li}}, \bibinfo {author} {\bibfnamefont {Jahyun}\ \bibnamefont {Koo}},
  \bibinfo {author} {\bibfnamefont {Wei}\ \bibnamefont {Ning}}, \bibinfo
  {author} {\bibfnamefont {Jinglei}\ \bibnamefont {Li}}, \bibinfo {author}
  {\bibfnamefont {Le}~\bibnamefont {Miao}}, \bibinfo {author} {\bibfnamefont
  {Lujin}\ \bibnamefont {Min}}, \bibinfo {author} {\bibfnamefont {Yulin}\
  \bibnamefont {Zhu}}, \bibinfo {author} {\bibfnamefont {Yu}~\bibnamefont
  {Wang}}, \bibinfo {author} {\bibfnamefont {Nasim}\ \bibnamefont {Alem}},
  \bibinfo {author} {\bibfnamefont {Cheng-Xiang}\ \bibnamefont {Liu}},  \emph
  {et~al.},\ }\bibfield  {title} {\enquote {\bibinfo {title} {Giant room
  temperature anomalous {Hall} effect and tunable topology in a ferromagnetic
  topological semimetal {Co$_2$MnAl}},}\ }\href {\doibase
  10.1038/s41467-020-17174-9} {\bibfield  {journal} {\bibinfo  {journal}
  {Nature Communications}\ }\textbf {\bibinfo {volume} {11}},\ \bibinfo {pages}
  {3476} (\bibinfo {year} {2020})}\BibitemShut {NoStop}%
\bibitem [{\citenamefont {Sakai}\ \emph {et~al.}(2018)\citenamefont {Sakai},
  \citenamefont {Mizuta}, \citenamefont {Nugroho}, \citenamefont {Sihombing},
  \citenamefont {Koretsune}, \citenamefont {Suzuki}, \citenamefont {Takemori},
  \citenamefont {Ishii}, \citenamefont {Nishio-Hamane}, \citenamefont {Arita},
  \citenamefont {Goswami},\ and\ \citenamefont {Nakatsuji}}]{Sakai2018bc}%
  \BibitemOpen
  \bibfield  {author} {\bibinfo {author} {\bibfnamefont {Akito}\ \bibnamefont
  {Sakai}}, \bibinfo {author} {\bibfnamefont {Yo~Pierre}\ \bibnamefont
  {Mizuta}}, \bibinfo {author} {\bibfnamefont {Agustinus~Agung}\ \bibnamefont
  {Nugroho}}, \bibinfo {author} {\bibfnamefont {Rombang}\ \bibnamefont
  {Sihombing}}, \bibinfo {author} {\bibfnamefont {Takashi}\ \bibnamefont
  {Koretsune}}, \bibinfo {author} {\bibfnamefont {Michi-to}\ \bibnamefont
  {Suzuki}}, \bibinfo {author} {\bibfnamefont {Nayuta}\ \bibnamefont
  {Takemori}}, \bibinfo {author} {\bibfnamefont {Rieko}\ \bibnamefont {Ishii}},
  \bibinfo {author} {\bibfnamefont {Daisuke}\ \bibnamefont {Nishio-Hamane}},
  \bibinfo {author} {\bibfnamefont {Ryotaro}\ \bibnamefont {Arita}}, \bibinfo
  {author} {\bibfnamefont {Pallab}\ \bibnamefont {Goswami}}, \ and\ \bibinfo
  {author} {\bibfnamefont {Satoru}\ \bibnamefont {Nakatsuji}},\ }\bibfield
  {title} {\enquote {\bibinfo {title} {Giant anomalous {Nernst} effect and
  quantum-critical scaling in a ferromagnetic semimetal},}\ }\href {\doibase
  10.1038/s41567-018-0225-6} {\bibfield  {journal} {\bibinfo  {journal} {Nature
  Physics}\ }\textbf {\bibinfo {volume} {14}},\ \bibinfo {pages} {1119 -- 1124}
  (\bibinfo {year} {2018})}\BibitemShut {NoStop}%
\bibitem [{\citenamefont {Guin}\ \emph {et~al.}(2019)\citenamefont {Guin},
  \citenamefont {Manna}, \citenamefont {Noky}, \citenamefont {Watzman},
  \citenamefont {Fu}, \citenamefont {Kumar}, \citenamefont {Schnelle},
  \citenamefont {Shekhar}, \citenamefont {Sun}, \citenamefont {Gooth},\ and\
  \citenamefont {Felser}}]{Guin2019Anomalous}%
  \BibitemOpen
  \bibfield  {author} {\bibinfo {author} {\bibfnamefont {Satya~N.}\
  \bibnamefont {Guin}}, \bibinfo {author} {\bibfnamefont {Kaustuv}\
  \bibnamefont {Manna}}, \bibinfo {author} {\bibfnamefont {Jonathan}\
  \bibnamefont {Noky}}, \bibinfo {author} {\bibfnamefont {Sarah~J.}\
  \bibnamefont {Watzman}}, \bibinfo {author} {\bibfnamefont {Chenguang}\
  \bibnamefont {Fu}}, \bibinfo {author} {\bibfnamefont {Nitesh}\ \bibnamefont
  {Kumar}}, \bibinfo {author} {\bibfnamefont {Walter}\ \bibnamefont
  {Schnelle}}, \bibinfo {author} {\bibfnamefont {Chandra}\ \bibnamefont
  {Shekhar}}, \bibinfo {author} {\bibfnamefont {Yan}\ \bibnamefont {Sun}},
  \bibinfo {author} {\bibfnamefont {Johannes}\ \bibnamefont {Gooth}}, \ and\
  \bibinfo {author} {\bibfnamefont {Claudia}\ \bibnamefont {Felser}},\
  }\bibfield  {title} {\enquote {\bibinfo {title} {Anomalous {Nernst} effect
  beyond the magnetization scaling relation in the ferromagnetic {Heusler}
  compound {Co$_2$MnGa}},}\ }\href {\doibase 10.1038/s41427-019-0116-z}
  {\bibfield  {journal} {\bibinfo  {journal} {NPG Asia Materials}\ }\textbf
  {\bibinfo {volume} {11}},\ \bibinfo {pages} {16} (\bibinfo {year}
  {2019})}\BibitemShut {NoStop}%
\bibitem [{\citenamefont {Nakatsuji}\ \emph {et~al.}(2015)\citenamefont
  {Nakatsuji}, \citenamefont {Kiyohara},\ and\ \citenamefont
  {Higo}}]{nakatsuji2015large}%
  \BibitemOpen
  \bibfield  {author} {\bibinfo {author} {\bibfnamefont {Satoru}\ \bibnamefont
  {Nakatsuji}}, \bibinfo {author} {\bibfnamefont {Naoki}\ \bibnamefont
  {Kiyohara}}, \ and\ \bibinfo {author} {\bibfnamefont {Tomoya}\ \bibnamefont
  {Higo}},\ }\bibfield  {title} {\enquote {\bibinfo {title} {Large anomalous
  {Hall} effect in a non-collinear antiferromagnet at room temperature},}\
  }\href@noop {} {\bibfield  {journal} {\bibinfo  {journal} {Nature}\ }\textbf
  {\bibinfo {volume} {527}},\ \bibinfo {pages} {212--215} (\bibinfo {year}
  {2015})}\BibitemShut {NoStop}%
\bibitem [{\citenamefont {Nayak}\ \emph {et~al.}(2016)\citenamefont {Nayak},
  \citenamefont {Fischer}, \citenamefont {Sun}, \citenamefont {Yan},
  \citenamefont {Karel}, \citenamefont {Komarek}, \citenamefont {Shekhar},
  \citenamefont {Kumar}, \citenamefont {Schnelle}, \citenamefont {Kübler},
  \citenamefont {Felser},\ and\ \citenamefont {Parkin}}]{nayak2016large}%
  \BibitemOpen
  \bibfield  {author} {\bibinfo {author} {\bibfnamefont {Ajaya~K.}\
  \bibnamefont {Nayak}}, \bibinfo {author} {\bibfnamefont {Julia~Erika}\
  \bibnamefont {Fischer}}, \bibinfo {author} {\bibfnamefont {Yan}\ \bibnamefont
  {Sun}}, \bibinfo {author} {\bibfnamefont {Binghai}\ \bibnamefont {Yan}},
  \bibinfo {author} {\bibfnamefont {Julie}\ \bibnamefont {Karel}}, \bibinfo
  {author} {\bibfnamefont {Alexander~C.}\ \bibnamefont {Komarek}}, \bibinfo
  {author} {\bibfnamefont {Chandra}\ \bibnamefont {Shekhar}}, \bibinfo {author}
  {\bibfnamefont {Nitesh}\ \bibnamefont {Kumar}}, \bibinfo {author}
  {\bibfnamefont {Walter}\ \bibnamefont {Schnelle}}, \bibinfo {author}
  {\bibfnamefont {Jürgen}\ \bibnamefont {Kübler}}, \bibinfo {author}
  {\bibfnamefont {Claudia}\ \bibnamefont {Felser}}, \ and\ \bibinfo {author}
  {\bibfnamefont {Stuart S.~P.}\ \bibnamefont {Parkin}},\ }\bibfield  {title}
  {\enquote {\bibinfo {title} {Large anomalous {Hall} effect driven by a
  nonvanishing {Berry} curvature in the noncolinear antiferromagnet
  {Mn$_3$Ge}},}\ }\href {\doibase 10.1126/sciadv.1501870} {\bibfield  {journal}
  {\bibinfo  {journal} {Science Advances}\ }\textbf {\bibinfo {volume} {2}},\
  \bibinfo {pages} {e1501870} (\bibinfo {year} {2016})}\BibitemShut {NoStop}%
\bibitem [{\citenamefont {Gonzalez~Betancourt}\ \emph
  {et~al.}(2023)\citenamefont {Gonzalez~Betancourt}, \citenamefont
  {Zub{\'a}{\v{c}}}, \citenamefont {Gonzalez-Hernandez}, \citenamefont
  {Geishendorf}, \citenamefont {{\v{S}}ob{\'a}{\v{n}}}, \citenamefont
  {Springholz}, \citenamefont {Olejn{\'\i}k}, \citenamefont {{\v{S}}mejkal},
  \citenamefont {Sinova}, \citenamefont {Jungwirth}, \citenamefont
  {Goennenwein}, \citenamefont {Thomas}, \citenamefont {Reichlov{\'a}},
  \citenamefont {{\v{Z}}elezn{\'y}},\ and\ \citenamefont
  {Kriegner}}]{gonzalezbetancourt2023spontaneous}%
  \BibitemOpen
  \bibfield  {author} {\bibinfo {author} {\bibfnamefont {R.~D.}\ \bibnamefont
  {Gonzalez~Betancourt}}, \bibinfo {author} {\bibfnamefont {J.}~\bibnamefont
  {Zub{\'a}{\v{c}}}}, \bibinfo {author} {\bibfnamefont {R.~J.}\ \bibnamefont
  {Gonzalez-Hernandez}}, \bibinfo {author} {\bibfnamefont {K.}~\bibnamefont
  {Geishendorf}}, \bibinfo {author} {\bibfnamefont {Z.}~\bibnamefont
  {{\v{S}}ob{\'a}{\v{n}}}}, \bibinfo {author} {\bibfnamefont {G.}~\bibnamefont
  {Springholz}}, \bibinfo {author} {\bibfnamefont {K.}~\bibnamefont
  {Olejn{\'\i}k}}, \bibinfo {author} {\bibfnamefont {L.}~\bibnamefont
  {{\v{S}}mejkal}}, \bibinfo {author} {\bibfnamefont {J.}~\bibnamefont
  {Sinova}}, \bibinfo {author} {\bibfnamefont {T.}~\bibnamefont {Jungwirth}},
  \bibinfo {author} {\bibfnamefont {S.~T.~B.}\ \bibnamefont {Goennenwein}},
  \bibinfo {author} {\bibfnamefont {A.}~\bibnamefont {Thomas}}, \bibinfo
  {author} {\bibfnamefont {H.}~\bibnamefont {Reichlov{\'a}}}, \bibinfo {author}
  {\bibfnamefont {J.}~\bibnamefont {{\v{Z}}elezn{\'y}}}, \ and\ \bibinfo
  {author} {\bibfnamefont {D.}~\bibnamefont {Kriegner}},\ }\bibfield  {title}
  {\enquote {\bibinfo {title} {Spontaneous anomalous {Hall} effect arising from
  an unconventional compensated magnetic phase in a semiconductor},}\ }\href
  {\doibase 10.1103/PhysRevLett.130.036702} {\bibfield  {journal} {\bibinfo
  {journal} {Physical Review Letters}\ }\textbf {\bibinfo {volume} {130}},\
  \bibinfo {pages} {036702} (\bibinfo {year} {2023})}\BibitemShut {NoStop}%
\bibitem [{\citenamefont {Zhou}\ \emph {et~al.}()\citenamefont {Zhou},
  \citenamefont {Yan}, \citenamefont {Rong}, \citenamefont {Zhao},
  \citenamefont {Xiao}, \citenamefont {Lai}, \citenamefont {Xi}, \citenamefont
  {Wang}, \citenamefont {Adhikari}, \citenamefont {Tiwari}, \citenamefont
  {Lin}, \citenamefont {Manuel}, \citenamefont {Orlandi}, \citenamefont
  {Khalyavin}, \citenamefont {Grutter}, \citenamefont {Liu}, \citenamefont
  {Yan},\ and\ \citenamefont {Chang}}]{zhou2026surface}%
  \BibitemOpen
  \bibfield  {author} {\bibinfo {author} {\bibfnamefont {Ling-Jie}\
  \bibnamefont {Zhou}}, \bibinfo {author} {\bibfnamefont {Zi-Jie}\ \bibnamefont
  {Yan}}, \bibinfo {author} {\bibfnamefont {Hongtao}\ \bibnamefont {Rong}},
  \bibinfo {author} {\bibfnamefont {Yufei}\ \bibnamefont {Zhao}}, \bibinfo
  {author} {\bibfnamefont {Pu}~\bibnamefont {Xiao}}, \bibinfo {author}
  {\bibfnamefont {Lok-Kan}\ \bibnamefont {Lai}}, \bibinfo {author}
  {\bibfnamefont {Zhiyuan}\ \bibnamefont {Xi}}, \bibinfo {author}
  {\bibfnamefont {Ke}~\bibnamefont {Wang}}, \bibinfo {author} {\bibfnamefont
  {Tibendra}\ \bibnamefont {Adhikari}}, \bibinfo {author} {\bibfnamefont
  {Ganesh~P.}\ \bibnamefont {Tiwari}}, \bibinfo {author} {\bibfnamefont
  {Zhong}\ \bibnamefont {Lin}}, \bibinfo {author} {\bibfnamefont {Pascal}\
  \bibnamefont {Manuel}}, \bibinfo {author} {\bibfnamefont {Fabio}\
  \bibnamefont {Orlandi}}, \bibinfo {author} {\bibfnamefont {Dmitry}\
  \bibnamefont {Khalyavin}}, \bibinfo {author} {\bibfnamefont {Alexander~J.}\
  \bibnamefont {Grutter}}, \bibinfo {author} {\bibfnamefont {Chao-Xing}\
  \bibnamefont {Liu}}, \bibinfo {author} {\bibfnamefont {Binghai}\ \bibnamefont
  {Yan}}, \ and\ \bibinfo {author} {\bibfnamefont {Cui-Zu}\ \bibnamefont
  {Chang}},\ }\href@noop {} {\enquote {\bibinfo {title} {Surface-state-driven
  anomalous {Hall} effect in altermagnetic {MnTe} films},}\ }\bibinfo
  {howpublished} {arXiv (Condensed Matter: Materials Science), submitted
  2026-02-10},\ \bibinfo {note} {{DOI}: 10.48550/arXiv.2602.09363 (accessed
  2026-09-24)}\BibitemShut {NoStop}%
\bibitem [{\citenamefont {Zhou}\ \emph {et~al.}(2020)\citenamefont {Zhou},
  \citenamefont {Chang}, \citenamefont {Wang}, \citenamefont {Gui},
  \citenamefont {Xu}, \citenamefont {Yin}, \citenamefont {Guguchia},
  \citenamefont {Zhang}, \citenamefont {Chang}, \citenamefont {Lin},
  \citenamefont {Xie}, \citenamefont {Hasan},\ and\ \citenamefont
  {Jia}}]{zhou2020enhanced}%
  \BibitemOpen
  \bibfield  {author} {\bibinfo {author} {\bibfnamefont {Huibin}\ \bibnamefont
  {Zhou}}, \bibinfo {author} {\bibfnamefont {Guoqing}\ \bibnamefont {Chang}},
  \bibinfo {author} {\bibfnamefont {Guangqiang}\ \bibnamefont {Wang}}, \bibinfo
  {author} {\bibfnamefont {Xin}\ \bibnamefont {Gui}}, \bibinfo {author}
  {\bibfnamefont {Xitong}\ \bibnamefont {Xu}}, \bibinfo {author} {\bibfnamefont
  {Jia-Xin}\ \bibnamefont {Yin}}, \bibinfo {author} {\bibfnamefont {Zurab}\
  \bibnamefont {Guguchia}}, \bibinfo {author} {\bibfnamefont {Songtian~S.}\
  \bibnamefont {Zhang}}, \bibinfo {author} {\bibfnamefont {Tay-Rong}\
  \bibnamefont {Chang}}, \bibinfo {author} {\bibfnamefont {Hsin}\ \bibnamefont
  {Lin}}, \bibinfo {author} {\bibfnamefont {Weiwei}\ \bibnamefont {Xie}},
  \bibinfo {author} {\bibfnamefont {M.~Zahid}\ \bibnamefont {Hasan}}, \ and\
  \bibinfo {author} {\bibfnamefont {Shuang}\ \bibnamefont {Jia}},\ }\bibfield
  {title} {\enquote {\bibinfo {title} {Enhanced anomalous {Hall} effect in the
  magnetic topological semimetal {Co$_3$Sn$_{2-x}$In$_x$S$_2$}},}\ }\href
  {\doibase 10.1103/PhysRevB.101.125121} {\bibfield  {journal} {\bibinfo
  {journal} {Phys. Rev. B}\ }\textbf {\bibinfo {volume} {101}},\ \bibinfo
  {pages} {125121} (\bibinfo {year} {2020})}\BibitemShut {NoStop}%
\bibitem [{\citenamefont {Lohani}\ \emph {et~al.}(2023)\citenamefont {Lohani},
  \citenamefont {Foulquier}, \citenamefont {Le~F{\`e}vre}, \citenamefont
  {Bertran}, \citenamefont {Colson}, \citenamefont {Forget},\ and\
  \citenamefont {Brouet}}]{lohani2023electronic}%
  \BibitemOpen
  \bibfield  {author} {\bibinfo {author} {\bibfnamefont {Himanshu}\
  \bibnamefont {Lohani}}, \bibinfo {author} {\bibfnamefont {Paul}\ \bibnamefont
  {Foulquier}}, \bibinfo {author} {\bibfnamefont {Patrick}\ \bibnamefont
  {Le~F{\`e}vre}}, \bibinfo {author} {\bibfnamefont {Fran{\c{c}}ois}\
  \bibnamefont {Bertran}}, \bibinfo {author} {\bibfnamefont {Doroth{\'e}e}\
  \bibnamefont {Colson}}, \bibinfo {author} {\bibfnamefont {Anne}\ \bibnamefont
  {Forget}}, \ and\ \bibinfo {author} {\bibfnamefont {V{\'e}ronique}\
  \bibnamefont {Brouet}},\ }\bibfield  {title} {\enquote {\bibinfo {title}
  {Electronic structure evolution of the magnetic {Weyl} semimetal
  {Co$_3$Sn$_2$S$_2$} with hole and electron doping},}\ }\href {\doibase
  10.1103/PhysRevB.107.245119} {\bibfield  {journal} {\bibinfo  {journal}
  {Phys. Rev. B}\ }\textbf {\bibinfo {volume} {107}},\ \bibinfo {pages}
  {245119} (\bibinfo {year} {2023})}\BibitemShut {NoStop}%
\bibitem [{\citenamefont {Matsuoka}\ \emph {et~al.}(2025)\citenamefont
  {Matsuoka}, \citenamefont {Fujishiro}, \citenamefont {Minami}, \citenamefont
  {Koretsune}, \citenamefont {Ueda}, \citenamefont {Kanazawa}, \citenamefont
  {Arita}, \citenamefont {Tokura},\ and\ \citenamefont
  {Iwasa}}]{matsuoka2025electron}%
  \BibitemOpen
  \bibfield  {author} {\bibinfo {author} {\bibfnamefont {Hideki}\ \bibnamefont
  {Matsuoka}}, \bibinfo {author} {\bibfnamefont {Yukako}\ \bibnamefont
  {Fujishiro}}, \bibinfo {author} {\bibfnamefont {Susumu}\ \bibnamefont
  {Minami}}, \bibinfo {author} {\bibfnamefont {Takashi}\ \bibnamefont
  {Koretsune}}, \bibinfo {author} {\bibfnamefont {Taiga}\ \bibnamefont {Ueda}},
  \bibinfo {author} {\bibfnamefont {Naoya}\ \bibnamefont {Kanazawa}}, \bibinfo
  {author} {\bibfnamefont {Ryotaro}\ \bibnamefont {Arita}}, \bibinfo {author}
  {\bibfnamefont {Yoshinori}\ \bibnamefont {Tokura}}, \ and\ \bibinfo {author}
  {\bibfnamefont {Yoshihiro}\ \bibnamefont {Iwasa}},\ }\bibfield  {title}
  {\enquote {\bibinfo {title} {Electron-doped magnetic {Weyl} semimetal
  {Co$_3$Sn$_2$S$_2$} by bulk gating},}\ }\href {\doibase 10.1103/ggjy-5569}
  {\bibfield  {journal} {\bibinfo  {journal} {Phys. Rev. Mater.}\ }\textbf
  {\bibinfo {volume} {9}},\ \bibinfo {pages} {064406} (\bibinfo {year}
  {2025})}\BibitemShut {NoStop}%
\bibitem [{\citenamefont {Wu}\ and\ \citenamefont
  {Qing}(2025)}]{wu2025midinfrared}%
  \BibitemOpen
  \bibfield  {author} {\bibinfo {author} {\bibfnamefont {Jun}\ \bibnamefont
  {Wu}}\ and\ \bibinfo {author} {\bibfnamefont {Ye~Ming}\ \bibnamefont
  {Qing}},\ }\bibfield  {title} {\enquote {\bibinfo {title} {Mid-infrared
  broadband nonreciprocal thermal radiation involving patterned {Weyl}
  semimetal},}\ }\href {\doibase 10.1063/5.0279087} {\bibfield  {journal}
  {\bibinfo  {journal} {Appl. Phys. Lett.}\ }\textbf {\bibinfo {volume}
  {126}},\ \bibinfo {pages} {252203} (\bibinfo {year} {2025})}\BibitemShut
  {NoStop}%
\end{thebibliography}

\clearpage
\onecolumngrid
\begin{center}
\textbf{TOC Graphic}\\[1em]
\includegraphics[width=3.25in]{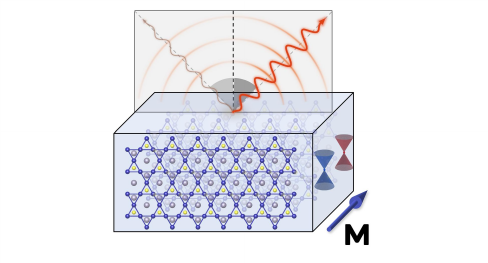}
\end{center}

\end{document}

\title{Supporting Information:\\ Designing Strong and Broadband Nonreciprocal Thermal Radiation in Magnetic Topological Materials}

\author{Yiyang Jiang}
\affiliation{Department of Physics, The Pennsylvania State University, University Park, Pennsylvania 16802, USA}
\author{Yufei Zhao}
\affiliation{Department of Physics, The Pennsylvania State University, University Park, Pennsylvania 16802, USA}
\affiliation{Department of Condensed Matter Physics, Weizmann Institute of Science, Rehovot 7610001, Israel}
\author{Linxiao Zhu}
\affiliation{Department of Mechanical Engineering, The Pennsylvania State University, University Park, Pennsylvania 16802, USA}
\author{Binghai Yan$^{*,}$}
\affiliation{Department of Physics, The Pennsylvania State University, University Park, Pennsylvania 16802, USA}
\affiliation{Department of Condensed Matter Physics, Weizmann Institute of Science, Rehovot 7610001, Israel}
\affiliation{The Center for Theory of Emergent Quantum Matter, The Pennsylvania State University, University Park, Pennsylvania 16802, USA}
\date{\today}

\begin{abstract}
\noindent$^{*}$Corresponding author. Email: binghai.yan@psu.edu\par\smallskip

This document provides the detailed derivations, tabulated material parameters, and supporting analysis for the main text.
\end{abstract}
\maketitle

\setcounter{section}{0}
\renewcommand{\thesection}{S\arabic{section}}
\renewcommand{\theequation}{S\arabic{equation}}
\renewcommand{\thefigure}{S\arabic{figure}}
\renewcommand{\thetable}{S\arabic{table}}

\section{$4\times 4$ Berreman Matrix Method for Bianisotropic and Nonreciprocal Media}\label{app:Maxwell}

To compute the nonreciprocal thermal radiation properties—namely the angle-resolved and frequency-dependent absorptivity $\alpha$, emissivity $e$, reflectivity $R$, and transmissivity $T$—of a multilayered system, we employ a generalized $4\times 4$ Berreman matrix formalism integrated with a Scattering Matrix (S-matrix) method. This approach robustly handles arbitrary optical anisotropy and bianisotropy, directly linking first-principles dielectric tensors to observable photonic quantities.

\subsection{The $4\times 4$ Eigenmode Formulation}

For an arbitrary homogeneous layer within the stratified medium, the electromagnetic response is governed by the generalized bianisotropic constitutive relations:
\begin{equation}
\begin{pmatrix} \mathbf{D}(\omega, \mathbf{r}) \\ \mathbf{B}(\omega, \mathbf{r}) \end{pmatrix} = \begin{pmatrix} \varepsilon(\omega) & \xi(\omega) \\ \zeta(\omega) & \mu(\omega) \end{pmatrix} \begin{pmatrix} \mathbf{E}(\omega, \mathbf{r}) \\ \mathbf{H}(\omega, \mathbf{r}) \end{pmatrix}
\end{equation}
where $\varepsilon, \mu, \xi$, and $\zeta$ are $3\times 3$ tensors representing the permittivity, permeability, and magneto-electric coupling. For the magnetic topological materials studied here, the primary driver of nonreciprocal behavior is the anomalous Hall conductivity originating from the band topology, which manifests as a non-Hermitian, fully populated permittivity tensor $\varepsilon(\omega)$, while $\mu = \mu_0 \mathbb{I}_{3\times3}$ and $\xi = \zeta = 0$.

Assuming continuous translational symmetry in the $x-y$ plane, the transverse momenta $k_x$ and $k_y$, along with the angular frequency $\omega$, are conserved quantities. We look for plane-wave solutions of the form $\Psi(z) \exp[i(k_x x + k_y y - \omega t)]$, where the vector $\Psi = (E_x, E_y, H_x, H_y)^T$ captures the tangential field components. Substituting this ansatz into Maxwell's equations yields the first-order differential eigenvalue problem:
\begin{equation}
\frac{\partial \Psi}{\partial z} = i \Delta \Psi
\end{equation}
where $\Delta$ is the $4\times 4$ Berreman matrix dependent on the material tensors and transverse momenta. Diagonalization of $\Delta$ yields four eigenvalues $k_{z; \nu\tau}$ and corresponding four-component eigenvectors $\Psi_{\nu\tau}$. The index $\nu \in \{u, d\}$ denotes the propagation direction (upward or downward with respect to the $z$-axis), and $\tau \in \{s, p\}$ denotes the polarization eigenstate. In gyrotropic media, the spatial dispersion degeneracy between $s$ (TE) and $p$ (TM) modes is lifted.

The total field within the layer $l$ is then expanded as a linear superposition of these eigenmodes:
\begin{equation}
\Psi^{(l)}(z) = \sum_{\tau \in \{s,p\}} \left( c_{u,\tau}^{(l)} \Psi_{u,\tau}^{(l)} e^{i k_{z;u,\tau}^{(l)} z} + c_{d,\tau}^{(l)} \Psi_{d,\tau}^{(l)} e^{i k_{z;d,\tau}^{(l)} z} \right)
\end{equation}
where $c_{\nu\tau}^{(l)}$ are complex amplitude coefficients.

\subsection{Scattering Matrix (S-Matrix) Propagation}

To cascade the field across multiple interfaces while avoiding the numerical instabilities typical of standard transfer-matrix methods, we utilize the S-matrix formalism. The electromagnetic boundary conditions enforce the continuity of the tangential fields $\Psi$ across an interface between layer $l$ and $l+1$:
\begin{equation}
\sum_{\nu\tau} c_{\nu\tau}^{(l)} \Psi_{\nu\tau}^{(l)} = \sum_{\nu\tau} c_{\nu\tau}^{(l+1)} \Psi_{\nu\tau}^{(l+1)}
\end{equation}
By sorting the components according to incoming (source) and outgoing (scattered) waves, we define the interface scattering matrix $S_{l,l+1}$:
\begin{equation}
\begin{pmatrix} \mathbf{c}_{d}^{(l)} \\ \mathbf{c}_{u}^{(l+1)} \end{pmatrix} = S_{l,l+1} \begin{pmatrix} \mathbf{c}_{u}^{(l)} \\ \mathbf{c}_{d}^{(l+1)} \end{pmatrix}
\end{equation}
where the amplitude vectors $\mathbf{c}_{\nu}$ group both $s$ and $p$ polarization states. The matrix elements of $S_{l,l+1}$ directly evaluate the polarization mixing ($s-p$ coupling) at the interface. Successive applications of the Redheffer star product over all interfaces yield the global S-matrix of the multilayer stack, relating the input fields from the exterior vacuum regions to the reflected and transmitted fields.

\subsection{Computation of Observables}

Energy conservation is rigorously enforced by normalizing the incident and scattered waves relative to the $z$-component of their respective Poynting vectors. The global S-matrix $S$ is decomposed into reflection and transmission blocks for upward and downward incidences.

The spectral reflectivity $R$ and transmissivity $T$ are obtained by squaring the absolute values of the elements within the corresponding S-matrix blocks, ensuring proper accounting for polarization conversion channels (e.g., $R_{ss}, R_{sp}, R_{ps}, R_{pp}$).

Once the S-matrix $S$ is computed, the directional absorptivity $\alpha$ and emissivity $e$ are evaluated through the generalized energy balance relations:
\begin{equation}
\alpha_\tau = \langle \mathbf{\Psi}_{u,\tau}^{(vac)} | \mathbb{I} - S^\dagger S | \mathbf{\Psi}_{u,\tau}^{(vac)} \rangle
\end{equation}
\begin{equation}
e_\tau = \langle \mathbf{\Psi}_{d,\tau}^{(vac)} | \mathbb{I} - S S^\dagger | \mathbf{\Psi}_{d,\tau}^{(vac)} \rangle
\end{equation}
where $\tau \in \{s, p\}$, and the vectors correspond to incident plane waves from the top and bottom ambient spaces, respectively. For unpolarized thermal radiation, the net nonreciprocal response is the arithmetic average over polarizations: $\bar{\alpha} = \frac{1}{2}(\alpha_s + \alpha_p)$ and $\bar{e} = \frac{1}{2}(e_s + e_p)$. The intrinsic thermal nonreciprocity is thus rigorously mapped as the contrast $\Delta(\omega, \theta) = e(\omega, \theta) - \alpha(\omega, \theta)$, capturing all off-diagonal material effects without approximations.

\section{Tabulated Material Parameters}\label{app:material_parameter}
All numerical values listed in the tables are rounded for clarity and uniform presentation.
All quantities are evaluated at an intrinsic optical scattering rate $\eta = 0.01$~eV.
The bandwidth $\Delta\omega$ is the frequency interval between the outermost points at which $|\Delta(\omega,\theta_{\max})|$ falls to half of its maximum value;
entries marked with $\dagger$ indicate that this support reaches the upper edge of the simulated
frequency window ($\omega = 0.5$~eV) and therefore represent lower bounds on the true bandwidth.

\subsection{Dependence on the chemical potential}

\label{par:si_mu_scan}
\begin{figure*}[tbp]
  \centering
  \includegraphics[width=0.7\textwidth]{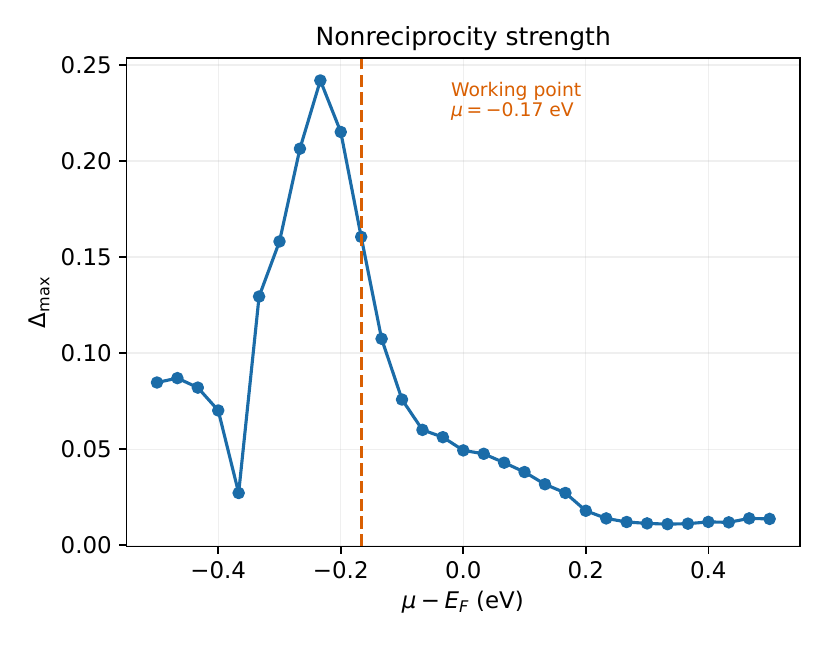}
  \caption{Chemical-potential dependence of the peak nonreciprocity $\Delta_{\max}$ in Co$_3$Sn$_2$S$_2$ at $\eta=10$~meV. The dashed line marks the main-text working point $\mu=-0.17$~eV.}
  \label{fig:si_mu}
\end{figure*}

Figure~\ref{fig:si_mu} shows the peak nonreciprocity of Co$_3$Sn$_2$S$_2$ under rigid shifts of the Fermi level at $\eta=10$~meV. At charge neutrality, $\Delta_{\max}=0.049$ and $\partial\Delta_{\max}/\partial\mu=-0.13$~eV$^{-1}$. At $\mu=-0.17$~eV, the strength rises to $0.16$ and the slope to $-1.6$~eV$^{-1}$, corresponding to a $10\%$ change per $10$~meV. The maximum strength is $0.24$ near $\mu=-0.23$~eV.

\label{par:si_mu_window}Within $\pm50$~meV of the working point, $\Delta_{\max}=0.11$--$0.22$. Electrostatic screening limits bulk tuning in Co$_3$Sn$_2$S$_2$, so chemical substitution is a possible route to explore. In Co$_3$Sn$_{2-x}$In$_x$S$_2$, In substitution lowers the Fermi level by about $60$~meV at $x\approx0.15$ (rigid-band estimate)~\cite{zhou2020enhanced} and by about $150$~meV at $x=0.44$ in ARPES, close to the working point~\cite{lohani2023electronic}, and electron doping has also been realized experimentally, with Li-ion bulk gating raising it by about $200$~meV~\cite{matsuoka2025electron}. The rigid-band sweep does not determine achievable doping precision or account for substitution-induced disorder and changes in the band structure.

Chemical potentials at which the dispersive permittivity exhibits no epsilon-near-zero crossing
within the simulated window---i.e.\ no well-defined $\omega_0$---do not enter the nonreciprocal-emission
scaling analysis and are omitted from the tables (shown as ``$-$''). For MnTe and InAs, whose ENZ response
is confined to a restricted doping window, only the chemical potentials supporting a well-defined $\omega_0$
are tabulated~\cite{nrtp_material_database_2026}.
\begin{table}[htbp]
\centering
\caption{Co$_3$Sn$_2$S$_2$~\cite{liu2018giant}}
\begin{tabular}{lccccccc}
\hline
$\mu$(eV) & -0.3 & -0.2 & -0.1 & 0.0 & 0.1 & 0.2 & 0.3 \\
\hline
$\beta$(eV${}^{-1}$) & 1280 & 981 & 798 & 1348 & 603 & 121 & - \\
$\omega_0$(eV) & 0.1088 & 0.1571 & 0.1540 & 0.1156 & 0.2058 & 0.4599 & - \\
$\mathrm{Im}\,\epsilon_{xx}$ & 17.17 & 12.64 & 55.73 & 104.10 & 70.93 & 40.69 & - \\
$\mathrm{Re}\,\epsilon_{xz}$ & -3.45 & -1.59 & -22.79 & -38.16 & -9.45 & -2.19 & - \\
$\mathrm{Im}\,\epsilon_{xz}$ & 13.43 & 15.04 & 19.81 & 45.01 & 14.68 & 4.40 & - \\
$\Delta_{\text{max}}$(\%) & 15.8 & 21.5 & 7.6 & 4.9 & 3.8 & 1.8 & - \\
$\Delta\omega$(eV) & 0.0294 & 0.0392 & 0.1470 & 0.1666 & 0.3234 & 0.3234$^\dagger$ & - \\
\hline
\end{tabular}
\end{table}
\begin{table}[htbp]
\centering
\caption{Eu$_3$In$_2$As$_4$ FMa~\cite{song2024topotaxial, zhao2024hybrid}}
\begin{tabular}{lccccccc}
\hline
$\mu$(eV) & -0.3 & -0.2 & -0.1 & 0.0 & 0.1 & 0.2 & 0.3 \\
\hline
$\beta$(eV${}^{-1}$) & 54 & 64 & 177 & 351 & 470 & 179 & 96 \\
$\omega_0$(eV) & 0.3013 & 0.2359 & 0.1158 & 0.0049 & 0.0502 & 0.0993 & 0.1645 \\
$\mathrm{Im}\,\epsilon_{xx}$ & 11.78 & 7.23 & 3.13 & 20.71 & 3.61 & 1.59 & 1.46 \\
$\mathrm{Re}\,\epsilon_{xz}$ & 2.18 & 2.96 & 0.62 & 0.77 & 0.23 & 0.29 & 0.28 \\
$\mathrm{Im}\,\epsilon_{xz}$ & 0.81 & -1.91 & -3.93 & -20.75 & -1.37 & -0.36 & -0.03 \\
$\Delta_{\text{max}}$(\%) & 7.9 & 19.2 & 35.4 & 17.3 & 15.3 & 13.8 & 6.8 \\
$\Delta\omega$(eV) & 0.1764$^\dagger$ & 0.1078 & 0.0588 & 0.0196 & 0.0196 & 0.0196 & 0.0294 \\
\hline
\end{tabular}
\end{table}
\begin{table}[htbp]
\centering
\caption{Eu$_3$In$_2$As$_4$ AFMa~\cite{song2024topotaxial, zhao2024hybrid}}
\begin{tabular}{lccccccc}
\hline
$\mu$(eV) & -0.3 & -0.2 & -0.1 & 0.0 & 0.1 & 0.2 & 0.3 \\
\hline
$\beta$(eV${}^{-1}$) & 38 & 61 & 152 & 50 & 742 & 220 & 126 \\
$\omega_0$(eV) & 0.4580 & 0.3135 & 0.1341 & 0.1352 & 0.0437 & 0.0920 & 0.1420 \\
$\mathrm{Im}\,\epsilon_{xx}$ & 4.15 & 3.30 & 7.97 & 3.94 & 5.35 & 1.97 & 1.17 \\
$\mathrm{Re}\,\epsilon_{xz}$ & -0.81 & 0.08 & 0.10 & -0.51 & -0.49 & -0.48 & -0.48 \\
$\mathrm{Im}\,\epsilon_{xz}$ & 0.33 & 0.12 & 0.31 & -0.01 & -0.08 & -0.04 & -0.04 \\
$\Delta_{\text{max}}$(\%) & 9.7 & 6.7 & 6.0 & 0.6 & 0.4 & 0.7 & 1.0 \\
$\Delta\omega$(eV) & 0.1764 & 0.2352 & 0.0392 & 0.0196 & 0.0784 & 0.0196 & 0.0098 \\
\hline
\end{tabular}
\end{table}
\begin{table}[htbp]
\centering
\caption{Eu$_3$In$_2$As$_4$ FMc~\cite{song2024topotaxial, zhao2024hybrid}}
\begin{tabular}{lccccccc}
\hline
$\mu$(eV) & -0.3 & -0.2 & -0.1 & 0.0 & 0.1 & 0.2 & 0.3 \\
\hline
$\beta$(eV${}^{-1}$) & 389 & 268 & 626 & 1123 & 485 & 183 & 101 \\
$\omega_0$(eV) & 0.1486 & 0.1147 & 0.0506 & 0.0225 & 0.0520 & 0.1024 & 0.1619 \\
$\mathrm{Im}\,\epsilon_{xx}$ & 10.82 & 8.27 & 10.58 & 10.09 & 3.91 & 1.66 & 1.31 \\
$\mathrm{Re}\,\epsilon_{xz}$ & -1.95 & -0.60 & -2.90 & 0.08 & 0.00 & -0.13 & -0.14 \\
$\mathrm{Im}\,\epsilon_{xz}$ & 7.11 & 6.13 & 10.03 & 1.02 & 0.89 & 0.42 & 0.14 \\
$\Delta_{\text{max}}$(\%) & 9.2 & 15.0 & 19.0 & 5.1 & 11.6 & 15.1 & 8.8 \\
$\Delta\omega$(eV) & 0.2156 & 0.1862 & 0.0686 & 0.1372 & 0.0196 & 0.0196 & 0.0392 \\
\hline
\end{tabular}
\end{table}
\begin{table}[htbp]
\centering
\caption{Mn$_3$Ge~\cite{nakatsuji2015large,nayak2016large,zhang2017strong}}
\begin{tabular}{lccccccc}
\hline
$\mu$(eV) & -0.3 & -0.2 & -0.1 & 0.0 & 0.1 & 0.2 & 0.3 \\
\hline
$\beta$(eV${}^{-1}$) & 549 & 1331 & 3212 & 2055 & 993 & 314 & 736 \\
$\omega_0$(eV) & 0.2823 & 0.1413 & 0.0814 & 0.1358 & 0.1924 & 0.2368 & 0.2902 \\
$\mathrm{Im}\,\epsilon_{xx}$ & 94.32 & 167.78 & 195.84 & 75.38 & 73.61 & 103.67 & 91.07 \\
$\mathrm{Re}\,\epsilon_{xz}$ & 2.39 & -11.72 & -32.26 & -1.12 & 2.53 & 5.23 & 3.89 \\
$\mathrm{Im}\,\epsilon_{xz}$ & 1.21 & 11.54 & 29.84 & 7.83 & 2.47 & 0.70 & 1.45 \\
$\Delta_{\text{max}}$(\%) & 0.3 & 0.6 & 0.8 & 0.8 & 0.8 & 0.4 & 0.3 \\
$\Delta\omega$(eV) & 0.4116$^\dagger$ & 0.2744 & 0.2352 & 0.2058 & 0.1764 & 0.2646 & 0.1764 \\
\hline
\end{tabular}
\end{table}
\begin{table}[htbp]
\centering
\caption{Mn$_3$Sn~\cite{nakatsuji2015large,nayak2016large,zhang2017strong}}
\begin{tabular}{lccccccc}
\hline
$\mu$(eV) & -0.3 & -0.2 & -0.1 & 0.0 & 0.1 & 0.2 & 0.3 \\
\hline
$\beta$(eV${}^{-1}$) & 817 & 794 & 1152 & 1688 & 1160 & 359 & 319 \\
$\omega_0$(eV) & 0.3493 & 0.2271 & 0.1458 & 0.1385 & 0.1721 & 0.2742 & 0.3715 \\
$\mathrm{Im}\,\epsilon_{xx}$ & 41.98 & 63.64 & 100.10 & 54.80 & 63.04 & 73.52 & 59.13 \\
$\mathrm{Re}\,\epsilon_{xz}$ & 1.09 & -1.25 & -5.09 & -0.84 & 1.47 & 3.19 & 1.63 \\
$\mathrm{Im}\,\epsilon_{xz}$ & 1.71 & 3.17 & 5.67 & 6.58 & 0.98 & 1.51 & 0.66 \\
$\Delta_{\text{max}}$(\%) & 0.7 & 1.6 & 0.7 & 0.6 & 0.5 & 0.5 & 0.4 \\
$\Delta\omega$(eV) & 0.2940 & 0.1372 & 0.1960 & 0.2156 & 0.3920 & 0.0882 & 0.2156$^\dagger$ \\
\hline
\end{tabular}
\end{table}
\begin{table}[htbp]
\centering
\caption{Co$_2$MnAl~\cite{Li2020Giant}}
\begin{tabular}{lccccccc}
\hline
$\mu$(eV) & -0.3 & -0.2 & -0.1 & 0.0 & 0.1 & 0.2 & 0.3 \\
\hline
$\beta$(eV${}^{-1}$) & 970 & 995 & 433 & 917 & 1213 & 723 & 675 \\
$\omega_0$(eV) & 0.2907 & 0.2655 & 0.1758 & 0.1578 & 0.2202 & 0.2797 & 0.3432 \\
$\mathrm{Im}\,\epsilon_{xx}$ & 57.21 & 64.49 & 167.91 & 169.61 & 62.21 & 44.42 & 36.18 \\
$\mathrm{Re}\,\epsilon_{xz}$ & -65.54 & -63.96 & -77.09 & -83.34 & -48.58 & -40.47 & -39.94 \\
$\mathrm{Im}\,\epsilon_{xz}$ & -12.16 & 2.43 & 18.88 & 11.30 & 7.33 & 4.52 & 1.99 \\
$\Delta_{\text{max}}$(\%) & 0.7 & 1.8 & 1.7 & 2.2 & 2.4 & 1.4 & 1.0 \\
$\Delta\omega$(eV) & 0.3528 & 0.2744 & 0.4018 & 0.3430 & 0.0686 & 0.1568 & 0.1764 \\
\hline
\end{tabular}
\end{table}
\begin{table}[htbp]
\centering
\caption{Co$_2$MnGa~\cite{Sakai2018bc, Guin2019Anomalous}}
\begin{tabular}{lccccccc}
\hline
$\mu$(eV) & -0.3 & -0.2 & -0.1 & 0.0 & 0.1 & 0.2 & 0.3 \\
\hline
$\beta$(eV${}^{-1}$) & 857 & 365 & 452 & 368 & 527 & 572 & 113 \\
$\omega_0$(eV) & 0.3047 & 0.3412 & 0.3238 & 0.2941 & 0.3086 & 0.3444 & 0.4317 \\
$\mathrm{Im}\,\epsilon_{xx}$ & 67.76 & 67.78 & 76.70 & 89.89 & 55.81 & 44.80 & 54.07 \\
$\mathrm{Re}\,\epsilon_{xz}$ & -69.17 & -65.19 & -60.43 & -55.90 & -47.71 & -43.49 & -41.92 \\
$\mathrm{Im}\,\epsilon_{xz}$ & -6.96 & -0.16 & 2.69 & 6.43 & 5.02 & 4.20 & 2.53 \\
$\Delta_{\text{max}}$(\%) & 0.7 & 1.8 & 1.5 & 2.3 & 1.6 & 1.2 & 1.0 \\
$\Delta\omega$(eV) & 0.4018$^\dagger$ & 0.0686 & 0.4312 & 0.0490 & 0.1176 & 0.1568 & 0.2450$^\dagger$ \\
\hline
\end{tabular}
\end{table}
\begin{table}[htbp]
\centering
\caption{MnTe~\cite{gonzalezbetancourt2023spontaneous, zhou2026surface}}
\begin{tabular}{lcccccc}
\hline
$\mu$(eV) & -0.233 & -0.200 & -0.167 & -0.133 & -0.100 & -0.067 \\
\hline
$\beta$(eV${}^{-1}$) & 34 & 54 & 57 & 97 & 146 & 695 \\
$\omega_0$(eV) & 0.4844 & 0.3961 & 0.3437 & 0.2897 & 0.2210 & 0.0624 \\
$\mathrm{Im}\,\epsilon_{xx}$ & 5.01 & 4.91 & 4.62 & 4.28 & 4.85 & 14.05 \\
$\mathrm{Re}\,\epsilon_{xz}$ & -1.51 & -1.47 & -1.17 & -1.71 & -1.74 & -1.45 \\
$\mathrm{Im}\,\epsilon_{xz}$ & 0.07 & 0.08 & 0.05 & 0.17 & 0.02 & 0.07 \\
$\Delta_{\text{max}}$(\%) & 3.1 & 4.3 & 4.9 & 3.3 & 2.3 & 0.7 \\
$\Delta\omega$(eV) & 0.1764 & 0.0196 & 0.0784 & 0.0882 & 0.0294 & 0.1666 \\
\hline
\end{tabular}
\end{table}
\begin{table}[htbp]
\centering
\caption{InAs ($B = 1$~T)~\cite{zhang2025observation, shayegan2024broadband}}
\begin{tabular}{lccccc}
\hline
$\mu$(eV) & 0.1 & 0.2 & 0.3 & 0.4 & 0.5 \\
\hline
$\beta$(eV${}^{-1}$) & 470 & 298 & 218 & 177 & 148 \\
$\omega_0$(eV) & 0.0469 & 0.0792 & 0.1081 & 0.1355 & 0.1620 \\
$\mathrm{Im}\,\epsilon_{xx}$ & 2.64 & 1.55 & 1.13 & 0.90 & 0.76 \\
$\mathrm{Re}\,\epsilon_{xz}$ & -0.36 & -0.10 & -0.04 & -0.02 & -0.01 \\
$\mathrm{Im}\,\epsilon_{xz}$ & -0.81 & -0.38 & -0.23 & -0.15 & -0.11 \\
$\Delta_{\text{max}}$(\%) & 14.9 & 13.1 & 11.0 & 9.3 & 8.0 \\
$\Delta\omega$(eV) & 0.0123 & 0.0147 & 0.0147 & 0.0172 & 0.0171 \\
\hline
\end{tabular}
\end{table}

\section{Surface-State Contribution}\label{app:surface}

\label{par:si_surface_dos}
\begin{figure*}[tbp]
  \centering
  \includegraphics[width=0.8\textwidth]{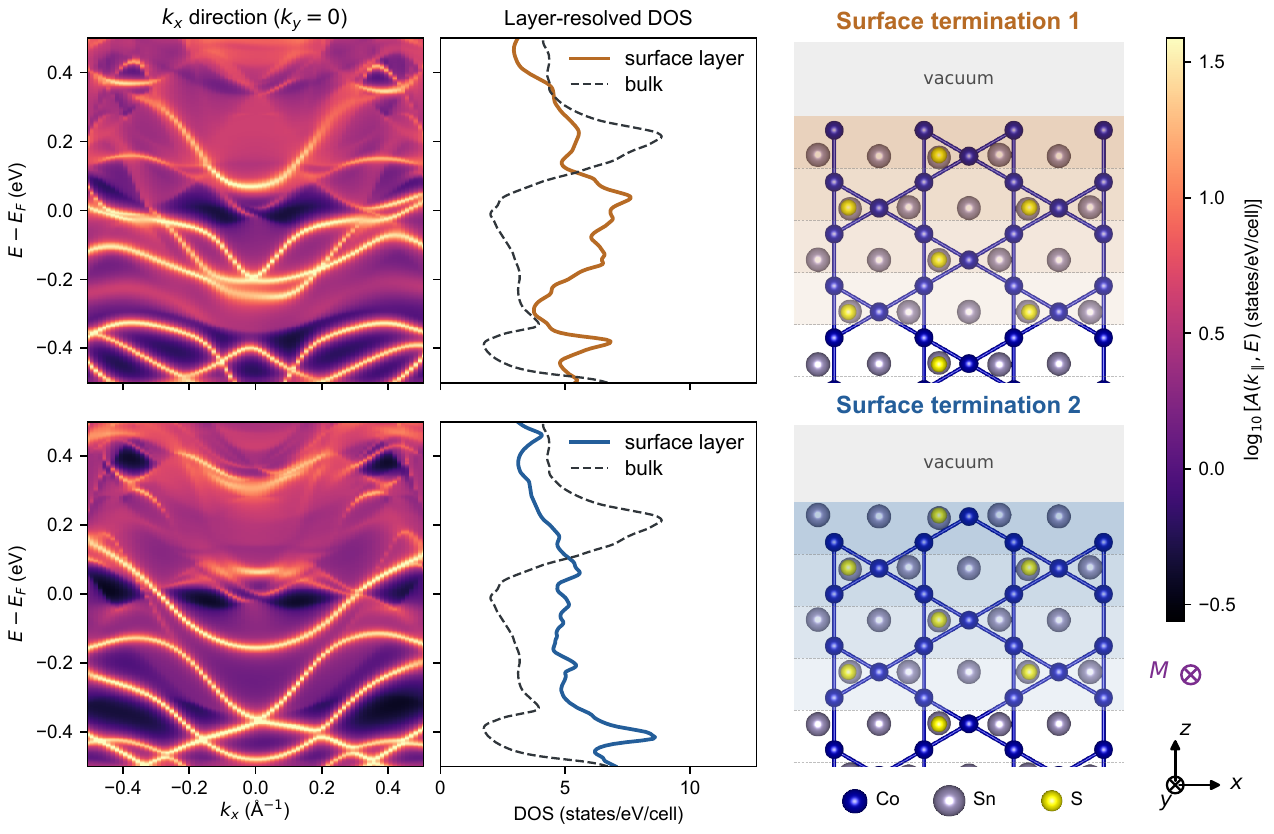}
  \caption{Surface electronic band structure of the Co$_3$Sn$_2$S$_2$ Voigt side surface from a semi-infinite surface Green's function, in the frame of Fig.~\ref{main-fig:schematic}(b) of the main text: $x$ in the surface plane, $z$ along the outward surface normal, and $\mathbf{M}$ along $y$, into the page. Top row: surface termination 1, whose outermost atomic plane contains only Co. Bottom row: surface termination 2, whose outermost plane contains Co, Sn, and S. Left: surface spectral function $A(k_x,E)$ along $k_x$ ($k_y=0$), $\eta=5$~meV. Middle: DOS of the outermost unit cell (solid) and of one bulk unit cell (dashed), integrated over the surface Brillouin zone ($64\times64$ $k$ points, $\eta=10$~meV). Right: one kagome layer viewed along $y$, with the adjacent S and Sn atoms; shading marks successive unit cells ($2.7$~\AA\ along $z$) below the surface, from the outermost cell (darkest) into the bulk.}
  \label{fig:si_surface}
\end{figure*}

We estimate the surface contribution for the Voigt side surface of Co$_3$Sn$_2$S$_2$ used in our calculations (surface normal along $z$, magnetization $\mathbf{M}$ along $y$ in the surface plane) with a semi-infinite surface Green's function built from the same Wannier Hamiltonian, for two ideal terminations (Fig.~\ref{fig:si_surface}). In the energy window probed at the ENZ frequency, the surface states raise the DOS of the outermost unit cell ($2.7$~\AA\ along the surface normal) to at most about three times the bulk value, and the enhancement decays within three to four unit cells, about $1$~nm.

\label{par:si_surface_depth}Light samples a much thicker region. At the ENZ wavelength $\lambda_0\approx7.8$~$\mu$m, $\mathrm{Re}\,\varepsilon_{xx}\approx0$ and $\varepsilon_d''=\mathrm{Im}\,\varepsilon_{xx}\approx24$, so $n\approx\sqrt{i\varepsilon_d''}=\sqrt{\varepsilon_d''/2}\,(1+i)$ and the absorbed power decays over $\delta\approx\lambda_0/(4\pi\sqrt{\varepsilon_d''/2})\approx180$~nm, about $670$ unit cells. The surface layer is therefore a negligible part of the absorbing region. This estimate assumes that the local optical response follows the local DOS; the surface optical conductivity itself is not computed.

\section{Permittivity Model for the InAs Benchmark}
\label{app:InAs}

To provide a quantitative comparison with the conventional platform,
we model the permittivity tensor of a heavily doped $n$-type
InAs-based semiconductor slab placed in an external magnetic field
$B = 1\,\mathrm{T}$.
Because the conduction band of InAs is narrow ($E_g \approx 0.35\,\mathrm{eV}$),
strong band-mixing between the conduction and valence bands causes the
electron dispersion to deviate significantly from the parabolic
approximation at the carrier concentrations ($n \sim 10^{18}\,\mathrm{cm}^{-3}$)
relevant to mid-infrared ENZ photonics. We therefore adopt the single-band non-parabolic effective-mass
model employed in Ref.~\cite{zhang2023broadband}, which
captures the dominant band-edge nonparabolicity through only the
bandgap $E_g$ and the band-edge mass $m_n$ of InAs.

\subsection*{Single-Band Non-Parabolic Effective Mass}

For a degenerate electron gas at carrier concentration $n$, the
optical effective mass at the Fermi surface follows from the
single-band non-parabolic dispersion
$E\bigl(1 + E/E_g\bigr) = \hbar^{2} k^{2}/(2 m_n)$, with Fermi
wavevector $k_F = (3\pi^{2} n)^{1/3}$. Closed-form integration
gives~\cite{zhang2023broadband}
\begin{equation}
    m^{*}(n) \;=\; m_n \sqrt{\,1
        + \frac{1}{2}\!\left(\dfrac{3}{\pi}\right)^{\!2/3}
          \dfrac{h^{2}}{E_g\,m_n}\, n^{2/3}\,},
    \label{eq:mstar_prapplied}
\end{equation}
with $h$ as Planck's constant. The corresponding chemical potential
$\mu$ is recovered from $\mu(1+\mu/E_g) = \hbar^{2}k_F^{2}/(2 m_n)$~\cite{Kane1957}.
As $n$ increases from $10^{18}$ to $4\times 10^{18}\,\mathrm{cm}^{-3}$,
$m^{*}$ rises smoothly from $\approx 0.037\,m_0$ to
$\approx 0.049\,m_0$, in agreement with experimentally extracted
effective-mass tables for heavily doped InAs
films~\cite{shayegan2024broadband}.

\subsection*{Magneto-Optical Drude Tensor}

Once $m^*$ is determined, we construct the permittivity tensor using
the magneto-optical Drude model.
For propagation in the $x$-$z$ plane with the external magnetic field
$\mathbf{B} = B\hat{y}$, the gyrotropic tensor takes the form

\begin{equation}
    \varepsilon(\omega) =
    \begin{pmatrix}
        \varepsilon_{xx} & 0 & \varepsilon_{xz} \\
        0 & \varepsilon_{yy} & 0 \\
        \varepsilon_{zx} & 0 & \varepsilon_{xx}
    \end{pmatrix},
    \qquad \varepsilon_{zx} = -\varepsilon_{xz},
\end{equation}

with components

\begin{align}
    \varepsilon_{xx}(\omega) &= \varepsilon_\infty
        - \frac{\omega_p^2\,(\omega + i\eta)}{\omega\bigl[(\omega + i\eta)^2 - \omega_c^2\bigr]},
    \label{eq:epsxx} \\[4pt]
    \varepsilon_{yy}(\omega) &= \varepsilon_\infty
        - \frac{\omega_p^2}{\omega(\omega + i\eta)},
    \label{eq:epsyy} \\[4pt]
    \varepsilon_{xz}(\omega) &=
        \frac{-i\,\omega_p^2\,\omega_c}{\omega\bigl[(\omega + i\eta)^2 - \omega_c^2\bigr]}.
    \label{eq:epsxz}
\end{align}

Here $\varepsilon_\infty$ is the high-frequency background permittivity,
$\eta$ is the electron scattering rate (half-linewidth), and
\begin{equation}
    \omega_p = \sqrt{\frac{n e^2}{\varepsilon_0\, m^*}}, \qquad
    \omega_c = \frac{eB}{m^*}
\end{equation}
are the plasma and cyclotron frequencies, respectively.
The gyrotropic component $\varepsilon_{xz}$ is the direct origin of
time-reversal breaking and nonreciprocal thermal emission in this system.
In the single-band Drude limit and for $\omega_c \ll \omega_p$, the
real part of $\varepsilon_{xx}$ crosses zero at the screened plasma
frequency $\omega_0 \approx \omega_p/\sqrt{\varepsilon_\infty}$,
producing the narrow nonreciprocal
peak seen in Fig.~2 of the main text.

\subsection*{Parameters and Validation}

The model parameters were chosen to match experimental effective-mass
tables from gradient-doped InAs films~\cite{shayegan2024broadband}, and are
summarized in Table~\ref{tab:InAs_params}.

For the representative carrier concentration
$n = 3.3\times 10^{18}\,\mathrm{cm}^{-3}$ used in
Fig.~2 — the doping employed in the
gradient-doped InAs films of
Ref.~\cite{shayegan2024broadband} — Eq.~\eqref{eq:mstar_prapplied}
yields an optical effective mass $m^{*} \approx 0.048\,m_0$,
a plasma frequency $\omega_p \approx 0.31\,\mathrm{eV}$ and a
cyclotron frequency $\omega_c \approx 2.4\,\mathrm{meV}$. The ENZ
crossing then sits at $\omega_0 \approx 0.087\,\mathrm{eV}$
($\lambda \approx 14\,\mu\mathrm{m}$), inside the mid-infrared range
of the magnetic-field-driven InAs experiments
of Refs.~\cite{shayegan2024broadband}.

We note that the intrinsic scattering rate $\eta = 0.01$~eV adopted for the
InAs benchmark is consistent with the broadening reported in recent
magneto-optical experiments on degenerate $n$-InAs. For instance, a broadening
$\hbar\Gamma \approx 3$~meV was extracted at a carrier density
$n \approx 1.5\times10^{18}$~cm$^{-3}$ in Ref.~\cite{shayegan2022nonreciprocal},
with comparable few-meV values across subsequent nonreciprocal emission and absorption studies
on the same $n$-InAs platform
\cite{shayegan2023direct,shayegan2024broadband,Liu2023Broadband}, and the
broadening grows with carrier density through impurity scattering.
Crucially, this is the same $\eta$ used in the DFT--Kubo evaluation of the
conductivity $\sigma_{ij}(\omega)$ [Eq.~(1)] and the
resulting dielectric tensor $\varepsilon_{ij}(\omega)$ for all the magnetic
topological materials, so that the InAs benchmark and the intrinsic magnetic
materials are compared on an equal footing, with no material assigned a
different broadening.

\begin{table}[htbp]
\centering
\caption{Single-band non-parabolic and Drude parameters used for the
InAs benchmark ($B = 1\,\mathrm{T}$).}
\label{tab:InAs_params}
\begin{tabular}{lll}
\hline
Parameter & Symbol & Value \\
\hline
Bandgap & $E_g$ & $0.35\,\mathrm{eV}$ \\
Band-edge effective mass & $m_n$ & $0.023\,m_0$ \\
High-freq.\ permittivity & $\varepsilon_\infty$ & $12.25$ \\
Scattering rate & $\eta$ & $0.01\,\mathrm{eV}$ \\
Magnetic field & $B$ & $1\,\mathrm{T}$ \\
\hline
\end{tabular}
\end{table}

\section{Analytical Derivations for Nonreciprocal Thermal Emission}
\label{app:factorization}

In this Supporting Information we provide the detailed derivations supporting the factorized scaling laws presented in the main text. We first derive the material--geometry factorization of the nonreciprocal contrast from the exact gyrotropic Fresnel reflectivity, then reduce the material kernel to the three-parameter ENZ model, and finally extract the closed-form strength and bandwidth scaling laws.

\subsection{Factorization of the Nonreciprocal Contrast}
\label{subsec:factorization_derivation}

The nonreciprocal directional contrast in $p$-polarized thermal emission is the absorptivity asymmetry under $\theta \to -\theta$,
\begin{equation}
\Delta(\omega,\theta) \;\equiv\;
\alpha(\omega,-\theta) - \alpha(\omega,\theta),
\label{eq:delta_def}
\end{equation}
the canonical observable for time-reversal-breaking thermal radiation~\cite{zhu2014near}. For the opaque semi-infinite half-space treated throughout this work the transmittance is identically zero and the absorptivity reduces to $\alpha(\omega,\theta) = 1 - R(\omega,\theta)$ with $R \equiv |r_{pp}|^2$, so Eq.~\eqref{eq:delta_def} rewrites as
\begin{equation}
\Delta(\omega,\theta)
= R(\omega,\theta) - R(\omega,-\theta)
= |r_{pp}(\omega,\theta)|^{2} - |r_{pp}(\omega,-\theta)|^{2}.
\label{eq:delta_from_R}
\end{equation}
All subsequent derivations in this appendix operate on the factorized form $\Delta(\omega,\theta) \approx 8\,f(\omega)\,G(\kappa,\theta)$ [main-text Eq.~(7), where the proportionality constant is made explicit]. In this subsection we derive that factorization from the exact Fresnel reflectivity of the gyrotropic half-space through a controlled sequence of five approximations, each with an explicit small parameter and quantified fidelity against the exact $\Delta$.

\paragraph*{Step 1: Exact reflectivity and parity identity.}
Solving Maxwell's equations in the semi-infinite half-space with $z$-component Poynting-vector normalization gives the gyrotropic reflection coefficient $r_{pp}(\theta) = [A - (B_1 + B_2)]/[A + (B_1 + B_2)]$, with the auxiliary amplitudes $A = (\varepsilon_d^2 - g^2)\cos\theta/\varepsilon_d$, $B_1 = \sqrt{(\varepsilon_d^2 - g^2)/\varepsilon_d - \sin^2\theta}$, and $B_2 = (i g/\varepsilon_d)\sin\theta$. Here $A$ and $B_1$ are even in $\theta$, while $B_2$ is odd. Writing $R(\theta) = |r_{pp}(\theta)|^2 = (\Sigma - \Pi)/(\Sigma + \Pi)$ with the reflectivity intermediates $\Sigma \equiv |A|^2 + |B|^2$, $\Pi \equiv 2\,\mathrm{Re}(A B^*)$, $B \equiv B_1 + B_2$, and separating each into parity-definite pieces $\Sigma = \Sigma_1 + \Sigma_2$, $\Pi = \Pi_1 + \Pi_2$ (subscript $1$ = even in $\theta$, subscript $2$ = odd), the parity decomposition follows directly from $A,B_1$ even and $B_2$ odd in $\theta$:
\begin{equation}
\begin{split}
\Sigma_1 &= |A|^2 + |B_1|^2 + |B_2|^2, \quad \Sigma_2 = 2\,\mathrm{Re}(B_1 B_2^*), \\
\Pi_1 &= 2\,\mathrm{Re}(A B_1^*), \qquad\;\;\; \Pi_2 = 2\,\mathrm{Re}(A B_2^*).
\end{split}
\label{eq:parity_pieces}
\end{equation}
Substituting Eq.~\eqref{eq:parity_pieces} into Eq.~\eqref{eq:delta_from_R} yields the \emph{exact} parity identity
\begin{equation}
\Delta(\omega,\theta)
\;=\; \frac{4\,(\Sigma_2\Pi_1 - \Sigma_1\Pi_2)}{(\Sigma_1 + \Pi_1)^{2} - (\Sigma_2 + \Pi_2)^{2}}.
\label{eq:exact_parity}
\end{equation}
Eq.~\eqref{eq:exact_parity} is exact, and no approximation has been made. The numerator vanishes at $\theta = 0$ because the odd pieces $\Sigma_2,\Pi_2$ are zero there, so $\Delta$ necessarily vanishes at normal incidence, where no transverse direction is distinguished. We stress that the field eigenmodes here are normalized to carry unit time-averaged $z$-directed Poynting (energy) flux, so that the reflectivity is $R = |r_{pp}|^2$ directly.

\paragraph*{Step 2: High-index + small-gyrotropy approximation.}
Two generic material inequalities hold for magnetic topological materials probed near the ENZ point: (A) $|n_v|^{2} \gg \sin^{2}\theta$ with $n_v \equiv \sqrt{(\varepsilon_d^{2} - g^{2})/\varepsilon_d}$, and (B) $|n_v|^{2} \gg |\gamma|^{2}$ with $\gamma \equiv i g/\varepsilon_d$. Condition (A) allows us to replace $B_1 \to n_v$, so that only $A\propto\cos\theta$ and $B_2\propto\sin\theta$ carry angular parity; condition (B) removes a subleading $|\gamma|^{2}\sin^{2}\theta$ term from $\Sigma_1$. Direct evaluation of the four parity pieces then gives $\Sigma_1 \approx |n_v|^{4}\cos^{2}\theta + |n_v|^{2}$, $\Sigma_2 = 2\,\mathrm{Re}(n_v\gamma^{*})\sin\theta$, $\Pi_1 = 2|n_v|^{2}\mathrm{Re}(n_v)\cos\theta$, and $\Pi_2 = 2\,\mathrm{Re}(n_v^{2}\gamma^{*})\sin\theta\cos\theta$. These reproduce the exact $\Delta$ through Eq.~\eqref{eq:exact_parity} to within an absolute deviation $\lesssim 10^{-2}$ over the experimentally relevant angular range $\theta \in [5^\circ, 85^\circ]$. Condition (A) fails only in the extreme-grazing skin $\theta \gtrsim \cos^{-1}(1/|n_v|)$ (typically $\theta > 88^\circ$ for the benchmark materials); condition (B) is equivalent to $|x| \equiv |g|/\varepsilon_d'' < 1$ and remains benign even where it is marginal, because the downstream scaling laws saturate.

\begin{figure*}[tbp] \centering \includegraphics[width=\textwidth]{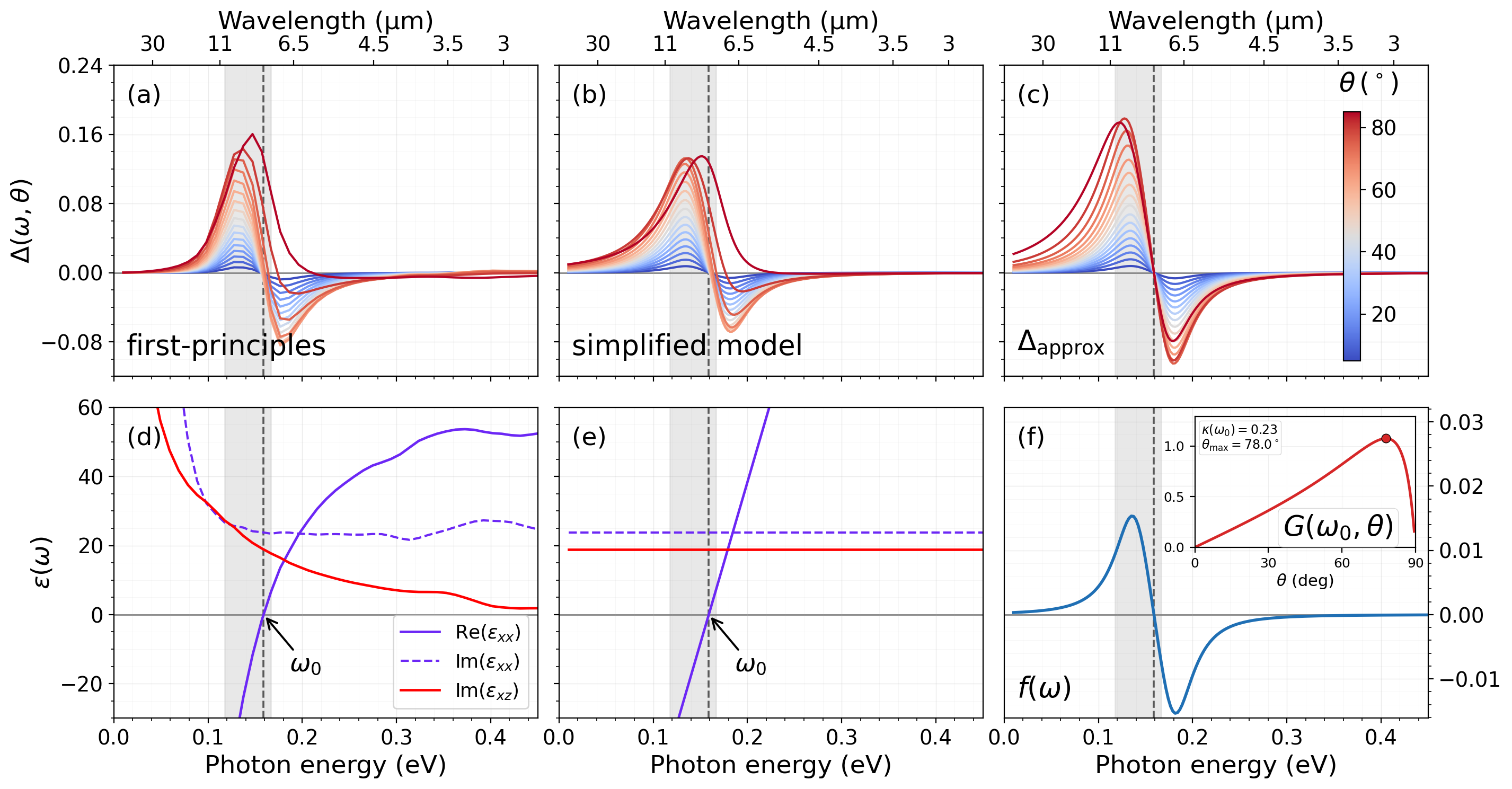} \caption{Numerical validation of the factorization $\Delta_{\text{approx}}(\omega,\theta) = 8\,f(\omega)\,G(\kappa,\theta)$ [Eq.~\eqref{eq:factorization_stable}] on Co$_3$Sn$_2$S$_2$ at $\mu = -0.17$~eV, $\eta = 0.01$~eV, with purely imaginary off-diagonal gyrotropy ($\mathrm{Re}\,\varepsilon_{xz} = 0$). \textbf{Top row:} $\Delta(\omega,\theta) = R(+\theta) - R(-\theta)$ computed from (a) the Berreman $4\times 4$ transfer-matrix solver on the first-principles dielectric tensor of panel~(d); (b) the same solver on the linearized 3-parameter ENZ model of panel~(e), with $\mathrm{Re}\,\varepsilon_d(\omega) = \beta(\omega - \omega_0)$ and $\mathrm{Im}\,\varepsilon_d$, $\varepsilon_{xz}$ frozen at $\omega_0$; (c) the analytical $\Delta_{\text{approx}}$ formula Eq.~\eqref{eq:factorization_stable} fed by the same simplified $\varepsilon$ as panel~(b). \textbf{Bottom row:} (d) first-principles $\varepsilon$ of Co$_3$Sn$_2$S$_2$; (e) simplified-model $\varepsilon$ used as input to panels (b) and~(c); (f) the material shape factor $f(\omega) = - \mathrm{Re}(n_v^2\gamma^{*})/|n_v|^4$ on the simplified $\varepsilon$, with the angular shape function $G(\kappa,\theta) = \sin\theta\cos\theta/(\cos\theta + \kappa)^2$ plotted as an inset ($\kappa \approx 0.23$ at $\omega_0$, analytic peak $\theta_{\max} \approx 78^\circ$). The dashed vertical line in panels (a)--(e) marks the ENZ frequency $\omega_0 \approx 0.159$~eV; the gray band marks the half-maximum interval $0.118$--$0.167$~eV, corresponding to $\Delta\omega=0.049$~eV for panel~(a). The colorbar inside panel~(c) encodes the incidence angle $\theta \in [5^\circ, 85^\circ]$ in $5^\circ$ steps. The three reductions preserve the spectral shape, the peak positions, and the bandwidth of the nonreciprocal feature. The peak amplitudes are $\max|\Delta_{\text{FP}}| = 0.161$~(a), $\max|\Delta_{\text{simplified}}| = 0.135$~(b), and $\max|\Delta_{\text{approx}}| = 0.179$~(c): the parameter-freezing step (a$\to$b) and the analytical factorization (b$\to$c) shift the peak amplitude in opposite directions and partially compensate, so the end-to-end analytical $\Delta_{\text{approx}}$ reproduces the first-principles peak to within $\sim 10\%$ on a realistic magnetic topological metal.}
\label{fig:fp_vs_approx_doublepeak}
\end{figure*}

\paragraph*{Step 3: Dominant-denominator reduction.}
In Eq.~\eqref{eq:exact_parity}, the even pieces $\Sigma_1,\Pi_1$ dominate the odd pieces $\Sigma_2,\Pi_2$ away from normal incidence, and the cross-term $\Sigma_1\Pi_2$ dominates $\Sigma_2\Pi_1$ in the numerator whenever $|n_v|^{2}\cos^{2}\theta > 1$. Dropping $(\Sigma_2 + \Pi_2)^{2}$ from the denominator and $\Sigma_2\Pi_1$ from the numerator,
\begin{equation}
\Delta(\omega,\theta) \;\approx\;
- \frac{4\,\Sigma_1\,\Pi_2}{(\Sigma_1 + \Pi_1)^{2}}.
\label{eq:Delta1_raw}
\end{equation}
This compact form tracks the exact $\Delta$ closely across $\theta \in [5^\circ, 85^\circ]$ --- reproducing the two-peak structure, the peak positions, and the bandwidth --- although the peak amplitude can differ by up to tens of percent (see Fig.~\ref{fig:fp_vs_approx_doublepeak} and the discussion at the end of Step~5). The angular dependence still lives in both numerator and denominator and the form is not yet factorized.

\paragraph*{Step 4: Linearization and factorization --- $\Delta_{\tan}$.}
Wherever $|n_v|^{2}\cos^{2}\theta \gg 1$, the subleading $|n_v|^{2}$ piece of $\Sigma_1$ is small, so $\Sigma_1 \to |n_v|^{4}\cos^{2}\theta$. In the same limit the term $\kappa \equiv 2\mathrm{Re}(n_v)/|n_v|^{2}$ is small, making $\Pi_1/\Sigma_1 = \kappa/\cos\theta \ll 1$, so the $\Pi_1$ term may also be dropped from the denominator of Eq.~\eqref{eq:Delta1_raw}. The residual factors $|n_v|^{4}\cos^{2}\theta$ in numerator and denominator cancel exactly, producing the completely factorized form

\begin{equation}
\Delta_{\tan}(\omega,\theta)
\;=\; 8\,f(\omega)\,\tan\theta,
\qquad
f(\omega) \;\equiv\;
-\frac{\mathrm{Re}\!\left[n_v^{2}(\omega)\,\gamma^{*}(\omega)\right]}{|n_v(\omega)|^{4}}.
\label{eq:Delta_tan}
\end{equation}
Equation~\eqref{eq:Delta_tan} is the algebraic linchpin of the entire analysis: the material and geometric variables \emph{separate}, and every downstream scaling-law statement operates on $f(\omega)$ alone. The price paid for this clean factorization is that $\tan\theta$ diverges at grazing --- an algebraic artefact of dropping $\Pi_1$, not a physical statement, since the exact Eq.~\eqref{eq:exact_parity} guarantees $\Delta \to 0$ at $\theta = \pi/2$.

\paragraph*{Step 5: Grazing-stabilized angular shape function.}
Restoring $\Pi_1$ in the denominator of Eq.~\eqref{eq:Delta1_raw} while keeping the linearization $\Sigma_1 \to |n_v|^{4}\cos^{2}\theta$ in the numerator gives
\begin{equation}
\begin{split}
\Delta(\omega,\theta)
&\;\approx\; -\frac{4\,|n_v|^{4}\cos^{2}\theta\,\Pi_2}{\bigl(|n_v|^{4}\cos^{2}\theta + \Pi_1\bigr)^{2}} \\
&\;=\; 8\,f(\omega)\,
        \frac{\sin\theta\,\cos\theta}{(\cos\theta + \kappa)^{2}},
\end{split}
\end{equation}
where $\kappa \equiv 2\,\mathrm{Re}(n_v)/|n_v|^{2}$ and the last equality follows from $\Pi_1/(|n_v|^{4}\cos^{2}\theta) = \kappa/\cos\theta$. Defining the dimensionless \emph{angular shape function}
\begin{equation}
\begin{split}
G(\kappa,\theta) &\;\equiv\;
  \frac{\sin\theta\,\cos\theta}{(\cos\theta + \kappa)^{2}}, \\
\Delta_{\text{approx}}(\omega,\theta) &\;\equiv\;
  8\,f(\omega)\,G(\kappa,\theta).
\end{split}
\label{eq:factorization_stable}
\end{equation}
Equation~\eqref{eq:factorization_stable} is the form cited as Eq.~(7) of the main text (there written $G(\theta)$, with the same $\kappa = 2\,\mathrm{Re}(n_v)/|n_v|^{2}$). It preserves the material--geometry separation of Eq.~\eqref{eq:Delta_tan} and vanishes at grazing, $G(\kappa,\pi/2) = 0$, simply because the numerator factor $\sin\theta\cos\theta \to 0$ while the denominator $\to \kappa^{2} > 0$; the stabilization parameter $\kappa$ thus removes the unphysical $\tan\theta$ divergence of Eq.~\eqref{eq:Delta_tan}. The reduction reproduces the \emph{spectral shape}, the peak \emph{positions}, and the \emph{bandwidth} of the exact $\Delta$ over the full angular window $\theta \in [5^\circ, 85^\circ]$; the peak \emph{amplitude} is reproduced to within tens of percent, with the residual coming from the parameter-freezing and factorization steps (quantified in Fig.~\ref{fig:fp_vs_approx_doublepeak}). The benchmark is Co$_3$Sn$_2$S$_2$ at $\mu = -0.17$~eV, which has purely imaginary off-diagonal gyrotropy ($\mathrm{Re}\,\varepsilon_{xz} = 0$). The stabilization parameter $\kappa$ is a material scalar evaluated at $\omega_0$; its weak $\omega$ dependence is subleading at the scaling-analysis order of interest.

Equations~\eqref{eq:Delta_tan}--\eqref{eq:factorization_stable} together establish that the nonreciprocal contrast $\Delta(\omega,\theta)$ cleanly separates into a material kernel $f(\omega)$ --- developed in the 3-parameter ENZ model of the next subsection --- and an angular shape function $G(\kappa,\theta)$ whose closed-form extremization is deferred to Sec.~\ref{subsec:geometric_extremization}.

\subsection{The 3-Parameter ENZ Model}
\label{subsec:enz_model}

To analytically track the macroscopic observables, we define a reduced 3-parameter material model centered around the epsilon-near-zero (ENZ) frequency $\omega_0$ where $\mathrm{Re}\,\varepsilon_{xx}(\omega_0) = 0$. The dielectric tensor response is fully parameterized by three scalars evaluated at $\omega_0$:
\begin{enumerate}
\item \emph{ENZ crossing slope} $\beta$: the dielectric dispersion $\beta = \left.\partial_\omega\,\mathrm{Re}\,\varepsilon_{xx}\right|_{\omega_0}$, governing the linear detuning $\varepsilon_d' \approx \beta(\omega-\omega_0)$.
\item \emph{Optical loss} $\varepsilon_d''$: the imaginary part of the diagonal permittivity, $\varepsilon_d'' = \mathrm{Im}\,\varepsilon_{xx}(\omega_0)$, treated as constant near ENZ.
\item \emph{Anomalous Hall gyrotropy} $g$: the off-diagonal term $\varepsilon_{xz} = ig$.
\end{enumerate}

\subsection{Derivation of the Material Figure of Merit}
\label{subsec:material_fom}

The nonreciprocal contrast factorizes as $\Delta \approx 8\,f(\omega)\,G(\kappa,\theta)$ [Eq.~\eqref{eq:factorization_stable}], so the material dependence is carried entirely by $f(\omega) = -\mathrm{Re}(n_v^2 \gamma^*)/|n_v|^4$. Evaluating this for real gyrotropy $g$ (so $\varepsilon_{xz} = ig$ is purely imaginary) gives the closed form
\begin{equation}
f(\varepsilon_d', \varepsilon_d'', g) =
-\frac{2 g \varepsilon_d' \varepsilon_d''}{(\varepsilon_d'^2 + \varepsilon_d''^2)^2 + g^4
       + 2g^2(\varepsilon_d''^2 - \varepsilon_d'^2)}.
\label{eq:fDP_main}
\end{equation}
Extremizing Eq.~\eqref{eq:fDP_main} with respect to the real detuning $\varepsilon_d'$ yields the optimal detuning condition
\begin{equation}
\varepsilon_{d, \text{opt}}' = \pm \frac{\varepsilon_d''}{\sqrt{3}}
\sqrt{x^2 - 1 + 2\sqrt{x^4 + x^2 + 1}},
\label{eq:peak_positions_main}
\end{equation}
where $x = g/\varepsilon_d''$ is the normalized gyrotropy (the full extremization is given in Sec.~\ref{subsec:bandwidth}). Substituting this root back into $f$ provides the absolute theoretical ceiling for the material figure of merit,
\begin{equation}
\begin{split}
f_{\max}(x, \varepsilon_d'') &\;=\; \frac{\Phi(x)}{\varepsilon_d''},
\qquad S \;\equiv\; \sqrt{x^4 + x^2 + 1}, \\
\Phi(x) &\;\equiv\; \frac{3\sqrt{3}\, x\, \sqrt{2S + x^2 - 1}}{4\,(1 + 4x^2 + x^4 + S - x^2 S)}.
\end{split}
\label{eq:fmax}
\end{equation}
The dimensionless envelope $\Phi(x)$ has two limits that control the strength scaling below. For \emph{moderate} gyrotropy it is linear in the Hall response,
\begin{equation}
\Phi(x) \;\xrightarrow{\,x \lesssim 1\,}\; \frac{3\sqrt{3}}{8}\,x,
\qquad\text{i.e.}\qquad
f_{\max} \;\approx\; \frac{3\sqrt{3}}{8}\,\frac{g}{(\varepsilon_d'')^{2}},
\label{eq:fmax_smallx}
\end{equation}
so that $f_{\max}$ carries an explicit factor of $g$; only in the strong topological limit does it saturate, $\Phi(x) \to 1/2$, i.e.\ $f_{\max}\,\varepsilon_d'' \to 0.5$ for $x \gg 1$, a limit in which the peak amplitude ceases to grow with the gyrotropy. The linear branch Eq.~\eqref{eq:fmax_smallx} is the one relevant to the screened materials and is what sources the $g$ in the strength scaling law of Sec.~\ref{subsec:geometric_extremization}.

\subsection{Derivation of the Bandwidth Scaling Law}
\label{subsec:bandwidth}

The peak amplitude of the previous subsection fixes how \emph{strong} the nonreciprocal response can be; the \emph{width} of the same peak in frequency fixes how \emph{broadband} it is. The geometric shape function $G(\kappa,\theta)$ is set almost entirely by the dielectric values in the immediate neighborhood of $\omega_0$ and may therefore be treated as approximately frequency-independent across the narrow nonreciprocal window, so that the spectral width and shape of the directional contrast $\Delta(\omega,\theta)$ are governed by the material dispersion function $f(\omega)$. (Retaining the residual $\omega$ dependence of $G$ through $\kappa(\omega)$ introduces only a slight imbalance between the two otherwise symmetric spectral peaks; this correction is small and does not affect the scaling conclusions below.) We derive the scaling law in three logically separable steps: first the \emph{reduction} of the general $f(\omega)$ to a three-parameter kernel, then the characteristic \emph{width} of that kernel in the detuning variable $\varepsilon_d'$, and finally the linear \emph{conversion} from $\varepsilon_d'$ to frequency $\omega$ through the ENZ-crossing slope $\beta$.

\paragraph*{Step 0: the 3-parameter reduction of $f(\omega)$.}
The factorized form derived in Sec.~\ref{subsec:factorization_derivation} gives $\Delta_{\mathrm{approx}}(\omega,\theta) = 8\, f(\omega)\, G(\kappa,\theta)$ with the general material kernel
\begin{equation}
\begin{split}
f(\omega) &\;=\; -\frac{\mathrm{Re}\!\left[n_v^{2}(\omega)\,\gamma^{*}(\omega)\right]}{\lvert n_v(\omega)\rvert^{4}}, \\
n_v &\;\equiv\; \sqrt{\frac{\varepsilon_d^{2} - g^{2}}{\varepsilon_d}},
\qquad
\gamma \;\equiv\; \frac{i\,g}{\varepsilon_d}.
\end{split}
\label{eq:f_general}
\end{equation}
Two independent reductions collapse Eq.~\eqref{eq:f_general} to the three scalars of Sec.~\ref{subsec:enz_model}.

\emph{(i) Real gyrotropy.} For the intrinsic anomalous-Hall response of the magnetic topological materials screened here, the off-diagonal permittivity is purely imaginary, $\varepsilon_{xz} = ig$ with $g \in \mathbb{R}$ (equivalently $\mathrm{Re}\,\varepsilon_{xz} = 0$). Substituting into Eq.~\eqref{eq:f_general} and expanding gives the reduced kernel
\begin{equation}
f(\varepsilon_d', \varepsilon_d'', g)
\;=\; -\frac{2\,g\,\varepsilon_d'\,\varepsilon_d''}{\bigl(\varepsilon_d'^{\,2} + \varepsilon_d''^{\,2}\bigr)^{\!2}
             + g^{4}
             + 2 g^{2}\!\left(\varepsilon_d''^{\,2} - \varepsilon_d'^{\,2}\right)}.
\label{eq:fDP}
\end{equation}

\emph{(ii) ENZ linearization.} Near the ENZ frequency $\omega_0$ defined by $\varepsilon_d'(\omega_0) = 0$, a Taylor expansion of $\mathrm{Re}\,\varepsilon_{xx}$ gives
\begin{equation}
\varepsilon_d'(\omega) \;\approx\; \beta\,(\omega - \omega_0),
\qquad
\beta \;\equiv\;
\partial_\omega\,\mathrm{Re}\,\varepsilon_{xx}\big|_{\omega_0},
\label{eq:enz_linearization}
\end{equation}
a one-to-one affine map between detuning and frequency on the Taylor-neighborhood of $\omega_0$ where the peaks sit. Over the same narrow window, $\varepsilon_d''$ and $g$ vary only at subleading order and are \emph{frozen} at their values at $\omega_0$. The full material kernel therefore reduces to a rational function of one detuning variable, controlled by three scalar parameters:
\begin{equation}
f(\omega) \;\longrightarrow\;
f\!\bigl(\beta(\omega - \omega_0),\,\varepsilon_d''(\omega_0),\,g(\omega_0)\bigr).
\label{eq:3param}
\end{equation}
This three-parameter model, Eq.~\eqref{eq:3param}, is the starting point for all closed-form scaling analysis in this appendix and in the main text. It is valid when the ENZ crossing is clean and isolated and the nonreciprocal feature is confined to a narrow window $\lvert\omega - \omega_0\rvert$ small compared with the dispersion scale of $\varepsilon_d''$ and $g$, conditions met by the screened materials at the dopings considered here.

\paragraph*{Step 1a: compact form of the reduced shape factor.}
The denominator of Eq.~\eqref{eq:fDP} simplifies once one collects the $\varepsilon_d'^{\,2}$-dependent terms as a single square plus a constant remainder. Writing
\begin{equation}
(\varepsilon_d'^{\,2} + \varepsilon_d''^{\,2})^2 + g^4
+ 2 g^2 (\varepsilon_d''^{\,2} - \varepsilon_d'^{\,2})
= \bigl(\varepsilon_d'^{\,2} + \varepsilon_d''^{\,2} - g^2\bigr)^{\!2}
   + 4 g^2 \varepsilon_d''^{\,2},
\label{eq:den_compact}
\end{equation}
which one verifies directly by expanding the square on the right, we arrive at the compact form
\begin{equation}
\begin{split}
f(\varepsilon_d', \varepsilon_d'', g)
&\;=\; -\frac{2\,g\,\varepsilon_d''\,\varepsilon_d'}{A(\varepsilon_d')^{\,2} + 4 g^2 \varepsilon_d''^{\,2}}, \\
A(\varepsilon_d') &\;\equiv\; \varepsilon_d'^{\,2} + \varepsilon_d''^{\,2} - g^2.
\end{split}
\label{eq:shape_compact}
\end{equation}
Two structural properties follow immediately: (i) $f$ is an \emph{odd} function of $\varepsilon_d'$ (since $A$ depends only on $\varepsilon_d'^{\,2}$), so every extremum has a partner of equal magnitude and opposite sign --- the origin of the symmetric two-peak pattern; (ii) $f$ vanishes at $\varepsilon_d' = 0$, i.e.\ precisely at the ENZ crossing $\omega = \omega_0$, so the nonreciprocal feature always sits \emph{off} resonance.

\paragraph*{Step 1b: extremum condition.}
Taking the derivative of Eq.~\eqref{eq:shape_compact} and using $\partial A/\partial\varepsilon_d' = 2\varepsilon_d'$,
\begin{equation}
\frac{\partial f}{\partial \varepsilon_d'}
= -\frac{2 g\, \varepsilon_d''}{\bigl[A^2 + 4 g^2 \varepsilon_d''^{\,2}\bigr]^{2}}
   \Bigl\{ A^2 + 4 g^2 \varepsilon_d''^{\,2}
          - 4\,\varepsilon_d'^{\,2} A \Bigr\}.
\end{equation}
Setting the braces to zero (for $g,\varepsilon_d''\neq 0$) gives the extremum condition
\begin{equation}
A^2 - 4 \varepsilon_d'^{\,2} A + 4 g^2 \varepsilon_d''^{\,2} = 0,
\label{eq:quadratic_in_A}
\end{equation}
a quadratic in $A$. Solving and re-substituting $A = \varepsilon_d'^{\,2} + \varepsilon_d''^{\,2} - g^2$ yields
\begin{equation}
\varepsilon_d''^{\,2} - g^2 - \varepsilon_d'^{\,2}
\;=\; \pm\, 2\sqrt{\,\varepsilon_d'^{\,4} - g^2 \varepsilon_d''^{\,2}\,}.
\label{eq:pre_square}
\end{equation}
Squaring Eq.~\eqref{eq:pre_square} eliminates the ambiguous sign and yields a clean quadratic in $u\equiv\varepsilon_d'^{\,2}$,
\begin{equation}
3 u^{2}
+ 2(\varepsilon_d''^{\,2} - g^2)\,u
- (\varepsilon_d''^{\,2} + g^2)^{2} \;=\; 0,
\label{eq:quadratic_in_u}
\end{equation}
whose discriminant collapses to a compact form:
\begin{equation}
\begin{split}
&\bigl[2(\varepsilon_d''^{\,2} - g^2)\bigr]^{2}
+ 12 \,(\varepsilon_d''^{\,2} + g^2)^{2} \\
&\qquad \;=\; 16\!\left(\varepsilon_d''^{\,4} + g^2\varepsilon_d''^{\,2} + g^4\right).
\end{split}
\end{equation}

\paragraph*{Step 1c: closed-form peak positions.}
The physical (positive) root of Eq.~\eqref{eq:quadratic_in_u} is
\begin{equation}
u = \varepsilon_d'^{\,2}
\;=\; \frac{1}{3}\!\left[\,(g^2 - \varepsilon_d''^{\,2})
            + 2\sqrt{\varepsilon_d''^{\,4} + g^2\varepsilon_d''^{\,2} + g^4}\,\right],
\end{equation}
which, in terms of the dimensionless figure of merit $x \equiv g/\varepsilon_d''$, cleans up to $u = (\varepsilon_d''^{\,2}/3)\left[x^2 - 1 + 2\sqrt{x^4 + x^2 + 1}\,\right]$. Taking the square root and keeping both signs (both physical extrema) reproduces Eq.~\eqref{eq:peak_positions_main},
\begin{equation}
\varepsilon_{d,\mathrm{opt}}'
\;=\; \pm\, \frac{\varepsilon_d''}{\sqrt{3}}
    \sqrt{x^{2} - 1 + 2\sqrt{x^{4} + x^{2} + 1}}.
\label{eq:peak_positions}
\end{equation}
Substituting Eq.~\eqref{eq:peak_positions} back into Eq.~\eqref{eq:shape_compact} gives the peak \emph{value} $f_{\max}(x)$ of Eq.~\eqref{eq:fmax}, closing the self-consistency check between this subsection and Sec.~\ref{subsec:material_fom}.

\paragraph*{Step 1d: peak-to-peak width and asymptotics.}
The two extrema in Eq.~\eqref{eq:peak_positions} sit symmetrically about the ENZ resonance. The peak-to-peak separation in detuning space is therefore
\begin{equation}
\begin{split}
\Delta\varepsilon_d'
&\;\equiv\; 2\,|\varepsilon_{d,\mathrm{opt}}'|
\;=\; \varepsilon_d''\, W(x), \\
W(x) &\;\equiv\; \frac{2}{\sqrt{3}}
         \sqrt{x^{2} - 1 + 2\sqrt{x^{4} + x^{2} + 1}},
\end{split}
\label{eq:width_detuning}
\end{equation}
where the dimensionless width function $W(x)$ collects all dependence on the normalized gyrotropy. Its two asymptotic regimes follow by direct expansion of the nested radical:
\begin{equation}
W(x)\;\longrightarrow\;
\begin{cases}
\dfrac{2}{\sqrt{3}}\!\left[\,1 + \mathcal{O}(x^{2})\,\right],
   & x \ll 1 \quad \text{(loss-set)},\\[6pt]
2x\,\!\left[\,1 + \mathcal{O}(x^{-2})\,\right],
   & x \gg 1 \quad \text{(gyrotropy-set)}.
\end{cases}
\label{eq:W_asymptotes}
\end{equation}
\label{par:si_W_limit}Thus $W(x)$ approaches a constant for $|x|\ll1$ and grows as $2|x|$ for $|x|\gg1$.

\paragraph*{Step 2: conversion to frequency.}
Applying the ENZ linearization of Eq.~\eqref{eq:enz_linearization} to Eq.~\eqref{eq:width_detuning} --- a one-to-one affine map --- the detuning width translates rigidly into the frequency width $\Delta\omega = \Delta\varepsilon_d'/\beta$, yielding the bandwidth scaling law
\begin{equation}
\Delta\omega \;=\; \frac{\varepsilon_d''\, W(x)}{\beta}
\;\propto\; \frac{\varepsilon_d''}{\beta},
\label{eq:bandwidth_scaling_appendix}
\end{equation}
which establishes Eq.~\eqref{main-eq:bandwidth_scaling} of the main text. The dimensionless factor $W(x)$ is bounded below by $2/\sqrt{3} \approx 1.155$ for any physical material and varies smoothly between the two asymptotes in Eq.~\eqref{eq:W_asymptotes}; it therefore never changes the sign or order-of-magnitude of the leading dependence on $\varepsilon_d''/\beta$.

\subsection{Geometric Extremization and Total Scaling Bounds}
\label{subsec:geometric_extremization}

The full macroscopic nonreciprocity combines the material maximum with the angular shape function $G(\kappa,\theta) = \sin\theta\cos\theta/(\cos\theta + \kappa)^2$. Setting $\partial G/\partial\theta = 0$ gives the optimal angle and the closed-form geometric maximum $G_{\max}(\kappa)$. Evaluated at the ENZ point ($\varepsilon_d' = 0$, $\varepsilon_d = i\varepsilon_d''$), the stabilization parameter is, \emph{exactly},
\begin{equation}
\kappa \;=\; \sqrt{\frac{2}{\varepsilon_d''(1 + x^2)}}.
\label{eq:kappa_enz}
\end{equation}
Thus $\kappa$ is small when the effective loss $\varepsilon_d''(1+x^2)$ is large and vice versa; at moderate gyrotropy this tracks $\varepsilon_d''$ itself.

The total peak contrast is $\Delta_{\max} = 8\,f_{\max}(x,\varepsilon_d'')\,G_{\max}(\kappa)$. Using the linear-in-$g$ branch Eq.~\eqref{eq:fmax_smallx}, $f_{\max} \approx \tfrac{3\sqrt{3}}{8}\,g/(\varepsilon_d'')^2$ --- the regime relevant to the screened materials, in which the amplitude has not yet saturated --- and bounding $G_{\max}(\kappa)$ in its two limits gives:
\begin{enumerate}
\item \emph{High-loss limit} ($\varepsilon_d'' \gg 1$, $\kappa \to 0$): $G_{\max} \approx 1/(4\kappa) \propto \sqrt{\varepsilon_d''}$. Combined with $f_{\max} \propto g/(\varepsilon_d'')^2$, the net peak contrast scales as
      \begin{equation}
      \Delta_{\max} \;\propto\;
        \sqrt{\varepsilon_d''}\,\cdot\,\frac{g}{(\varepsilon_d'')^{2}}
        \;=\; \frac{g}{(\varepsilon_d'')^{3/2}}.
      \label{eq:strength_highloss}
      \end{equation}
\item \emph{Low-loss limit} ($\varepsilon_d'' \ll 1$, $\kappa \to \infty$): $G_{\max} \approx 1/(2\kappa^2) \propto \varepsilon_d''$. Combined with the same $f_{\max} \propto g/(\varepsilon_d'')^2$, one power of the loss cancels, leaving
      \begin{equation}
      \Delta_{\max} \;\propto\;
        \varepsilon_d''\,\cdot\,\frac{g}{(\varepsilon_d'')^{2}}
        \;=\; \frac{g}{\varepsilon_d''}.
      \label{eq:strength_lowloss}
      \end{equation}
\end{enumerate}
Note that it is the explicit $g$ of the unsaturated branch Eq.~\eqref{eq:fmax_smallx} --- not the saturated value $f_{\max}\varepsilon_d'' \to 1/2$ --- that carries the anomalous Hall response through to the result; in the saturated strong-gyrotropy regime the amplitude no longer grows with $g$ and these proportionalities cease to hold. Equations~\eqref{eq:strength_highloss}--\eqref{eq:strength_lowloss} prove that the macroscopic nonreciprocity scales as
\begin{equation}
\Delta_{\max} \;\propto\; \frac{g}{(\varepsilon_d'')^{p}},
\qquad p \in [1,\, 3/2],
\label{eq:strength_bounds}
\end{equation}
which is Eq.~(8) of the main text: suppressing optical loss is essential for strong nonreciprocal emission, the exponent interpolating between the low-loss semiconductor limit ($p \to 1$) and the high-loss metal limit ($p \to 3/2$).

Comparing Eq.~\eqref{eq:strength_bounds} with the bandwidth law Eq.~\eqref{eq:bandwidth_scaling_appendix} exposes the fundamental asymmetry between the two performance targets: optical loss \emph{degrades} the peak amplitude but \emph{enlarges} the operational bandwidth ($\Delta\omega \propto \varepsilon_d''/\beta$). Crucially, the third parameter $\beta$ --- set by the band structure and doping level, and largely decoupled from $g$ and $\varepsilon_d''$ --- is an independent design knob: tuning $\beta$ downward broadens the response without any penalty in peak amplitude, providing the design freedom needed to break the apparent strength-versus-bandwidth trade-off in realistic magnetic topological materials.

\section{Outliers of the Scaling Laws}\label{app:outliers}

\begin{figure*}[tbp]
  \centering
  \includegraphics[width=\linewidth]{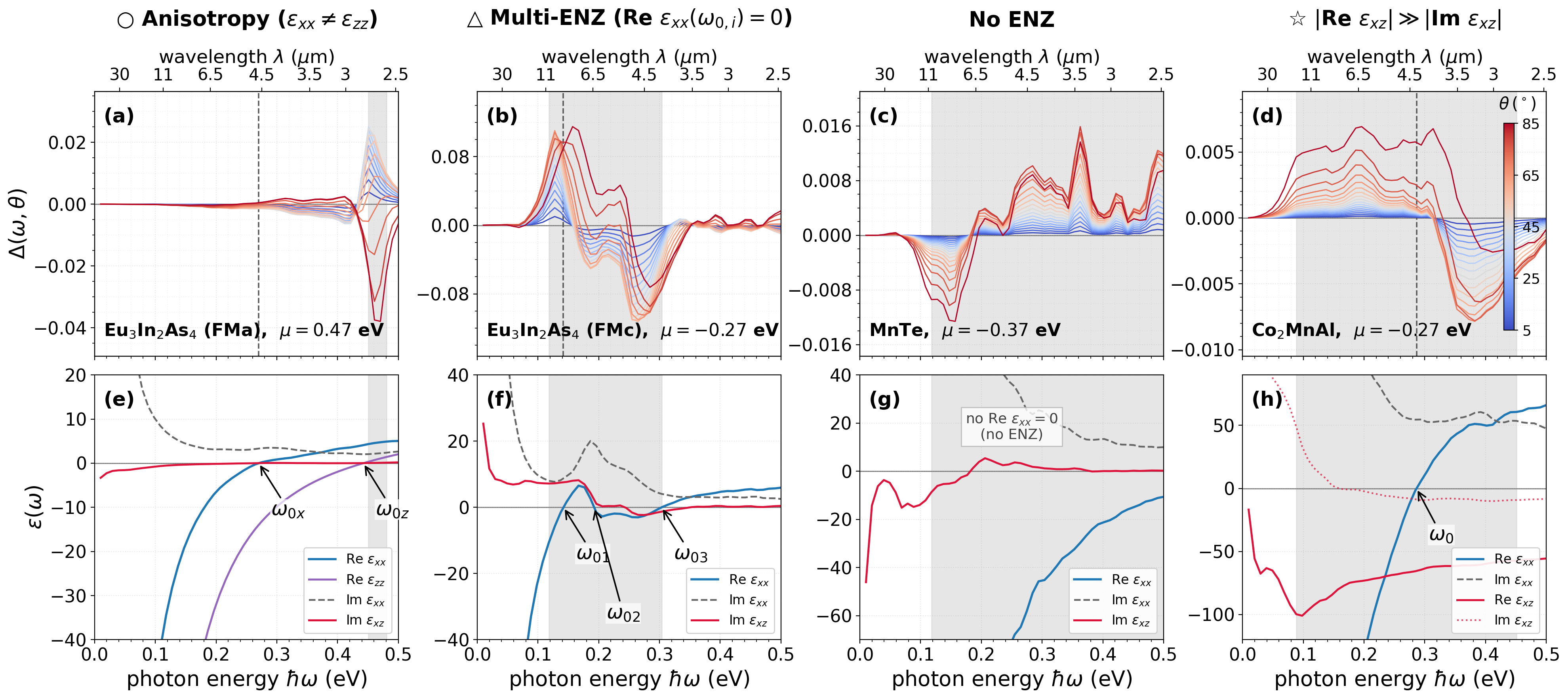}
  \caption{Representative outliers of the scaling collapse, one per deviation mechanism, in the same format as Fig.~2. Top row: angle-resolved nonreciprocity $\Delta(\omega,\theta)$ (color encodes incidence angle $5^\circ$--$85^\circ$); bottom row: the corresponding dielectric tensor with $\mathrm{Re}\,\varepsilon_{xx}$ (blue), the optical loss $\mathrm{Im}\,\varepsilon_{xx}$ (dashed grey), and the off-diagonal gyrotropy (crimson). Short arrows mark the ENZ crossings. (a,e)~Eu$_3$In$_2$As$_4$ (FMa), $\mu=+0.47$~eV --- diagonal optical anisotropy: the overlaid $\mathrm{Re}\,\varepsilon_{zz}$ (purple) crosses zero at $\omega_{0z}\approx 0.44$~eV, distinct from $\omega_{0x}\approx 0.27$~eV, and the nonreciprocity band ($0.45$--$0.48$~eV) sits near the former. (b,f)~Eu$_3$In$_2$As$_4$ (FMc), $\mu=-0.27$~eV --- multiple ENZ: $\mathrm{Re}\,\varepsilon_{xx}$ crosses zero three times ($\omega_{01},\omega_{02},\omega_{03}=0.142,0.193,0.304$~eV). (c,g)~MnTe, $\mu=-0.37$~eV --- no ENZ: $\mathrm{Re}\,\varepsilon_{xx}$ stays negative throughout, so no $\omega_0$ exists. (d,h)~Co$_2$MnAl, $\mu=-0.27$~eV --- real-dominated gyrotropy: the off-diagonal response is carried by $\mathrm{Re}\,\varepsilon_{xz}$ (solid crimson), with $\mathrm{Im}\,\varepsilon_{xz}$ (faint dotted) nearly an order of magnitude smaller at $\omega_0\approx 0.29$~eV. Each column title is tagged with the open symbol ($\bigcirc$, $\triangle$, \ding{73}) that marks the corresponding point in Fig.~3; the no-ENZ column (c) is untagged, having no scaling coordinate.}
  \label{fig:outlier_spectra}
\end{figure*}

The strength and bandwidth scaling laws of the main text [Eqs.~(8) and~(9)] follow from the minimal three-parameter gyrotropic model of Eq.~(5), which rests on three idealizing assumptions: (i) the off-diagonal anomalous-Hall response is \emph{purely imaginary}, $\varepsilon_{xz}=ig$ with $g=\mathrm{Im}\,\varepsilon_{xz}\in\mathbb{R}$ (the double-peak regime), so that the strength figure of merit collapses onto $x=g/\varepsilon_d''=\mathrm{Im}\,\varepsilon_{xz}/\mathrm{Im}\,\varepsilon_{xx}$; (ii) the diagonal response is \emph{optically isotropic} in the relevant plane, $\varepsilon_{xx}=\varepsilon_{zz}$, so that a single epsilon-near-zero (ENZ) frequency $\omega_0$ governs the resonance; and (iii) $\mathrm{Re}\,\varepsilon_{xx}(\omega)$ has a \emph{single, clean} ENZ crossing inside the radiation window, defining a unique $\omega_0$ and a finite dispersion slope $\beta$. The large majority of the screened materials satisfy these conditions and collapse onto the slope-one reference lines of Fig.~3. The minority that deviate do so through one of four physically transparent mechanisms, each violating a single assumption above, and presented here in the order of the columns of Fig.~\ref{fig:outlier_spectra}: (a)~diagonal optical anisotropy, (b)~multiple ENZ crossings, (c)~the absence of an ENZ, and (d)~a real-part-dominated off-diagonal response. Three of them---(a), (b), and (d)---carry well-defined scaling coordinates and are flagged with matching red open symbols ($\bigcirc$, $\triangle$, \ding{73}) in Fig.~3; the no-ENZ case (c) has no coordinate and is omitted from that figure. These deviations are not failures of the underlying physics but signatures of richer optical response that the minimal three-parameter model does not capture.

\paragraph*{(a) Diagonal optical anisotropy, $\varepsilon_{xx}\neq\varepsilon_{zz}$ ($\bigcirc$).}
The reduced model assigns a single diagonal permittivity $\varepsilon_d$ to all three Cartesian directions, so that one ENZ frequency---the zero of $\mathrm{Re}\,\varepsilon_{xx}$---sets the resonance. In magnetically and structurally anisotropic crystals the in-plane and out-of-plane diagonal components cross zero at \emph{different} frequencies. The hybrid topological semimetal Eu$_3$In$_2$As$_4$ in its FMa phase at $\mu=+0.47$~eV is representative: $\mathrm{Re}\,\varepsilon_{xx}$ vanishes at $\omega_{0x}\approx 0.27$~eV while $\mathrm{Re}\,\varepsilon_{zz}$ vanishes only at $\omega_{0z}\approx 0.44$~eV, and the nonreciprocity band ($0.45$--$0.48$~eV) sits near the latter [Fig.~\ref{fig:outlier_spectra}(a,e)]. Because the model anchors $\omega_0$ at the $\varepsilon_{xx}$ crossing, where the off-diagonal response is small, the plotted figure of merit $x=\mathrm{Im}\,\varepsilon_{xz}/\mathrm{Im}\,\varepsilon_{xx}$ is anomalously small even though the realized $\Delta_{\max}\approx 3.8\%$ is moderate; the point therefore floats well above the strength line in Fig.~3(a). Anisotropy thus signals that the single-axis figure of merit, evaluated at the wrong ENZ, underestimates the true nonreciprocal capacity of the material.

\paragraph*{(b) Multiple ENZ crossings of $\mathrm{Re}\,\varepsilon_{xx}$ ($\triangle$).}
Even within a single diagonal component, strong dielectric dispersion from overlapping interband features can drive $\mathrm{Re}\,\varepsilon_{xx}(\omega)$ through zero more than once inside the radiation window. The Eu$_3$In$_2$As$_4$ family is the clearest example: in the FMc phase at $\mu=-0.27$~eV, $\mathrm{Re}\,\varepsilon_{xx}$ crosses zero at $0.142$, $0.193$, and $0.304$~eV [Fig.~\ref{fig:outlier_spectra}(b,f)], and related multi-crossing structure recurs across the Eu phases and dopings. Each crossing seeds its own nonreciprocal feature, so the effective response is a superposition of resonances whose combined width far exceeds that of the single ENZ on which the reduced model anchors $\omega_0$; the point consequently lies above the bandwidth line in Fig.~3(b). Multi-ENZ structure is the dominant bandwidth-enhancing deviation in our set and identifies the Eu$_3$In$_2$As$_4$ family as a route to intrinsically broadband nonreciprocity.

\paragraph*{(c) Absence of an ENZ crossing (unmarked).}
At sufficiently high hole doping the diagonal permittivity of some materials---e.g.\ the altermagnet MnTe at $\mu=-0.37$~eV---remains negative across the \emph{entire} simulated window: $\mathrm{Re}\,\varepsilon_{xx}$ (and $\mathrm{Re}\,\varepsilon_{zz}$) never cross zero [Fig.~\ref{fig:outlier_spectra}(c,g)]. With no genuine plasma edge there is no well-defined $\omega_0$ and hence no dispersion slope $\beta$, so the reduced-model coordinates $x$ and $\varepsilon_d''/\beta$ are simply undefined. Such points still exhibit a finite nonreciprocity ($\Delta_{\max}\approx 1.6\%$ for MnTe), generated by the off-diagonal $\varepsilon_{xz}$ through the full Berreman response rather than by an ENZ resonance; they are physical but lie outside the ENZ-anchored framework and are therefore omitted from both panels of Fig.~3.

\paragraph*{(d) Real-dominated gyrotropy, $|\mathrm{Re}\,\varepsilon_{xz}|\gg|\mathrm{Im}\,\varepsilon_{xz}|$ (\ding{73}).}
The strength abscissa $x=\mathrm{Im}\,\varepsilon_{xz}/\mathrm{Im}\,\varepsilon_{xx}$ keeps only the imaginary part of the off-diagonal component, because the double-peak model assumes a purely imaginary $\varepsilon_{xz}=ig$. In the Heusler ferromagnets Co$_2$MnAl and Co$_2$MnGa the off-diagonal response near $\omega_0$ is instead dominated by its \emph{real} part---the single-peak regime $g=-i\varepsilon_o'$ with $\varepsilon_o'=-\mathrm{Re}\,\varepsilon_{xz}$ [cf.\ Sec.~\ref{app:factorization}]. For Co$_2$MnAl at $\mu=-0.27$~eV, $|\mathrm{Re}\,\varepsilon_{xz}|\approx 65$ at $\omega_0\approx 0.29$~eV, an order of magnitude larger than $|\mathrm{Im}\,\varepsilon_{xz}|\approx 9$ [Fig.~\ref{fig:outlier_spectra}(d,h)]. The imaginary-only abscissa therefore mismeasures the true off-diagonal drive, and the amplitude is governed by the single-peak kernel $f_{\rm SP}$ of Sec.~\ref{app:factorization} rather than the double-peak law; combined with the high metallic loss ($\varepsilon_d''\gtrsim 60$), $\Delta_{\max}$ is suppressed and the point falls below the strength line in Fig.~3(a), while its large loss places it among the broadest-band materials on panel~(b). Treating these materials in the single-peak regime, with the full $|\varepsilon_{xz}|$, restores their description.

\paragraph*{Summary.}
The four mechanisms map directly onto the three model assumptions. Diagonal anisotropy~(a) and real-dominated gyrotropy~(d) act on the \emph{strength} panel: (a) mis-anchors $\omega_0$ in an optically anisotropic crystal and displaces the point above the line, while (d) measures only the imaginary part of a real-dominated off-diagonal response and, with high loss, falls below it. Multiple ENZ crossings~(b) act on the \emph{bandwidth} panel, broadening the response above the line, and the absence of an ENZ~(c) removes the point from the collapse entirely. Every deviation is traceable to a specific, identifiable breakdown of assumptions~(i)--(iii), so the scaling laws of Eqs.~(8)--(9) remain robust pre-screening filters whose residuals are themselves physically interpretable diagnostics of where richer optical response can be exploited.

\section{Strength--Bandwidth Trade-off}\label{app:tradeoff}

\label{par:si_product}In the weak-gyrotropy limit, Eqs.~\eqref{main-eq:strength_scaling} and~\eqref{main-eq:bandwidth_scaling} combine into $\Delta_{\max}\Delta\omega\propto(g/\beta)(\varepsilon_d'')^{1-p}$. Optical loss therefore enlarges the product for $p<1$ and reduces it for $p>1$. Whether loss acts as a bandwidth resource or as a net penalty thus depends on the exponent, which the analytic bracket $1\le p\le3/2$ does not fix.

\label{par:si_peff}We fit the exponent $p$ of each material phase from the full Berreman calculation, at four chemical potentials with a single ENZ crossing. Table~\ref{tab:peff} lists the median and range of the fitted $p_{\mathrm{eff}}$ for each phase; over all $36$ cases the median is $1.15$ and the interquartile range $0.97$--$1.43$. The phases with median $p_{\mathrm{eff}}$ outside the analytic bracket, Eu$_3$In$_2$As$_4$ (FMa and FMc) and Co$_2$MnAl, are the outliers of Sec.~\ref{app:outliers}, which violate the assumptions of the reduced model.

\begin{table}[tbp]
\centering
\caption{Fitted exponent $p_{\mathrm{eff}}$ for each material phase, from four chemical potentials with a single ENZ crossing at $\eta=10$~meV.}
\label{tab:peff}
\begin{tabular}{lcc}
\hline
Material & median $p_{\mathrm{eff}}$ & range \\
\hline
Co$_3$Sn$_2$S$_2$ & 1.35 & 1.04--1.79 \\
Eu$_3$In$_2$As$_4$ (FMa) & 0.82 & 0.63--1.13 \\
Eu$_3$In$_2$As$_4$ (AFMa) & 1.03 & 0.59--2.33 \\
Eu$_3$In$_2$As$_4$ (FMc) & 2.08 & 0.73--2.57 \\
Co$_2$MnAl & 0.77 & 0.23--1.02 \\
Co$_2$MnGa & 1.02 & 0.35--1.60 \\
Mn$_3$Ge & 1.19 & 1.08--1.29 \\
Mn$_3$Sn & 1.49 & 1.38--1.72 \\
MnTe & 1.19 & 1.11--1.34 \\
\hline
\end{tabular}
\end{table}

\begin{figure*}[tbp]
  \centering
  \includegraphics[width=0.7\textwidth]{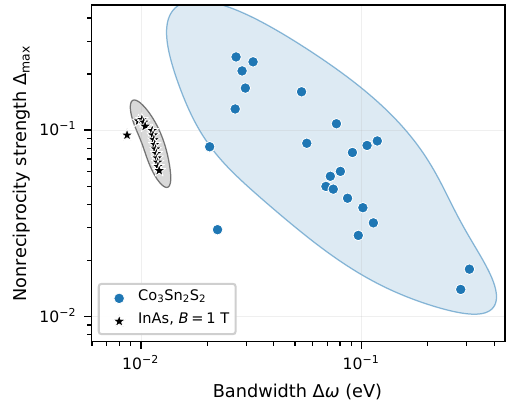}
  \caption{Strength--bandwidth trade-off of Co$_3$Sn$_2$S$_2$ (circles) and InAs under an external magnetic field of $1$~T (stars) across chemical potentials. For both materials $\Delta_{\max}$ decreases as $\Delta\omega$ increases; Co$_3$Sn$_2$S$_2$ is broader than InAs at every chemical potential and, near the working point, also stronger.}
  \label{fig:si_tradeoff}
\end{figure*}

\label{par:si_tradeoff_data}Figure~\ref{fig:si_tradeoff} compares Co$_3$Sn$_2$S$_2$ with InAs under an external magnetic field of $1$~T in the strength--bandwidth plane. Each point is one chemical potential at $\eta=10$~meV; chemical potentials without an ENZ crossing are omitted, leaving $23$ points for Co$_3$Sn$_2$S$_2$ and $15$ for InAs. $\Delta\omega$ is the half-maximum bandwidth at $\theta_{\max}$ defined in the main text. It is evaluated on a $0.05$~meV frequency grid for InAs and, for Co$_3$Sn$_2$S$_2$, on a $0.25$~meV grid after cubic-spline interpolation of the Kubo permittivity. Shading marks $80\%$ density contours in logarithmic coordinates. Within each material, $\Delta_{\max}$ falls as $\Delta\omega$ grows, with Spearman rank correlations of $-0.58$ for Co$_3$Sn$_2$S$_2$ and $-0.92$ for InAs; this is the trade-off expressed by Eqs.~\eqref{main-eq:strength_scaling} and~\eqref{main-eq:bandwidth_scaling}.

\label{par:si_tradeoff_compare}Between the two materials, Co$_3$Sn$_2$S$_2$ is broader at every chemical potential: its $\Delta\omega$ spans $0.021$--$0.31$~eV, whereas InAs stays within $0.0086$--$0.0121$~eV. From $\mu=-0.33$ to $-0.17$~eV, which includes the main-text working point, Co$_3$Sn$_2$S$_2$ is also stronger than every InAs point, with $\Delta_{\max}$ up to $0.25$ compared with at most $0.11$ for InAs. Consequently, the product $\Delta_{\max}\Delta\omega$ of Co$_3$Sn$_2$S$_2$ is about $5.0$ times that of InAs in median and $9.0$ times at best, and at the working point it is $7.5$ times the best InAs value.

\begin{figure*}[tbp]
  \centering
  \includegraphics[width=\textwidth]{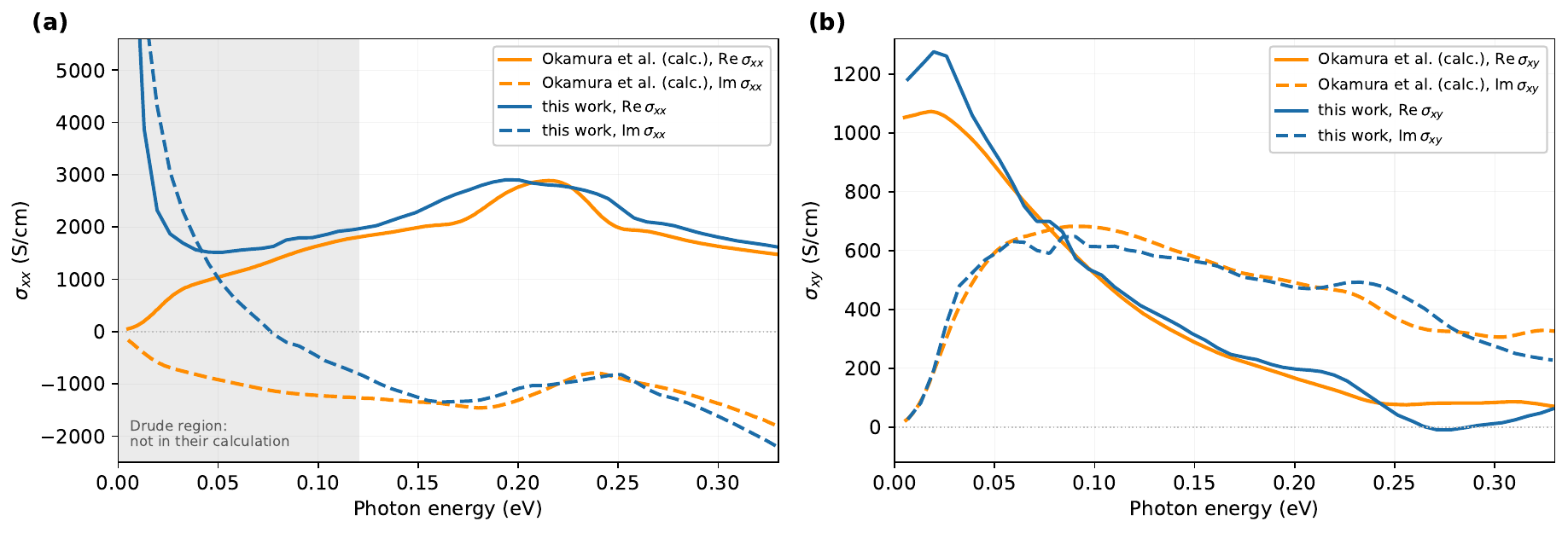}
  \caption{Calculated optical conductivity of Co$_3$Sn$_2$S$_2$ at nominal composition: (a) $\sigma_{xx}$ and (b) $\sigma_{xy}$. Blue: this work ($\eta=10$~meV); orange: the calculation of Ref.~\cite{okamura2020giant}, digitized and replotted. Solid and dashed lines denote real and imaginary parts. Our photon energies are divided by $1.52$ for this comparison only. The shaded region in (a) marks the Drude range, which their interband-only calculation does not include. Orange curves adapted from Ref.~\cite{okamura2020giant}. Available under a CC-BY 4.0 license. Copyright 2020 Y. Okamura et al.}
  \label{fig:si_benchmark}
\end{figure*}

\section{Computational Details}\label{app:computational}

The spin-polarized DFT calculations use VASP with the projector augmented-wave method~\cite{kresse1999from}, PBE exchange and correlation~\cite{perdew1996generalized}, spin--orbit coupling, and a $k$-point density of $0.1$~\AA$^{-1}$. Dudarev corrections~\cite{dudarev1998electron} use $U=4$~eV for Mn~$d$ in MnTe and $U=7$~eV for Eu~$f$ in Eu$_3$In$_2$As$_4$. Wannier90 constructs the maximally localized Wannier functions~\cite{souza2001maximally,mostofi2014wannier90}. The Co$_3$Sn$_2$S$_2$ model contains $58$ functions on $543$ lattice vectors, with $a=5.367$~\AA{}, $c=13.172$~\AA{}, and $E_F=8.3433$~eV relative to the archived Hamiltonian energy zero.

The Kubo conductivity is evaluated on a $96^3$ $k$-mesh with $\eta=10$~meV, which plays the role of the finite-temperature quasiparticle scattering rate and thus carries the thermal broadening of the optical response. Chemical potentials span $\mu-E_F\in[-0.5,+0.5]$~eV in $31$ steps. Conductivity-to-permittivity conversion follows the main text; the optical response is calculated with the Berreman method of Sec.~\ref{app:Maxwell}.

\section{Comparison with Published Optical Responses}\label{app:benchmark}

Figure~\ref{fig:si_benchmark} compares our longitudinal and Hall optical conductivities at nominal composition ($\mu-E_F=0$, $\eta=10$~meV) with the calculation of Okamura \emph{et al.}~\cite{okamura2020giant}. Their calculation divides photon energies by a renormalization factor of $1.52$ to account for electron correlation. We do not apply this factor elsewhere, so that all materials are compared on the same unrenormalized DFT energy scale; only in this figure is our photon-energy axis divided by $1.52$ to place our curve on theirs, with conductivity amplitudes unchanged. The rms differences in $\mathrm{Re}\,\sigma_{xy}$ and $\mathrm{Im}\,\sigma_{xy}$ are $48$ and $47$~S/cm over $0.12$--$0.33$~eV on that common axis.

\label{par:si_benchmark_xx}Figure~\ref{fig:si_benchmark}(a) shows the longitudinal response. Their calculation contains only interband transitions, whereas ours also includes the intraband (Drude) term, so below about $0.12$~eV our $\sigma_{xx}$ rises toward the Drude limit while theirs vanishes. Above this range the two calculations share the same spectral features, with the $\mathrm{Re}\,\sigma_{xx}$ maximum at $0.19$ and $0.21$~eV respectively, and agree in value to $279$ and $195$~S/cm rms for the real and imaginary parts, about $15\%$ of the signal.

Agreement with measured spectra is at the level of magnitude rather than lineshape. The measured $\mathrm{Re}\,\sigma_{xy}$ changes sign near $0.21$~eV, which neither calculation reproduces. Our scans over $\eta=1$--$100$~meV and $\mu-E_F=\pm0.5$~eV do not reproduce this feature either. This comparison therefore does not establish quantitative agreement with the measured Hall spectrum or validate the predicted nonreciprocal emission.

Our material predictions retain the original DFT energy scale. The unrescaled bands agree with the unrenormalized calculation in Ref.~\cite{okamura2020giant} to $29$~meV rms. The main-text working point at $\mu=-0.17$~eV remains a rigid-band extrapolation from nominal composition.

\section{Bare-Interface Contrast, Application Metrics, and Photonic Structuring}\label{app:device_scope}

Near-complete nonreciprocal contrast has been predicted for InAs photonic structures under an external magnetic field~\cite{zhu2014near,Zhao2019Near}. These results illustrate the role of structural design: a low-loss dielectric resonance, its overlap with the magnetic material, and its radiative coupling jointly determine the response. Consequently, comparing our bare substrates directly with such optimized InAs devices would conflate material properties with different photonic geometries. To compare materials on a common footing, we instead use the same bare, semi-infinite geometry, without gratings, cavities, or additional photonic layers, for both the magnetic topological materials and the InAs benchmark. The resulting strength and bandwidth trends characterize this reference geometry. Our results identify candidate materials and material-selection principles.

For approaching the thermodynamic limits in energy harvesting, such as solar cells, thermophotovoltaics, and harvesting of outgoing thermal radiation, a unity contrast $|\Delta|=1$ between emissivity and absorptivity is needed; for nonreciprocity-based sensing and infrared camouflage, a high $|\Delta|$ is desired. These applications motivate the material study. Reaching $|\Delta|\approx1$ requires the emissivity or absorptivity to approach unity, which typically calls for critical coupling, where the intrinsic decay rate from material loss equals the external decay rate~\cite{zhu2014near}. Promising magnetic topological materials may be used with photonic structures such as gratings, guided-mode resonators, and thin films to achieve near-unity nonreciprocity contrast~\cite{zhao2020axion,wu2025midinfrared,Wu2021Nearcomplete,zhang2023broadband}.

\renewcommand{\bibsection}{\section*{References}}%
%\bibliography{reference_photonics,references-chiral,nonlinear-literature,nonlinear-reference}